\documentclass{mn2e}

\usepackage{amsfonts}
\usepackage{amsmath}
\usepackage{graphicx}

\title[Environmental Dependence of Bars and Bulges]
      {Galaxy Zoo: The Environmental Dependence of Bars and Bulges in Disc Galaxies}%\thanks{This publication has been made possible by the participation of more than 100,000 volunteers in the Galaxy Zoo project.  Their contributions are individually acknowledged at \texttt{http://www.galaxyzoo.org/Volunteers.aspx}.}
\author[R. A. Skibba, K. L. Masters, R. C. Nichol, et al.]
 {Ramin A. Skibba$^1$\thanks{E-mail: rskibba@as.arizona.edu}, Karen L. Masters$^{2,3}$, 
Robert C. Nichol$^{2,3}$, Idit Zehavi$^4$, \newauthor 
Ben Hoyle$^{5,6}$, Edward M. Edmondson$^2$, Steven P. Bamford$^7$, Carolin N. Cardamone$^{8,9}$, 
\newauthor 
 William C. Keel$^{10}$, Chris Lintott$^{11,12}$, Kevin Schawinski$^{13}$\\ %, Michael J. Williams$^{14}$\\ 
  $^{1}$Steward Observatory, University of Arizona, 933 N. Cherry Ave., Tucson, AZ 85721, USA\\
  $^{2}$Institute of Cosmology and Gravitation, University of Portsmouth, 
        Dennis Sciama Building, Burnaby Road, Portsmouth PO1 3FX, UK\\
  $^{3}$SEPnet, South East Physics Network, ({\tt www.sepnet.ac.uk})\\
  $^{4}$Department of Astronomy \& CERCA, Case Western Reserve University, 
        10900 Euclid Ave., Cleveland, OH 44106, USA\\
  $^{5}$Institute for Sciences of the Cosmos (ICC-UB, IEEC), University of Barcelona, 
        Marti i Franques 1, Barcelona, 08024, Spain\\
  $^{6}$Helsinki Institute of Physics, P.O. Box 64, FIN-00014 University of Helsinki, Finland\\
  $^{7}$School of Physics and Astronomy, University of Nottingham, University Park, Nottingham NG7 2RD, UK\\
  $^{8}$Department of Physics, Massachusetts Institute of Technology, 77 Massachusetts Avenue, Cambridge, MA 02139, USA\\
  $^{9}$The Sheridan Center for Teaching \& Learning, Brown University, Box 1912, 96 Waterman St., Providence, RI 02912, USA\\
  $^{10}$Department of Physics \& Astronomy, 206 Gallalee Hall, 514 University Blvd., University of Alabama, Tuscaloosa, AL 35487, USA\\
  $^{11}$Oxford Astrophysics, Department of Physics, University of Oxford, Denys Wilkinson Building, Keble Road, Oxford OX1 3RH, UK\\
  $^{12}$Adler Planetarium, 1300 S. Lakeshore Drive, Chicago, IL 60605, USA\\
  $^{13}$Yale Center for Astronomy and Astrophysics, Yale University, 
        P.O. Box 208121, New Haven, CT 06520, USA%\\
}

\newcounter{appfig}

\begin{document}

\pagerange{\pageref{firstpage}--\pageref{lastpage}}

\maketitle
\label{firstpage}

\begin{abstract}
We present an analysis of the environmental dependence of bars and bulges in disc galaxies, 
using a volume-limited catalogue of 15810 galaxies at $z<0.06$ from the Sloan Digital Sky Survey 
with visual morphologies from the Galaxy Zoo~2 project.  We find %confirm 
that the likelihood of having a bar, or bulge, in disc galaxies increases when 
the galaxies have redder (optical) colours and larger stellar masses, and observe a transition 
in the bar and bulge likelihoods at $M_\ast=2\times10^{10}\,M_\odot$, 
such that massive disc galaxies %with high stellar mass surface densities, old stellar populations, and low star formation rates and gas masses 
are more likely to host bars and bulges. 
%However, we find that bar and bulge likelihood are not monotonically correlated with each other.
%while disc galaxies with low bulge likelihoods are not like to have bars, disc galaxies with high bulge likelihoods have a wide range of bar likelihoods.   
In addition, while some barred and most bulge-dominated galaxies are %massive and
on the ``red sequence" of the colour-magnitude diagram, 
we see a wider variety of colours for galaxies that host bars. 
We use galaxy clustering methods to demonstrate \textit{statistically significant environmental 
correlations of barred, and bulge-dominated, galaxies}, from projected separations of 
$150\,\mathrm{kpc}/h$ to $3\,\mathrm{Mpc}/h$. 
These environmental correlations appear to be independent of each other: 
i.e., bulge-dominated disc galaxies exhibit a significant bar-environment correlation, and barred disc galaxies show a bulge-environment correlation.
As a result of sparse sampling tests---our sample is nearly twenty times larger than those used previously---we argue that previous studies that did not detect a bar-environment correlation were likely inhibited by small number statistics. 
%
%> The authors should distinguish between the contribution to the environmental dependence 
% > of bars due to colour in the galaxy sample and the one derived from mock catalogues.
We demonstrate that approximately half of the bar-environment correlation can be explained by the fact that more massive dark matter haloes host redder disc galaxies, which are then more likely to have bars; 
this fraction is estimated to be $50\pm10~\%$ from a mock catalogue analysis and $60\pm5~\%$ from the data. 
%This is shown with two independent tests: (1) by shuffling the bar likelihood at a given optical colour; and %(2) normalized mark test 
%(2) using an SDSS-like mock galaxy catalogue in which bar likelihoods are assigned based on the model galaxy colour.  
%
%the environmental dependence of stellar mass explains less (10-40~\%) of the bar-environment correlation.
%Conversely, a large fraction of the environmental dependence of barred galaxies is \textit{not} due to colour or stellar mass, and hence must be due to some other galaxy property. 
Likewise, we show that the environmental dependence of stellar mass can only explain 
a smaller fraction ($25\pm10~\%$) of the bar-environment correlation.  
Therefore, \textit{a significant fraction of our observed environmental dependence of barred galaxies is not due to colour or stellar mass dependences}, and hence must be due to another galaxy property, such as gas content, or to environmental influences. 
Finally, by analyzing the projected clustering of barred and unbarred disc galaxies with halo occupation models, we argue that barred galaxies are in slightly higher-mass haloes than unbarred ones, and some of them (approximately 25\%) are satellite galaxies in groups. 
%
%Using a SDSS-like mock galaxy catalogue, we argue that the environmental dependence of bar 
%likelihood is mostly due to the fact that bars are likely to be found in redder galaxies, 
%which tend to be hosted by more massive dark matter haloes.  By analyzing the projected 
%clustering of barred and unbarred disc galaxies with halo occupation models, 
%we argue that unbarred galaxies are dominated by central galaxies in low-mass haloes, 
%while the larger-than-average satellite fraction of barred galaxies is due 
%to satellites in more massive haloes. 
%The main point is that many central galaxies in underdense regions and satellite galaxies 
%in groups and clusters form strong and stable bars. 
We discuss the implications of our results on the effects of minor mergers and interactions on bar formation in disc galaxies.
\end{abstract}


\section{Introduction}\label{sec:intro}

% bar simulation papers: O'Neill \& Dubinski (2003); Valenzuela \& Klypin (2003); Berentzen et al. (2007); Heller et al. (2007); Foyle, Courteau \& Thacker (2008); Scannapieco et al. (2010)

%%%%% THE FIRST FOUR PARAGRAPHS SET THE CONTEXT %%%%%
%%%%% DISCUSS IMPORTANCE OF GALAXY 'ENVIRONMENT' LATER IN THE INTRO. %%%%%i
%on merging, could also cite: Kartaltepe J. S., et al.,  2010, ApJ, 721, 98
% and Stewart K., Bullock J. S., Barton E. J., Wechsler R. H., 2009, ApJ, 702, 1005
%With a growing understanding that major mergers are rare (e.g., Hopkins et al.\ 2010b; Darg et al.\ 2010; Lotz et al.\ 2011), and may not play as important a role in galaxy evolution as had previously been thought (Dav{\'e} et al.\ 2011), there is a resurgence of interest in the way ``secular" processing, and bars in particular, are related to galaxy formation (e.g., Weinzirl et al.\ 2009; Schawinski et al.\ 2010; Emsellem et al.\ 2011). 
In recent years, there has been a resurgence in interest in the ``secular" processes that could affect galaxy evolution (e.g., Weinzirl et al.\ 2009; Schawinski et al.\ 2010; Emsellem et al.\ 2011), driven by the growing understanding that major mergers are rare (e.g., Hopkins et al.\ 2010b; Darg et al.\ 2010; Lotz et al.\ 2011), and may not play as important a role in galaxy evolution as had previously been thought (Parry et al.\ 2009; Dav{\'e} et al.\ 2011). 
In particular, bars have been found to be common structures in disc galaxies, and are thought to 
affect the evolution of galaxies (e.g., Sellwood \& Wilkinson 1993; Kormendy \& Kennicutt 2004) and the dark matter haloes that host them (e.g., Debattista \& Sellwood 2000; Weinberg \& Katz 2002). 
The abundance and properties of barred galaxies have been analyzed in low and 
high-redshift surveys (e.g., Jogee et al.\ 2005; Sheth et al.\ 2005, 2008; Barazza et al.\ 2008; Aguerri et al.\ 2009; Nair \& Abraham 2010; Cameron et al.\ 2010; Masters et al.\ 2011; Hoyle et al.\ 2011; Ellison et al.\ 2011) 
% could also cite Barazza et al.\ 2008; Aguerri et al.\ 2009; Masters et al.\ 2011
and have been modeled with detailed numerical simulations, including their interactions with the host dark matter haloes (e.g., Valenzuela \& Klypin 2003; O'Neill \& Dubinski 2003; Debattista et al.\ 2006; Heller et al.\ 2007; Weinberg \& Katz 2007).  

Bars are extended linear structures in the central regions of galaxies, which 
form from disc instabilities and angular momentum redistribution within the disc 
(e.g., Athanassoula 2003; Berentzen et al.\ 2007; Foyle et al.\ 2008).  
Bars are efficient at driving gas inwards, perhaps sparking central star formation 
(e.g., Friedli et al.\ 1994; Ellison et al.\ 2011), and thus may help to grow a central bulge 
component in galaxy discs (e.g., Dalcanton et al.\ 2004; Debattista et al.\ 2006; Gadotti 2011). 
% bulge: centrally concentrated stellar distribution
Such bulges are sometimes referred to as ``pseudo-bulges", to distinguish them from ``classical" 
bulges, which are often thought to have formed from the hierarchical merging of smaller 
objects (e.g., Kormendy \& Kennicutt 2004; Drory \& Fisher 2007; De Lucia et al.\ 2011; Fontanot et al.\ 2011). %also could cite Zhao Y., 2011, A\&SS, accepted (arXiv:1109.4492) 

Bars and (classical) bulges may also be related structures and in some cases could form simultaneously. 
Galaxies with earlier-type morphologies, which have more prominent bulges, tend to have more, 
and longer, bars (Elmegreen \& Elmegreen 1985; Weinzirl et al.\ 2009; Masters et al.\ 2011; Hoyle et al.\ 2011; Elmegreen et al.\ 2011; cf., Barazza et al.\ 2008). 
In addition, at least in some galaxies, the bars and bulges have similar stellar populations 
(S\'{a}nchez-Bl\'{a}zquez et al.\ 2011).  
%Karen: perhaps cite increase in bar fraction as bulge dec. in late-types (Barazza+,Aguerri+,KK04)
Nonetheless, there are some barred galaxies that lack bulges and many bulge-dominated 
galaxies that lack bars (e.g., Laurikainen et al.\ 2007; %Weinzirl et al.\ 2009; 
P\'{e}rez \& S\'{a}nchez-Bl\'{a}zquez 2011).
%see also Martin 1995 & Chapelon+ 1999 on effect of bar on SF & Z (refs in Ellison+ 2011 paper)

%mention papers discussing bars and bulges, and connecting their evolution...
%theory papers discussing bars \& bulges: Heller et al.\ (2007); Foyle et al.\ (2008); De Lucia et al. (2011). 
%observational ones: Kormendy \& Kennicutt (2004); Bureau et al.\ (2006); Laurikainen et al. (2007); Perez \& Sanchez-Blazquez (2011).

%pseudo-bulges versus classical bulges (e.g., Kormendy \& Kennicutt 2004; Drory \& Fisher 2007; Gadotti 2009; Fisher \& Drory 2011; Fontanot et al. 2011).

Various classification methods have been developed to observationally identify bars, either 
visually or using automated techniques, such 
as ellipse-fitting of isophotes and Fourier decomposition of surface brightness distributions 
(e.g., Erwin 2005; Aguerri et al.\ 2009; Gadotti 2009). 
These have yielded similar, but not always consistent, bar fractions 
(see discussions in Sheth et al.\ 2008; Nair \& Abraham 2010; Masters et al.\ 2011). 
All bar identification methods are affected by issues such as 
inclination, spatial resolution, wavelength dependence, surface brightness limits, 
and selection biases (e.g., Men\'{e}ndez-Delmestre et al. 2007). %also mention S/N issue 

%briefly introduce Galaxy Zoo~2 (Lintott et al., in prep.), providing detailed visual classifications of hundreds of thousands of galaxies in the SDSS...
In this paper, we use data from the Galaxy Zoo~2 project (see Masters et al.\ 2011), which 
provides detailed visual classifications of $\sim250,000$ galaxies in the 
Sloan Digital Sky Survey (SDSS; York et al.\ 2000). 
Galaxy Zoo yields a relatively large catalogue of galaxies with reliable classifications 
in a variety of environments.  It is particularly suited for analyses of the 
environmental dependence of the morphological and structural properties of galaxies 
across a range of scales, %and is therefore less affected by cosmic variance than other catalogues. 
as the large volume and sample size makes it less affected by cosmic variance than other catalogues. 

It has long been known that galaxy morphologies are correlated with the 
environment, such that spiral galaxies tend to be located in low-density 
regions and early-type galaxies in denser regions (e.g., %Hubble \& Humason 1931
Dressler 1980; Postman \& Geller 1984; %Blanton et al.\ 2005). 
and confirmed by Galaxy Zoo: Bamford et al.\ 2009; Skibba et al.\ 2009). 
There are a variety of ways to assess the correlation between galaxy 
properties and the environment, such as fixed aperture counts and 
distances to nearest neighbors (see reviews by Haas et al.\ 2012; 
Muldrew et al.\ 2012).  We follow Skibba \& Sheth (2009) and Skibba et al.\ 
(2009) by using two-point galaxy clustering.

%\textbf{the goal of this paper}: determining the environmental dependence of galaxy bars and bulges (and hopefully disentangling them, as well as color or stellar mass dependence).
There has been some recent work focused specifically on the environmental dependence of 
barred galaxies (van den Bergh 2002; Li et al.\ 2009; Aguerri et al.\ 2009; M{\'e}ndez-Abreu et al.\ 2010; Giordano et al.\ 2011). 
All of these studies argue that there is little to no dependence of galaxy bars on the environment. 
%although Marinova et al.\ (2009) find a slightly larger bar fraction in the cluster core of Abell 901/2.  
Contrary to these results, Barazza et al.\ (2009) and Marinova et al.\ (2009, 2012) detect a slightly larger bar fraction in the cores of galaxy clusters, but of weak statistical significance, and 
Barway et al.\ (2011) find a higher bar fraction of faint S0s in group/cluster environments. 
There is as yet no consensus on the environmental dependence of galaxy bars. 
These studies have been hampered by small number statistics, having typically a few hundred to a thousand galaxies at most. %however, with typically between a few hundred and less than a thousand galaxies. 
%\textbf{[mention Martinez \& Muriel and Lee et al. here rather than later?]} 
We improve upon this work by analyzing the environmental dependence 
galaxy bars and bulges in Galaxy Zoo~2, using a volume-limited catalogue consisting of 
15810 disc galaxies in the SDSS.  
%Contrary to these previous studies, we do detect a statistically significant correlation between galaxy bars and the environment. 
%\textbf{[Is this sufficient or should we be more specific about the goal(s) of the paper here?]}

%\textbf{[moved from intro.]} 
%During the final stages of this work, Mart{\'i}nez \& Muriel (2011) in a related study 
%found that the bar fraction does not significantly depend on group mass or 
%luminosity, or on the distance to the nearest neighbour.  Their sample is 
%also smaller than ours, however, and is apparent magnitude-limited rather than volume-limited. 
%In addition, they use bar classifications from Nair \& Abraham (2010), which 
%include weaker bars than Galaxy Zoo~2 (see Masters et al.\ 2011), which are bars that tend 
%to be found in bluer galaxies (and hence in less dense environments). 
%In another recent paper, Lee et al.\ (2011) also analyze the environmental dependence of 
%bars, using bar classifications consistent with Nair \& Abraham (2010), and 
%claim that the bar fraction does not depend on the environment at fixed colour or 
%central velocity dispersion (contrary to our results later in this paper).  
%However, a crucial difference between these two analyses and ours is that they use 
%environment measures that mix environments at different scales, while we analyze 
%the environmental correlations as a function of galaxy separation, as we will explain below.

This paper is organized as follows. 
In the next section, we describe the Galaxy Zoo~2 data and our volume-limited catalogue. 
We introduce mark clustering statistics, and in particular, the marked correlation function, 
in Section~\ref{sec:markstats}. 
In Section~\ref{sec:galprops}, we show the distributions and correlations between measures of bars and bulges. 
Then in Section~\ref{sec:clustering}, we present some of our main results, about the environmental dependence of barred and bulge-dominated galaxies, 
and we interpret the results with mock catalogues and halo occupation models in Section~\ref{sec:interpretation}.  
We end with a discussion of our results in Section~\ref{sec:discuss}.

\section{Data}\label{sec:data}

\subsection{Morphological Information from Galaxy Zoo}\label{sec:GZ}

To identify bars in local galaxies we use classifications provided by members of 
the public through the Galaxy Zoo website\footnote{\texttt{http://www.galaxyzoo.org}}.  
Specifically, we use classifications from the second phase of Galaxy Zoo 
(hereafter GZ2) which ran for 14 months (between 9th 
Feb 2009 and 22nd April 2010)\footnote{This version of the website is archived 
at \texttt{http://zoo2.galaxyzoo.org}}.  In GZ2, volunteers were asked to provide 
detailed classifications for the brightest (in terms of flux) 250,000 galaxies in the SDSS Data 
Release~7 (DR7; Abazajian et al.\ 2009) Main Galaxy Sample (MGS; Strauss et al.\ 
2002).  The selection criteria for GZ2 are $m_r<17$ and $r_{90}>3$\arcsec, where 
$m_r$ is the $r$-band Petrosian magnitude, and $r_{90}$ is the radius containing 
90\% of this flux.  Additionally, where the galaxy has a measured redshift, the 
selection $0.0005<z<0.25$ is applied. 

Following the method of the original Galaxy Zoo (Lintott et al.\ 2008, 2011; hereafter GZ1),  
volunteers were asked to classify galaxies from the $gri$ composite images.  The 
complete GZ2 decision tree is shown in Fig.~1 of Masters et al.\ (2011).  Users 
were presented each question in turn, with their progress down the tree 
depending on their previous answers.  In the 14 months GZ2 ran, 60 million 
individual classifications were collected, with each galaxy in GZ2 having been 
classified by a median of 40 volunteers (i.e., 40 people answering the question 
at the top of the tree).  As was discussed for GZ1 in Lintott et al.\ (2008) and 
Bamford et al.\ (2009), the conversion of these raw clicks into a unique 
classification for each galaxy is a process similar to data reduction that must 
take into account possible spurious classifications and other problems.  An 
iterative weighting scheme is used to remove the influence of unreliable users, 
and a cleaning procedure is applied to remove multiple classifications of the 
same galaxy by the same user. 
%From M11: "As has been discussed extensively for GZ1 data (e.g. Lintott et al. 2008; Bamford et al. 2009), there are many ways to go from ‘clicks to classifications’. This simple choice means that no galaxy is left unclassified, and gives equal weight to all users. Alternative threshold classifications were explored, and were found to make no qualitative difference to the results."

In what follows we will call the total number of cleaned and weighted 
classifications for a given question, $N_X$, and the fraction of positive 
answers to a given question $p_X$.  For example, we will discuss $N_{\rm total}$, 
which is the total (weighted) number of users classifying a given galaxy; 
$p_{\rm features}$, the (weighted) fraction of such users identifying the galaxy 
as having features; $N_{\rm bar}$, the total (weighted) number of 
classifications to the ``bar question"; and $p_{\rm bar}$, the (weighted) 
fraction of users who indicated that they saw a bar. 

\begin{figure}
 \includegraphics[width=\hsize]{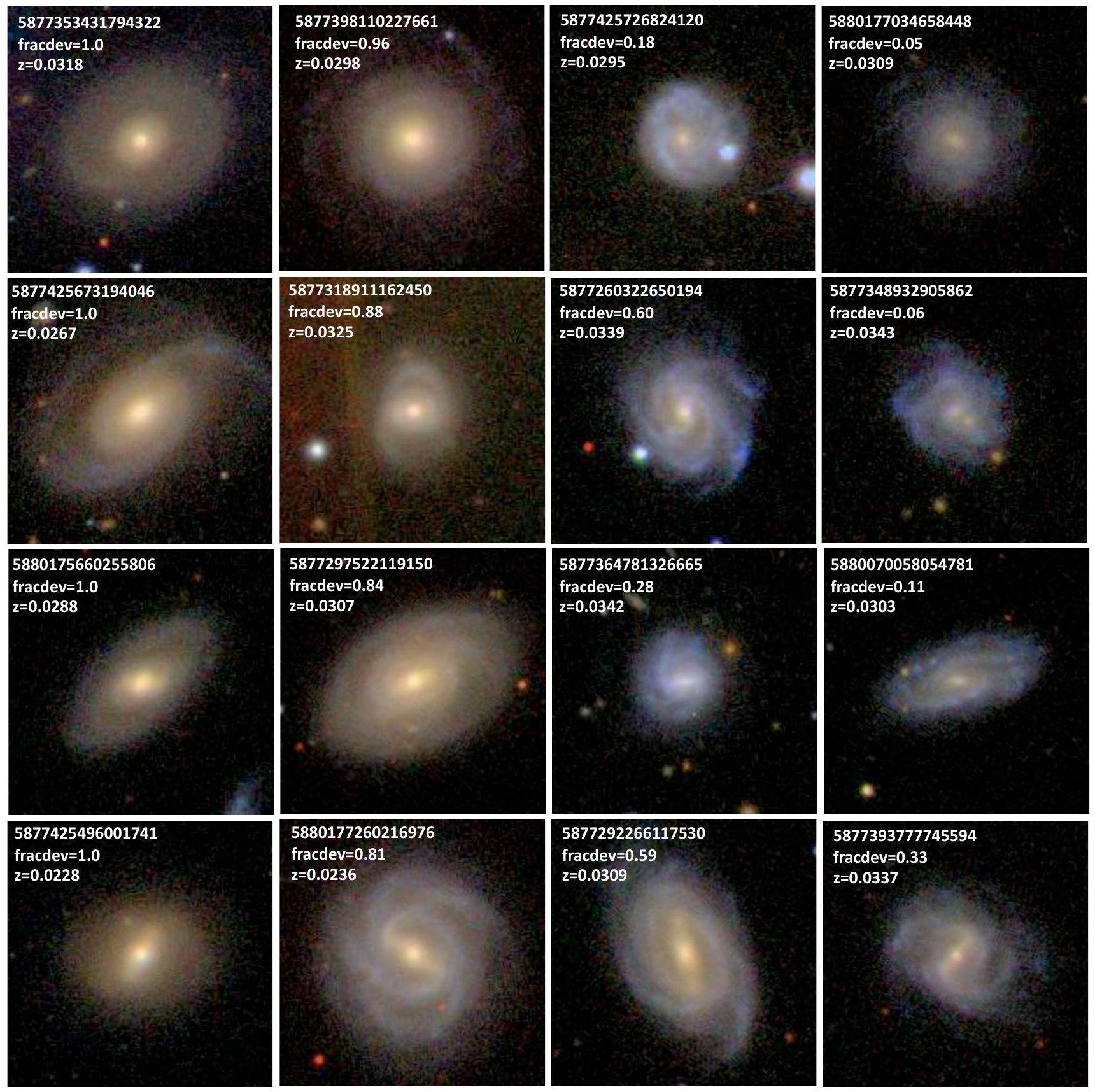} %{fig1.pdf} %{barexamples2.ps} 
 %\vspace{-4cm}
 \caption{Example images of disc galaxies with a range of values of $p_{\rm bar}$ and fracdeV. The top row shows four randomly selected galaxies with $p_{\rm bar}=0.0$ (and a range of values of fracdeV, as indicated). The second row shows galaxies with $p_{\rm bar}=0.2$, then $p_{\rm bar}=0.5$ in the row below, and finally $p_{\rm bar}=1.0$ in the bottom row. 
% A few of the galaxies in the bottom row appear larger only because they are at slightly lower redshift than the other galaxies.
         }
 \label{barexamples}
\end{figure}

Examples of disc galaxies with a range of $p_{\rm bar}$ are shown in Figure~\ref{barexamples}. 
The top row shows four randomly selected galaxies with $p_{\rm bar}=0.0$ (and a range of values of fracdeV, which is used to indicate the bulge size; see Sec.~\ref{sec:fracdeV} for details). 
The lower rows show galaxies with larger values of $p_{\rm bar}$; the galaxies in the bottom row clearly have strong bars.
%The second row shows  $p_{\rm bar}=0.2$, then $p_{\rm bar}=0.5$ the row below, and finally $p_{\rm bar}=1.0$ in the bottom row. 
%

\subsection{Bulge Sizes}\label{sec:fracdeV}

While bulge size identification was present in the GZ2 classification scheme (in 
the question of ``How prominent is the central bulge?"), %which has the answers (1) 
%``no bulge", (2) ``just visible", (3) ``obvious" and (4) ``dominant"), 
we choose instead to follow Masters et al.\ (2010a) and use the SDSS parameter ``fracdeV", 
which is a continuous indicator of bulge sizes in disc galaxies 
(see Kuehn \& Ryden 2005; Bernardi et al.\ 2010) and is strongly correlated 
with the GZ2 bulge classification (Masters et al.\ 2011). 
In the SDSS pipeline (Subbarao et al.\ 2002), galaxy light profiles are fit with 
both an exponential and de Vaucouleurs profile (de Vaucouleurs 1948), and the model magnitude comes 
from the best fit linear combination of these two profiles.  The parameter 
fracdeV indicates the fraction of this model $r$-band magnitude that is contributed by  
the de Vaucouleur profile (Vincent \& Ryden 2005).  It is expected to have the 
value fracdeV$=1$ in elliptical galaxies, and also bulge-dominated disc galaxies 
whose central light is dominated by a spheroidal bulge component.  In pure disc 
galaxies with no central light excess over an exponential disc, fracdeV$=0$ is 
expected.  
As we will see in Section~\ref{sec:galprops}, many galaxies have either fracdeV$=0$ or 1.

The fracdeV parameter is likely to be most effective at identifying 
classical bulges, although any central excess of light over
an exponential disc will drive fracdeV away from a zero value (Masters et al.\ 2010a). 
The S\'{e}rsic index, %(S\'{e}rsic 1968), 
which is closely related to fracdeV (Vincent \& Ryden 2005), 
is correlated with the bulge-to-total luminosity ratio, but with some scatter (Gadotti 2009). 
Gadotti (2009) also shows that the S\'{e}rsic index can be used to distinguish 
between classical and pseudo-bulges, although it cannot perfectly separate them 
(see also Graham 2011). 
The concentration $r_{90}/r_{50}$, the ratio of the radii containing 
90\% and 50\% of a galaxy's light in the $r$-band, is another morphological indicator 
(Strateva et al.\ 2001), and we have found that it exhibits a qualitatively similar 
clustering dependence as fracdeV.  
A comparison between fracdeV, concentration, and GZ1 spiral and early-type classifications is shown in Masters et al.\ (2010a).

\subsection{Other Galaxy Properties}\label{sec:otherprops}

In addition to morphological classifications and light profile shapes, we use 
various other parameters from the SDSS, including redshifts, $(g-r)$ and $(g-i)$ 
colours (from the model magnitudes), $M_r$ and $M_i$ total magnitudes (for which 
we use the Petrosian magnitudes), and axial ratio, $\mathrm{log}(a/b)$ (from the 
exponential model axial ratio fit in the $r$-band). All magnitudes and colours 
are corrected for Galactic extinction using the maps of Schlegel, Finkbeiner \& 
Davis (1998) and are $K$-corrected to $z=0.0$ using \texttt{kcorrect v4\_2} 
(Blanton \& Roweis 2007). 
% shall we omit the $^{0.0}$ superscript?

We compute stellar masses using the Zibetti et al.\ (2009) stellar mass 
calibration, which is based on the total magnitude ${^{0.0}M_i}$ and $^{0.0}(g-i)$ 
colour (the superscript ``0.0" refers to the redshift of the $K$-correction), 
with an absolute solar magnitude of $M_{i,\odot}=4.52$ (Blanton et al.\ 2001), 
and assuming a Chabrier initial mass function (Chabrier et al.\ 2003).  
%In particular, Zibetti et al.\ 
%(2009) use an updated version of the Bruzual \& Charlot (2003) stellar 
%population models with simple prescriptions for dust attenuation.  The model 
%assumes a two component star formation history, consisting of a continuous, 
%exponentially declining mode with random bursts superimposed. 
We refer the reader to Zibetti et al.\ (2009) for details of the model. 

The stellar mass-to-light ratios typically have $0.2-0.3$~dex scatter.
A more accurate method to estimate stellar masses would have been to
apply stellar population models (e.g., Maraston 2005) directly to the 
SDSS photometry in all five optical passbands.  The Zibetti et al.\ (2009)
calibration (which uses an updated version of Bruzual \& Charlot (2003)
models) is nonetheless consistent with Maraston (2005), with no 
systematic offsets between their masses and with discrepancies only at young stellar ages, 
and is sufficient for the analysis in this paper.

%from my KINGFISH paper:
%We estimate the stellar masses from Zibetti et al.\ (2009), using optical and 
%near-IR colors with $H$-band luminosity.  In particular, they combine 
%stellar population synthesis (SPS) models with simple prescriptions for dust 
%attenuation.  They use an updated version of the Bruzual \& Charlot (2003) 
%SPS models, which include revised prescriptions for the thermally pulsing 
%asymptotic giant branch (TP-AGB) evolutionary phase, with a two component 
%star formation history, consisting of a continuous, exponentially declining 
%mode with random bursts superimposed.  The Zibetti et al.\ (2009) model 
%outputs stellar mass-to-light ratios $\mathrm{log}\,M_\ast/L_H(B-V,V-H)$, and for the LVL 
%galaxies, which have $ugriz$-band SDSS data, we use $\mathrm{log}\,M_\ast/L_H(g-i,i-H)$.
%As a function of two colors, the model's mass-to-light ratios typically have 
%0.1-0.2 dex scatter. 

\subsection{Volume-Limited Disc Galaxy Sample}\label{sec:cat}

We perform our analysis on a volume-limited sample. %large-scale structure sample (Blanton et al.\ 2005a). 
Our catalogue is a subsample of the GZ2 catalogue, with limits $-23.5<{^{0.0}M_r}-5\mathrm{log}(h)\leq-19.4$ and $0.017\leq z<0.060$.  
%The superscript of ${^{0.0}M_r}$ refers to the $K$-correction. 
%refers to the fact that the absolute magnitudes are $k$-corrected to $z=0.0$ using \texttt{kcorrect v4\_2} (Blanton \& Roweis 2007). 
%The magnitudes have also been corrected for Galactic dust extinction (Schlegel, Finkbeiner \& Davis 1998). 
This catalogue is similar to the volume-limited catalogue used in Skibba et al.\ (2009; hereafter S09), but it has a slightly fainter absolute magnitude threshold 
because it is limited to slightly lower redshifts where we expect the bar identification in GZ2 to be most reliable (see Masters et al.\ 2011, hereafter M11, and Hoyle et al.\ 2011 for further discussion of this choice); 
in addition, GZ2 has a slightly brighter flux limit than GZ1.  

The absolute magnitude threshold of our volume-limited catalogue approximately corresponds to 
$M_r<M^\ast+1$, where $M^\ast$ is the Schechter function break in the
$r$-band luminosity function (Blanton et al. 2001). 
It also corresponds to an approximate stellar mass threshold of 
$\approx4\times10^9\,M_\odot$ (Zibetti et al.\ 2009) %maybe clarify this, or mention Baldry+ Mstar
and a halo mass threshold of $\approx5\times10^{11}\,h^{-1}\,M_\odot$ (Skibba \& Sheth 2009), 
although there is substantial scatter between galaxy luminosity and stellar and halo masses. 

The absolute magnitude and redshift limits result in a catalogue of 45581 galaxies. 
We limit the sample further to $\mathrm{log}(a/b)<0.3$, 
which is approximately an inclination of $60^\circ$, in order to select face-on or nearly face-on galaxies.  
%where $a/b$ is the axis ratio and $i$ is the inclination angle, 
%because identifying bars in highly inclined disc galaxies is challenging. 
This is a comparable inclination cut to other recent studies of bars (e.g., 
Barazza et al.\ 2008; Sheth et al.\ 2008; Aguerri et al.\ 2009) and 
identical to the cut used by M11 to study the bar fraction of GZ2. 
After this inclination cut, the catalogue is reduced to 32019 galaxies.

We require a reasonable number of answers to the bar identification question in 
GZ2.  As can been seen in the GZ2 classification tree, in order to identify the 
presence of a bar in a galaxy, the volunteer must first identify the galaxy as 
``having features", and in addition answer ``no" to the question ``Could this be 
an edge-on disc?"  We therefore limit the sample to $N_{\rm bar}\geq N_{\rm total}/4$ 
(which is equivalent to $p_{\rm features}p_{\rm not edge-on}\geq 0.25$), 
resulting in a catalogue of 15989 galaxies.  We also 
remove a small number (179) of objects with $N_{\rm bar}<10$ which may have bar 
identifications dominated by a small number of classifiers. 
% Idit: Is the pre-requisite question on 'having features' bias the sample against unbarred galaxies?  (i.e., aren't we missing unbarred galaxies among the many featureless galaxies?)
% I don't think so, except in the sense that we will miss some bar-less early-types (especially ellipticals, which we're deliberately excluding).

\begin{figure}
 \includegraphics[width=\hsize]{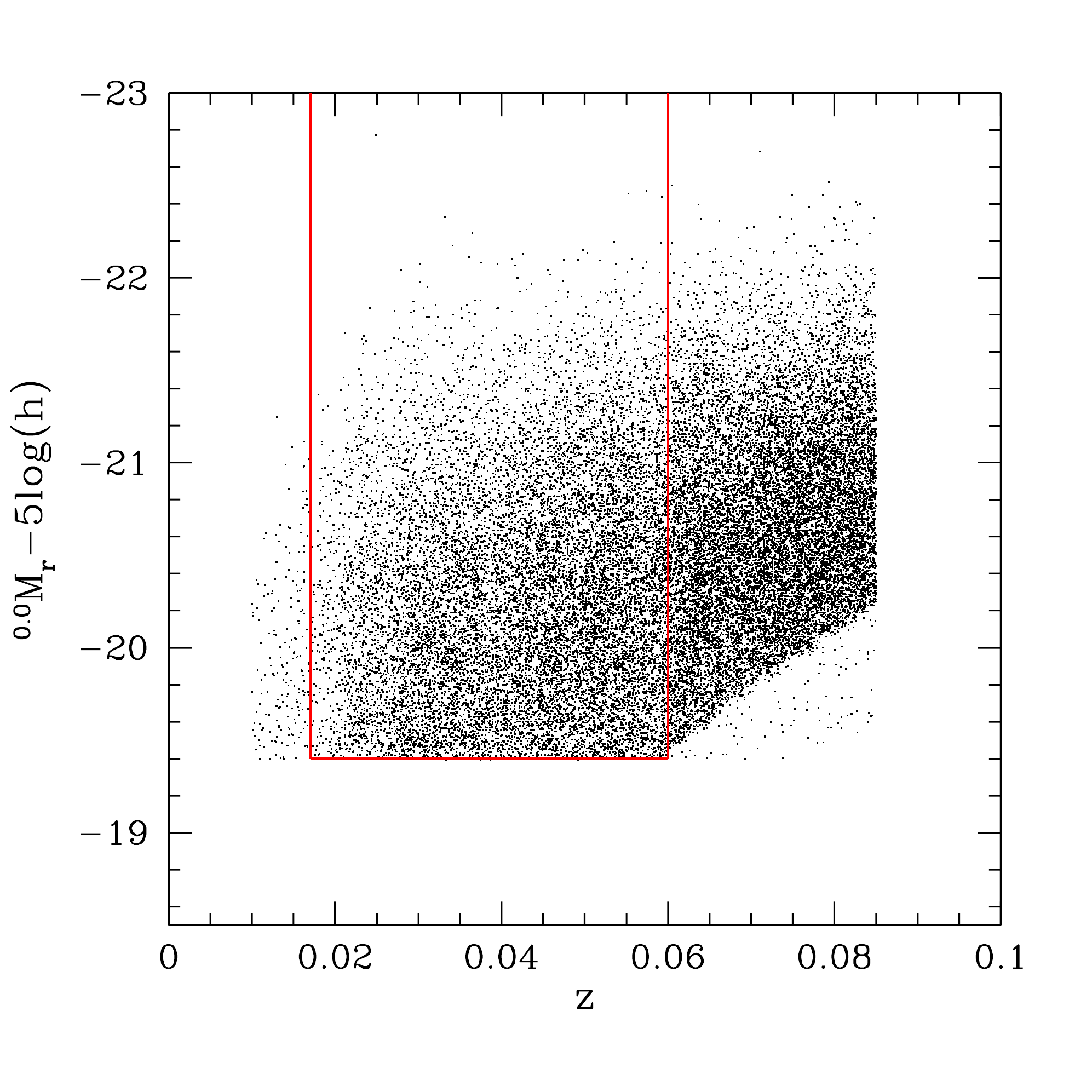} %{newbarsample_Mrzplot_Mr194Galextcorr060.ps} %{newbarsample_Mrzplot_petroMr194_kcorzeq0h.ps}
 \caption{Distribution in redshift and Petrosian $r$-band absolute magnitude, with the selection 
          criteria of the volume-limited catalogue: $-23.5<{^{0.0}M_r}-5\mathrm{log}(h)\leq-19.4$ and $0.017\leq z<0.060$. 
          %\textbf{Let's compare this to the distribution of rejected galaxies (e.g., $<10$ bar classifications.)}
          Without the $N_\mathrm{bar}$ classification cut (selecting galaxies with sufficient classifications) and $a/b$ axis ratio cut (selecting nearly face-on disc galaxies), the magnitude-redshift distribution is virtually identical; these cuts do not bias the catalogue.
          %(zmax,Ngals): (0.063, 17765), (0.062, 17113), (0.061,16397), (0.060,15810)
         }
 \label{fig1}
\end{figure}

Our resulting volume-limited catalogue comprises 15810 nearly face-on 
disc galaxies with reliable bar classifications.
% the number is 16378 galaxies when NOT using extinction-corrected magnitudes.
% the number is 17765 galaxies when using Galactic extinction-corrected magnitudes.  and 11229 (rather than 10396) in the lss files for the clustering measurements.
The galaxy distribution in redshift and magnitude, and the cuts 
used to define the catalogue, are shown in Figure~\ref{fig1}. %... 
%
%The parts constitute the whole; the whole comprises the parts.”
Note that, because of the $N_\mathrm{bar}$ cuts, elliptical galaxies are excluded from the sample. 
We emphasize that only disc galaxies (i.e., spiral galaxies and S0s) constitute the sample, including ``bulge-dominated" disc galaxies and ``pure disc" galaxies, to which we often refer as ``disc-dominated" galaxies. 
% We attempt to avoid the terms ``late-type" and ``early-type" galaxies here. 
We will distinguish between bulge-dominated and disc-dominated galaxies with the fracdeV parameter (described in Sec.~\ref{sec:fracdeV}).
% referee: say something about enviro dep of S0s here?

% remaining revisions from Bob: lines 265, 510, 992 (labels), 1143.

%Bob: since we require ten answers of bar, could we introduce environmental dependence? (i.e. by excluding certain galaxies in an enviro-dep way)
In principle, some of the selection criteria, such as the $N_\mathrm{bar}$ requirements, 
could bias our results by excluding certain galaxies in an 
environmentally dependent way.  Nevertheless, M11 have tested this by comparing 
the luminosity, colour, axial ratio, and redshift distributions with and without 
this requirement, and have found no significant differences.  Therefore, it is 
unlikely that there are any significant biases introduced; on the contrary, 
these criteria should eliminate biases by removing contaminating objects, such 
as elliptical galaxies or mergers with unreliable bar classifications. 

We have also tested our clustering measurements as a function of the axial 
ratio $a/b$, and confirmed that they are not affected by the inclination cut. 
In particular, the correlation functions for different inclinations are within 
0.03~dex ($7\%$), well within the error bars, and 
the mark correlation functions, described in the next section, are within 
$2\%$, except at separations of $r_p<500~\mathrm{kpc}/h$, 
where they still agree within $10\%$. 
%where the $p_\mathrm{bar}$ measurement without the inclination cut 
%is likely affected by inaccurate classifications of edge-on discs. 
% wp(rp) affected by <0.03 dex (<7\%), pbar mcf nearly identical except at rp<500 kpc/h.  for some reason, the drop off in Mbar(rp) at small scales is less dramatic when the edge-on discs are included.

%Idit: Re fiber collision, I think you need to be more careful there.  The collided galaxies are just missing, right?  (i.e., without a nearest neighbor correction)  Fiber collisions affect *all* scales (Z02 and Guo et al.), just with a more dramatic effect below 0.1 Mpc. (If you do a n.n. correction than you are fine above 0.1 with projected statistics.) Not observing the downturn on small scales is thus *not* an indication for a negligble effect.  Refer to Guo, Zehavi & Zheng, in prep. (to be sub soon) for details.  I agree that the trends won't change by this, but the systematics is still there in all w_p's.
%
In addition, ``fiber-collided'' galaxies are not included in the catalogue. %(\textbf{right?}). 
The thickness of the spectroscopic fibers means that some galaxies closer than 
$55\arcsec$ on the sky will be missing spectra and redshifts.  
This fiber-collision constraint is partly alleviated by the fact that neighbouring plates have overlap regions, but it still results in $7\%$ %$\sim7\%$ 
of targeted galaxies not having a measured redshift (Zehavi et al.\ 2005) 
and could significantly affect clustering measurements, especially at separations smaller than $100~\mathrm{kpc}/h$ (Guo, Zehavi \& Zheng, 2011). 
%shall we try nearest neighbor correction?
%Fiber collisions weakly affect correlation functions at scales of $s\sim100\,\mathrm{kpc}/h$ in redshift-space, and mark correlation functions as well, if galaxy spectra are required for accurate marks (Skibba et al.\ 2006).
Nevertheless, we focus our analysis on \textit{marked} correlation functions, in which
the effects of fiber collisions are expected to cancel out (see Eqn.~\ref{markedwp} in Section~\ref{sec:markstats}, where we describe the marked correlation functions). 
Moreover, the fiber assignments were based solely on target positions, and in
cases where multiple targets could only have a single fiber assigned, the target
selected to be observed was chosen randomly---hence independently of galaxy properties.
%For the projected correlation functions $w_p(r_p)$, we do not observe any downturn at small 
%scales, which one might expect, due to the higher fraction of missing galaxies. 
%\textbf{[Shall we attempt to correct the small-scale $w_p$ though, or is it not necessary?]} 
%old%In addition, we have performed tests with similar catalogs with different minimum redshifts, and have also compared some measurements to those of Skibba \& Sheth (2008), in which fiber-collided galaxies were included and were given the redshift of their nearest neighbour; in both cases, the correlation functions were consistent at small scales.
Therefore, we argue that the effects of fiber collisions are likely to be negligible 
for the marked correlation functions. 
%We conclude that the effects of fiber collisions are negligible for both the projected correlation functions and marked projected correlation functions.
%
% Chris: It would be good to cite the first Darg et al. paper where we showed that GZ was particularly effective at separating individual and merging systems, much more so than just excluding close pairs.

Throughout this paper we assume a spatially flat cosmology with 
$\Omega_m=0.3$ and $\Omega_\Lambda=1-\Omega_m$. %and $\sigma_8=0.9$,
We write the Hubble constant as $H_0=100h$~km~s$^{-1}$~Mpc$^{-1}$. %except when stated otherwise.
%XXX we assumed $h=0.7$ for the $k$-corrections, while the galaxy spatial scales are expressed in terms of $h^{-1}\,\mathrm{Mpc}$. XXX i k-corrected the mags myself, to z=0.1, with h=1.

\section{Mark Clustering Statistics and Environmental Correlations}\label{sec:markstats}

%\textit{[This is basically a repeat of Section~3 of our 2009 paper.]}

We characterize galaxies by their properties, or ``marks", such as their 
luminosity, colour, morphological type, stellar mass, star formation rate, etc. 
In most galaxy clustering analyses, a galaxy catalogue is cut into subsamples 
based on the mark, and the two-point clustering in the subsample is studied by 
treating each galaxy in it equally (e.g., %Norberg et al. 2002, 
Madgwick et al.\ 2003, Zehavi et al.\ 2005, %Li et al. 2006, 
Tinker et al.\ 2008).
These studies have shown that galaxy properties are correlated with the 
environment, such that elliptical, luminous, and redder galaxies tend to be 
more strongly clustered than spiral, fainter, and bluer galaxies.
%For example, Norberg et al. (2002) divided their catalogue by luminosity
%and spectral type into subsamples, measured their correlation functions,
%and found that luminous and quiescent galaxies have larger clustering strengths.

Nonetheless, the galaxy marks in these studies are used to define the subsamples for the 
analyses, but are not considered further.
This procedure is not ideal because the choice of critical threshold for 
dividing galaxy catalogues is somewhat arbitrary, and because throwing away the 
actual value of the mark represents a loss of information.
In the current era of large galaxy surveys, one can now measure not only galaxy clustering as a function of their properties, but the spatial correlations of the galaxy properties themselves.
We do this with ``marked statistics", in which we weight each galaxy by a particular mark,
rather than simply count galaxies as ``one" or ``zero".

Marked clustering statistics have been applied to a variety of astrophysical datasets by
Beisbart \& Kerscher (2000), Gottl\"{o}ber et al. (2002), and %Faltenbacher et al. (2002).
Mart{\'i}nez et al. (2010). 
Marked statistics are well-suited for identifying and quantifying correlations 
between galaxy properties and their environments (Sheth, Connolly \& Skibba 2005). 
They relate traditional unmarked galaxy clustering to the clustering in which each galaxy is weighted by a particular property.
Marked statistics are straightforward to measure and interpret: 
if the weighted and unweighted clustering are significantly different at a particular scale,
then the galaxy mark is correlated (or anti-correlated) with the environment at that scale,
and the degree to which they are different quantifies the strength of the correlation.
In addition, issues that plague traditional clustering measurements, such as incompleteness and complicated survey geometry, do not significantly affect measurements of marked statistics, as these effects cancel out to some extent, since the weighted and unweighted measurements are usually similarly affected. 
Mark correlations have recently been measured and analyzed in galaxy and dark matter halo catalogues (e.g., Sheth \& Tormen 2004; Sheth et al.\ 2006; Harker et al.\ 2006; Wetzel et al.\ 2007; Mateus et al.\ 2008; White \& Padmanabhan 2009; S09). 
Finally, the halo model framework has been used to interpret the correlations of luminosity and colour marks in terms of the correlation between halo mass and environment (Skibba et al. 2006, Skibba \& Sheth 2009).
We focus on morphological marks here, in particular, the likelihood of galaxies 
having a bar or bulge component. 
%and although they could be similarly analyzed with the halo model, perhaps by building on these luminosity and colour mark models, that is beyond the scope of this paper.

There are a variety of marked statistics, %(Beisbart \& Kerscher 2000, Sheth 2005),  
but the easiest to measure and interpret is the marked two-point correlation function.
The marked correlation function is defined as the following: 
\begin{equation}
  M(r) \,\equiv\, \frac{1+W(r)}{1+\xi(r)},
 \label{markedXi}
\end{equation}
where $\xi(r)$ is the two-point correlation function, 
the sum over galaxy pairs separated by $r$, in which all galaxies are ``weighted" by unity.
$W(r)$ is the same sum over galaxy pairs separated by $r$, but 
now each member of the pair is weighted by the ratio of its mark to 
the mean mark of all the galaxies in the catalogue 
(e.g., Stoyan \& Stoyan 1994).
That is, for a given separation $r$, $\xi(r)$ receives a count of 1 for each galaxy pair,
and $W(r)$ receives a count of $W_i W_j$ for $W(r)$.
The fact that the real-space (not redshift-distorted) marked statistic $M(r)$ 
can be approximately estimated by the simple pair count ratio $WW/DD$ 
(where $DD$ are the counts of data-data pairs and $WW$ are the weighted counts), 
without requiring a random galaxy catalogue, 
implies that the marked correlation function is less sensitive than the
unmarked correlation function to the effects of the survey edges (Sheth, Connolly \& Skibba 2005).
In effect, the denominator in Eqn.~\ref{markedXi} divides out the contribution to the 
weighted correlation function which comes from the spatial 
contribution of the points, leaving only the contribution from the 
fluctuations of the marks.  The mark correlation function
measures the clustering of the marks themselves, in environments of a given scale.

In practice, in order to obviate issues involving redshift distortions,
we use the projected two-point correlation function
\begin{equation}
  w_p(r_p)\,=\, \int {\mathrm d}r\,\xi({r_p},\pi)\, 
            = \,2\, \int_{r_p}^\infty \,{\mathrm d}r\,
                         \frac{r\,\xi(r)}{\sqrt{r^2-{r_p}^2}},
\end{equation}
where $r=\sqrt{{r_p}^2+\pi^2}$, $r_p$ and $\pi$ are the galaxy 
separations perpendicular and parallel to the line of sight, and 
we integrate up to line-of-sight separations of $\pi =40\,\mathrm{Mpc}/h$.
We estimate $\xi({r_p},\pi)$ using the Landy \& Szalay (1993) estimator
\begin{equation}
  \xi({r_p},\pi) \,=\, \frac{DD-2DR+RR}{RR},
\end{equation}
where $DD$, $DR$, and $RR$ are the normalized counts of data-data, 
data-random, and random-random pairs at each separation bin.
Similarly, the weighted projected correlation function is measured by
integrating along the line-of-sight the analogous weighted statistic 
\begin{equation}
  W(r_p,\pi) \,=\, \frac{WW-2WR+RR}{RR},
\end{equation}
where $W$ refers to a galaxy weighted by some property (for example, 
$p_\mathrm{bar}$ or fracdeV; see Section~\ref{sec:galprops}), and $R$ now refers to an
object in the catalogue of random points, weighted by a mark chosen randomly from its distribution.

We then define the marked projected correlation function:
\begin{equation}
  M_p(r_p)\,=\, \frac{1\,+\,W_p(r_p)/r_p}{1\,+\,w_p(r_p)/r_p} \, ,
 \label{markedwp}
\end{equation}
which makes $M_p(r_p)\approx M(r)$ on scales larger than a few Mpc, 
in the linear regime. 
The projected correlation functions $w_p$ and $W_p(r_p)$ are normalized
by $r_p$, so as to be made unitless.  
On large scales both the real-space and projected marked correlation
functions (Eqns~\ref{markedXi} and \ref{markedwp}) will approach unity, 
because at increasing scale the correlation functions $\xi(r)$ and $W(r)$ 
(or $w_p(r_p)$ and $W_p(r_p)$) become small as the universe appears nearly homogeneous.
The simple ratio of the weighted to the unweighted correlation function $W(r)/\xi(r)$
(or $W_p(r_p)/w_p(r_p)$) approaches unity similarly, provided that 
there are sufficient number statistics and the catalogue's volume is sufficiently large.

%mention Scranton's jackrandompolygon for the random and jack-knife catalogues. 
For the correlation functions and error measurements, which 
require random catalogues, we use the 
hierarchical pixelization scheme \texttt{SDSSPix}\footnote{\texttt{http://dls.physics.ucdavis.edu/\~{}scranton/SDSSPix}}, 
%\footnote{\texttt{http://lahmu.phyast.pitt.edu/$\sim$scranton/SDSSPix}}, 
which characterizes the survey geometry, including edges and holes 
from missing fields and areas near bright stars. 
This pixelization scheme has been used for clustering analyses (Scranton et al.\ 2005, 
Hansen et al.\ 2009) and lensing analyses (Sheldon et al.\ 2009). 
%and it is complementary to \texttt{MANGLE} (Swanson et al.\ 2008). 
%
%We use the Scranton et al. \texttt{jack\_random\_polygon} with window files from 
%SDSS DR5 (Adelman-McCarthy et al. 2007), which reduces the sample size by an additional 
%$\sim20$ per cent, to a final size of 72135 galaxies. 
%For the random catalogues, we used at least ten times as many random
%points as in the data, and for each of the error calculations, 
%we used thirty jack-knife sub-catalogues.
We use the Scranton et al.\ code, \texttt{jack\_random\_polygon}, to construct the 
catalogues, and we use at least twenty %thirty
times as many random points as in the data for all of the clustering measurements. 
%For each of the error calculations, we use thirty jack-knife sub-catalogues. 

We estimate statistical errors on our measurements using ``jack-knife" resampling. 
We define $N_\mathrm{sub}=30$ spatially contiguous subsamples of the full dataset, 
%(see Section~\ref{sec:cat}), 
and the jack-knife subsamples are then created by omitting 
each of these subsamples in turn. 
The scatter between the clustering measurements from the jack-knife samples is used 
to estimate the error on the clustering statistics, $w_p$, $W_p$, and $M_p$. 
The jack-knife covariance matrix is then, 
\begin{equation}
  \mathrm{Covar}(x_i,x_j) \,=\, \frac{N_\mathrm{sub}-1}{N_\mathrm{sub}}\,
     \displaystyle\sum_{k=1}^{N_\mathrm{sub}}(x_i^k-{\bar x}_i)(x_j^k-{\bar x}_j) ,
 \label{covar}
\end{equation}
where ${\bar x}_i$ is the mean value of the statistic $x$ measured in the $i^\mathrm{th}$ 
radial bin in all of the samples (see Zehavi et al.\ 2005; Norberg et al.\ 2009). 
As shown by these authors (see also McBride et al.\ 2011), however, the jack-knife technique 
only recovers a noisy realization of the error covariance matrix, as measured from mock 
catalogues, but in any case, our results are not sensitive to correlated errors 
in the clustering measurements.

\section{Results: Distributions and Correlations of Bar and Bulge Properties}\label{sec:galprops}

%%%%%%%%%%%%%%%%%%%%%%%%%%%%%%%%%%%%%%%%%%%%%%%%%%%%%%%%%%%%%%%%%%%%%%%%%%%%%%%%
% shall I update the plots, especially clustering ones, with Galactic extinction-corrected ones?
% for Figure 6: wp_Mp_Mr194Galextcorr_Pbarmark_new.ps wp_Mp_Mr194Galextcorr_fracdeVmark_new.ps; Fig.7: wp_Mp_Mr194Galextcorr_Pbarrescaled_fracdeVcuts, wp_Mp_Mr194Galextcorr_fracdeVrescaled_Pbarcuts02
% since the results are almost exactly the same, let's start by just updating clustering plots.
%%%%%%%%%%%%%%%%%%%%%%%%%%%%%%%%%%%%%%%%%%%%%%%%%%%%%%%%%%%%%%%%%%%%%%%%%%%%%%%%

The structural galaxy properties that we examine in this paper are the bar fraction or probability, 
$p_\mathrm{bar}$, and fracdeV, which quantifies bulge strength (see Sections~\ref{sec:GZ} 
and \ref{sec:fracdeV}). 
%To be precise, the value of $p_\mathrm{bar}$ of an individual galaxy 
%is the fraction of bar classifications by Galaxy Zoo classifiers. 
%is its `bar weighted fraction', votes for a bar feature.  
Using $p_\mathrm{bar}$, we can compute the bar fraction of galaxies in the 
volume-limited catalogue, such as with a $p_\mathrm{bar}$ threshold, 
%mention strong versus weak bar issue?
though as noted by M11, Galaxy Zoo tends to identify bars that are consistent with 
optically-identified strong bars (i.e., SB types). 
%
%If one conservatively counts barred galaxies as those with $p_\mathrm{bar}>0.8$, 
%the bar fraction is $\approx12\%$, while if one counts those with $p_\mathrm{bar}>0.5$, 
%the fraction is $\approx26\%$. 
%Karen: Where you cite the bar fraction can you compare to Masters et al. (2011) value of $29.4+/-0.5\%$. Can we quote the actual fraction, not the approximate number and figure out if 26\% is statistically different to the 29.4\% we had previously. 
If one generously counts barred galaxies as those with $p_\mathrm{bar}>0.2$, the bar fraction 
is $f_\mathrm{bar}=48.8\pm0.5\%$ (where the error is estimated with bootstrap resampling), 
while if one counts those with $p_\mathrm{bar}>0.5$, the fraction is $25.3\pm0.4\%$. 
The latter can be compared to M11, who obtain a fraction of $29.4\pm0.5\%$ for their sample.  
Our slightly lower fraction may be due to the different selection criteria, 
such as the fact that the M11 sample has slightly more lower luminosity galaxies. 
%Our slightly lower fraction is likely due to the different selection criteria, including more lower luminosity galaxies, which tend to have bars less often than higher luminosity galaxies, as we will see below.
For a comparison of bar classifications in GZ2, RC3, Barazza et al.\ (2008), and Nair \& Abraham (2010), we refer the reader to Masters et al.\ (2012).

% for improved bulge+disc decompositions (more accurate than fracdeV), see Simard et al. (2011)
%The SDSS parameter `fracdeV' is used as a proxy for bulge size in spirals (Kuehn \& Ryden 2005; Masters et al.\ 2010a). 
%fracdeV is the fraction of the best-fit $r$-band light profile that can be explained by a 
%de Vaucouleurs profile (de Vaucouleurs 1948), as opposed to an exponential profile. 
%In bright spirals, fracdeV is dominated by the inner light profile and increases in the 
%presence of a large bulge component. 
In this section and Section~\ref{sec:clustering}, we will analyze $p_\mathrm{bar}$ and fracdeV, their environmental 
dependence, and their relation to each other and to galaxy colour and stellar mass.  
%We use the Galactic extinction-corrected $^{0.0}(g-r)$ colour, and we use the Zibetti et al. (2009) 
%stellar mass calibration, based on extinction-corrected $i$-band luminosity ($^{0.0}M_i$) and 
%$^{0.0}(g-i)$ colour, assuming a Chabrier initial mass function (Chabrier 2003). 
%(As stated in Section~\ref{sec:cat}, the superscript `0.0' of the magnitudes and colours refers 
%to the $K$-correction to $z=0.0$; henceforth, we omit the superscript.) 
%
%Karen: Think we need a short paragraph on the choice of pbar>0.2 which splits
%the sample 50-50 and pbar>0.5 which seems to pick up strong bars.
%Could suggest that 0.2<pbar<0.5 may equate to weak bars and that we're
%looking into this by comparing with Nair & Abraham and Barazza et al.
%IDs. Could add lines to Fig 3 illustrating these 2 cuts offs?
%\textbf{[more about $p_\mathrm{bar}=0.2$ vs 0.5 threshold.]} 
We will later (Sec.~\ref{disentangleMCFs} and \ref{sec:HODmodels}) compare the clustering of barred and unbarred disc galaxies. At that point, we separate the barred and unbarred galaxies as those with $p_\mathrm{bar}>0.2$ and $<0.2$, respectively, because it approximately splits the sample in half. The $p_\mathrm{bar}>0.5$ threshold appears to identify strong bars, while we associate the range $0.2<p_\mathrm{bar}<0.5$ with weak bars (Fig.~\ref{barexamples}).  
%To better assess this, the Galaxy Zoo bar classifications are being compared to those of Nair \& Abraham (2010) and Barazza et al.\ (2008) in a separate study. 

\begin{figure}
 \includegraphics[width=\hsize]{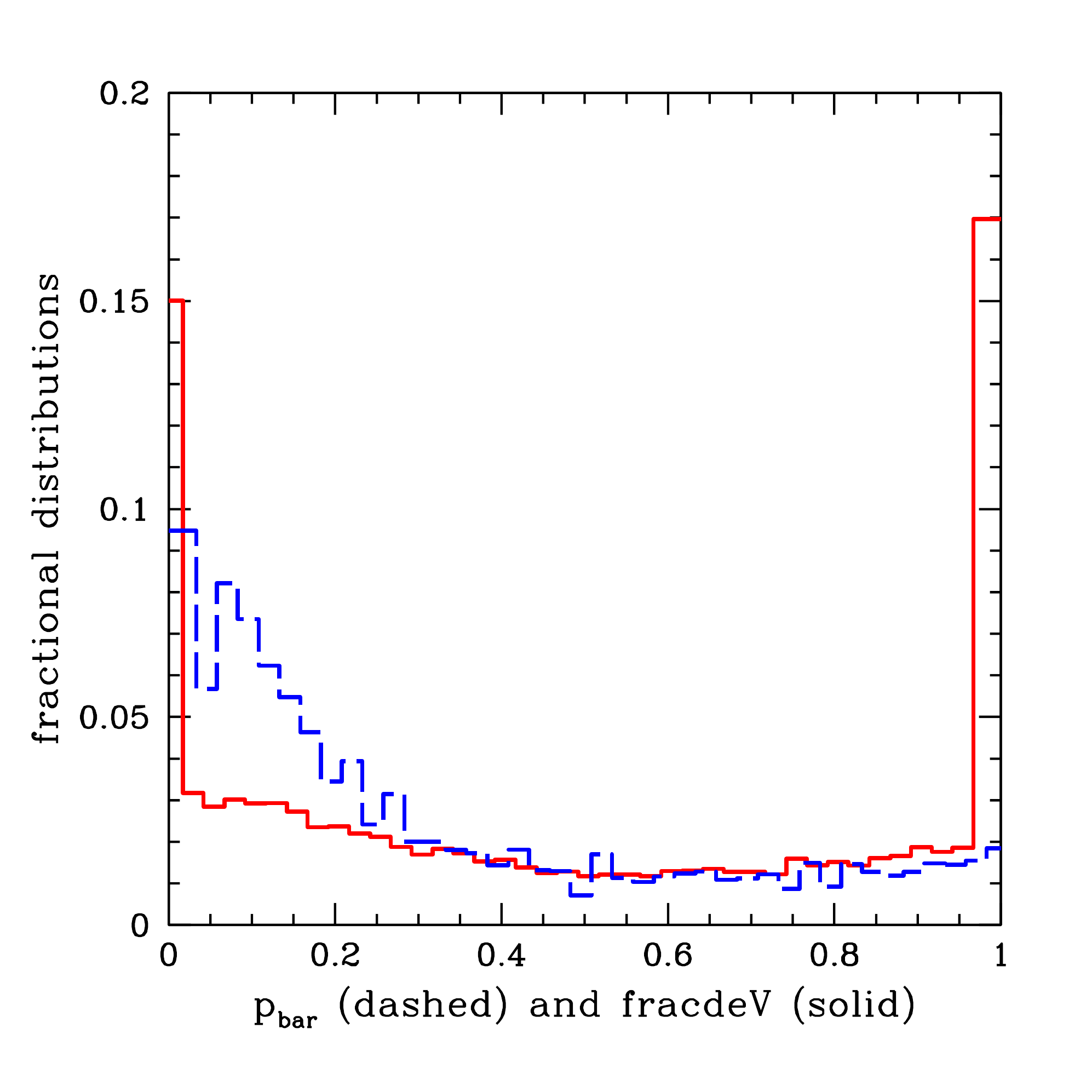} %{newsample_Mr194_PbarfracdeVdists_2.ps} %{newsample_Mr194_PbarfracdeVdists.ps}
 \caption{Distribution of bar likelihood ($p_\mathrm{bar}$, blue dashed histogram) and bulge strength (fracdeV, red solid histogram) 
          of galaxies in the volume-limited catalogue.  The histograms have been slightly offset from each other, for clarity.}
 \label{fig2}
\end{figure}

We first show the $p_\mathrm{bar}$ and fracdeV distributions Figure~\ref{fig2}. 
The $p_\mathrm{bar}$ distribution is smooth, with most galaxies in the catalogue ($62\%$) 
having $p_\mathrm{bar}<0.3$.  
In contrast, the fracdeV distribution is peaked near 0 and 1; namely, 
most galaxies are either distinctly disc-dominated or bulge-dominated. 
Recall though that, because of our selection criteria, all of the bulge-dominated galaxies 
in the catalogue have spiral arms or discs (i.e., elliptical galaxies are excluded). 
We have tested that the clustering dependence of fracdeV is not very sensitive to its distribution: rescaling it to have a smoother distribution (such as that of $p_\mathrm{bar}$) yields clustering measurements within 5\% of the results shown in the next section. %Section~\ref{MCFs}.

%is there a better way to plot this?
\begin{figure}
 \includegraphics[width=\hsize]{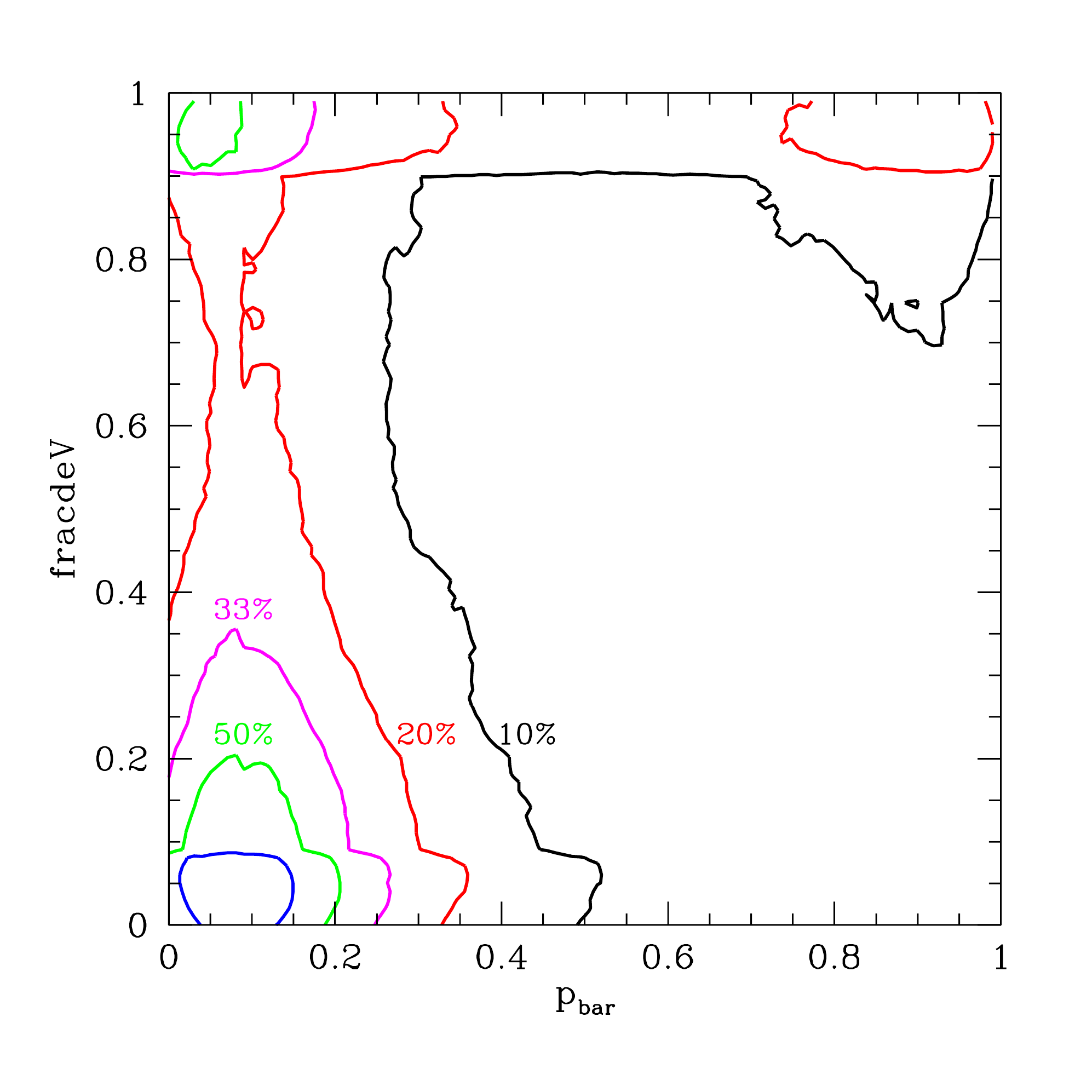}%{fig4.pdf} %{newsample_Mr194_PbarfdeVcontour_2.ps} %{newsample_Mr194_PbarfdeVcontour.ps}
 \caption{Distribution of $p_\mathrm{bar}$ versus fracdeV, with contours indicating 10, 20, 33, 50, and 75\% of the maximum counts.}
 \label{fig3}
\end{figure}

% if doing cross-correlation functions, I should show the CMD of the `parent' sample too. 
\begin{figure}
 \includegraphics[width=\hsize]{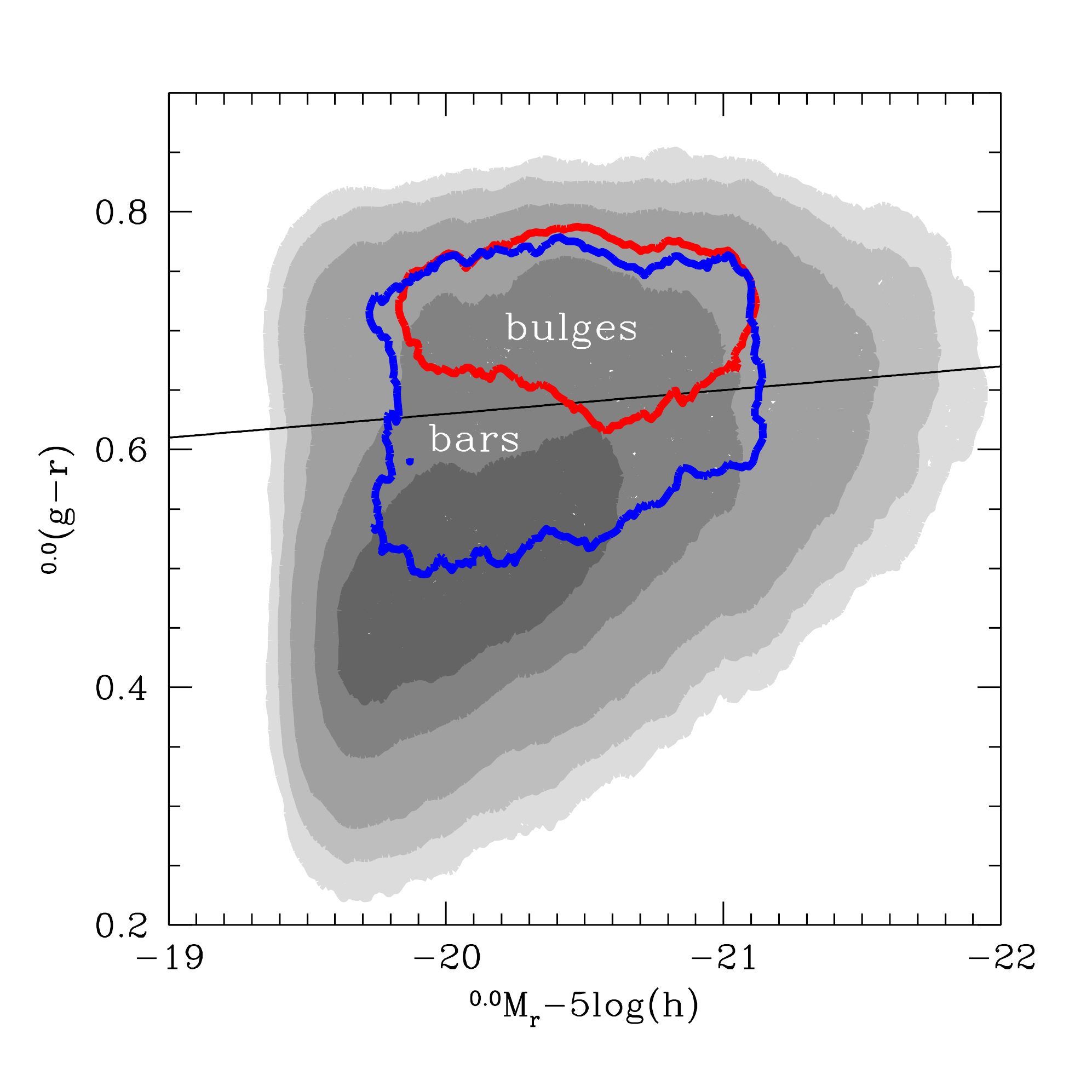} %{fig5.pdf} %{newsample_Mr194_origmagsCMDcontour_smooth_filledcontours2f.ps}%{newsample_Mr194_origmagsCMDcontour_smooth_v2b.ps}
 \caption{The optical colour-magnitude diagram (CMD), using extinction-corrected $g-r$ colour and $r$-band magnitude, with filled gray contours indicating 5, 10, 20, 50, and 75\% of the maximum counts. 
          %Note that since h=0.7 for these magnitudes, we plot $M_r-5\mathrm{log}(h)$. 
          The majority of the barred disc galaxies (with $p_\mathrm{bar}\geq0.5$) and bulge-dominated disc galaxies (with fracdeV$\geq0.9$) are indicated by the blue and red contours, respectively. 
          $p_\mathrm{bar}\geq0.5$ and fracdeV$\geq0.9$ select a similar number of galaxies in the catalogue. The black line indicates the red sequence separator ($g-r=0.63-0.02\,(M_r-5\mathrm{log}(h)+20)$), used to identify red spirals in Masters et al.\ (2010b). }
      %31%, 48%, and 74% of all, barred, and bulge-dominated galaxies are on the red sequence.
 \label{fig4}
\end{figure}

As discussed in Section~\ref{sec:intro}, some authors have argued that the formation 
and evolution of bars and bulges could be related, depending on the type of bulge, 
gas content, and angular momentum distribution (e.g., Debattista et al.\ 2006; 
Laurikainen et al.\ 2007). 
Nevertheless, we find that $p_\mathrm{bar}$ and fracdeV are not simply, or 
monotonically, correlated, as can be seen from the distribution of $p_\mathrm{bar}$ 
versus fracdeV in Figure~\ref{fig3}. 
%
% Bob thinks the interpretation of figure 3 is confusing here.  
We find that a large fraction of bulge-dominated galaxies are barred (in the upper right corner 
of the figure) and a large fraction are unbarred (upper left corner): for example, of those 
with fracdeV$>0.7$, $27\%$ have $p_\mathrm{bar}>0.7$ and $49\%$ have $p_\mathrm{bar}<0.3$. 
On the other hand, disc-dominated galaxies are mostly unbarred ($76\%$ of those with fracdeV$<0.3$ have $p_\mathrm{bar}<0.3$), and only a few per cent % $6\%$ have $p_\mathrm{bar}>0.7$ 
have bars---the lower right quadrant of the figure is empty. 
These results are consistent with M11, who showed that the bar fraction of disc galaxies increases with fracdeV, which is clearly the case for galaxies with $p_\mathrm{bar}>0.5$ on the right half of the figure. %Karen: Might be worth qualifying that there are few *strong* bars in disc-dominated galaxies (lower right of Figure 4) in case it's anyone from the Barazza et al. or Aguerri et al. papers refereeing.
%see GZ2barenviro_v1_KarensComments2.txt % end of Sec. 2 of M11 paper

% 31%, 48%, and 74% of all, barred, & bulged galaxies are on red seq. (Karen's def.)
We show the colour-magnitude distribution of the catalogue in Figure~\ref{fig4}. 
Many of the galaxies in the catalogue are disc-dominated, and the majority of them are located in the ``blue cloud", the bluer mode of the bimodal colour distribution (e.g., Skibba 2009).  
Applying the colour-magnitude separator used for red spiral galaxies (Masters et al.\ 2010b; similar to that of S09), we find that only $31\%$ of the galaxies in the whole sample are on the red sequence, while $74\%$ of the bulge-dominated galaxies (fracdeV$\geq0.9$, red contour) are on the red sequence (and are not highly inclined, so they are likely to have older stellar populations, rather than being dust reddened).  %see Masters et al. 2010a
%\textbf{[i should also show contours for subset of barred and subset of bulged galaxies.]} 
Barred galaxies ($p_\mathrm{bar}\geq0.5$, blue contour) have a much wider range of colours, with $48\%$ of them on the red sequence. 
In addition, barred galaxies have a bimodal colour distribution based on bar length, such that galaxies with longer bars are on the red sequence (Hoyle et al.\ 2011). 
% i used fracdeV>=0.9 (3916 galaxies, or 24%), and Pbar>=0.5 (4306 galaxies, or 26%)

%use running median or independent bins?
\begin{figure*}
 \includegraphics[width=0.497\hsize]{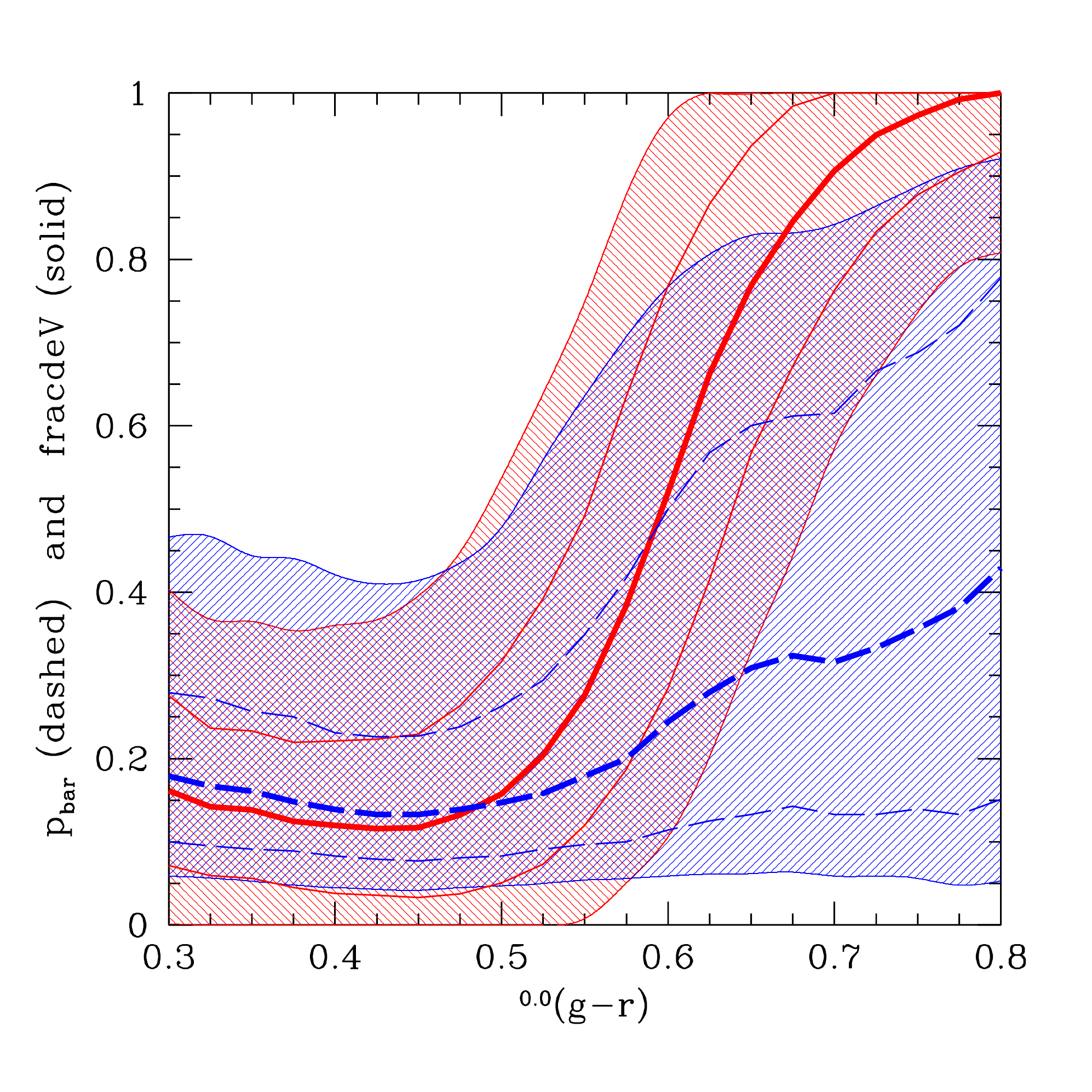}%{fig6a.pdf} %{newsample_Mr194_colorcorrs_hatched.ps} %{newsample_Mr194_colorcorrs2b.ps}
 \includegraphics[width=0.497\hsize]{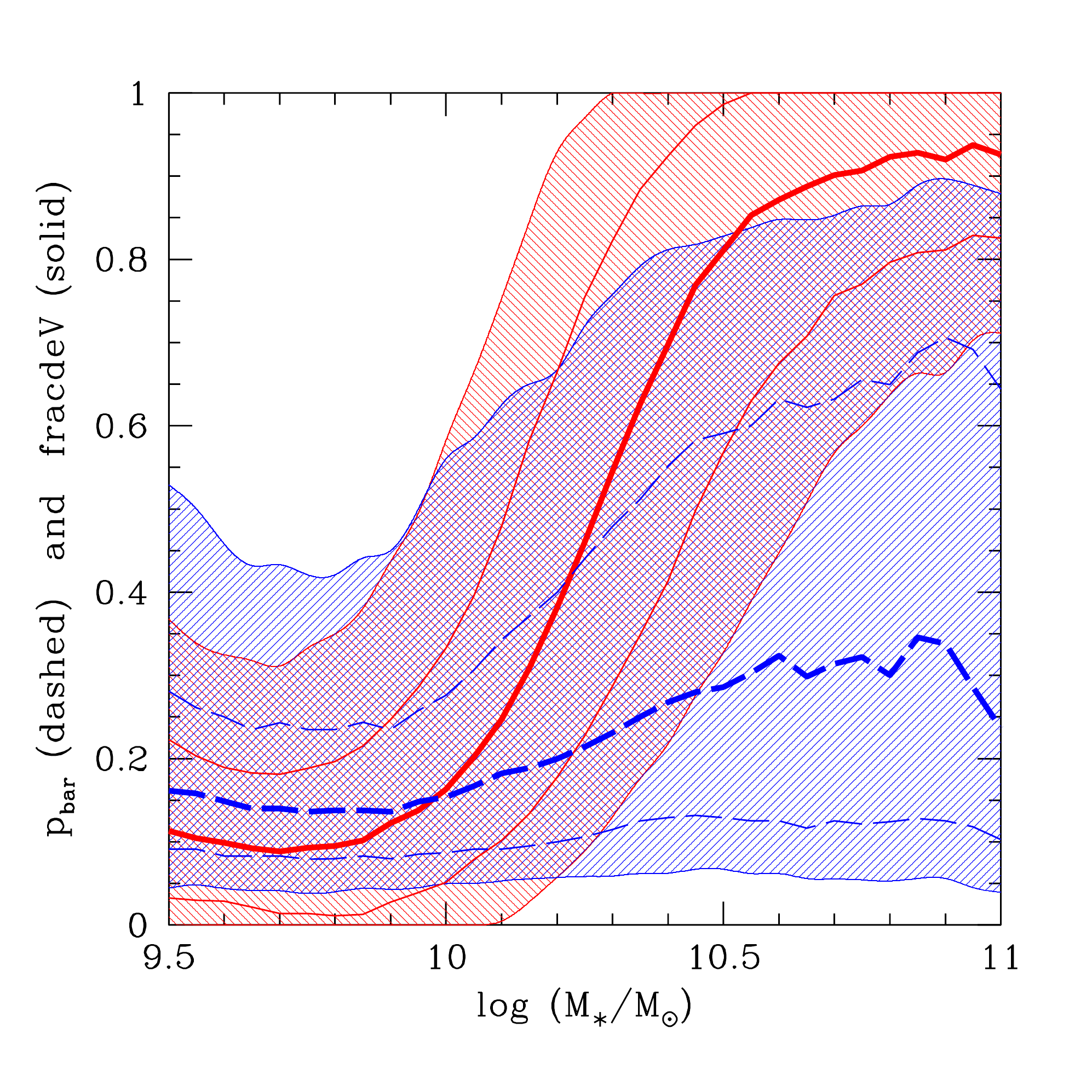}%{fig6b.pdf} %{newsample_Mr194_Mstarcorrs_hatched.ps} %{newsample_Mr194_Mstarcorrs2b.ps} 
 \caption{Left: $p_\mathrm{bar}$ (blue dashed line) and fracdeV (red solid line) as  
          a function of $g-r$ extinction-corrected colour.  Right: $p_\mathrm{bar}$ (blue 
          dashed line) and fracdeV (red solid line) as a function of stellar mass. 
          Running medians are shown as the thicker lines, 
          the $^1/_2$-$\sigma$ range is shown by the thinner lines, and the hatched regions 
          indicate the 1-$\sigma$ range between the 16 and 84 percentiles. 
          %Left: $p_\mathrm{bar}$ (blue long-dashed line) and fracdeV (red solid line) as 
          %a function of $g-r$ extinction-corrected color.  Right: $p_\mathrm{bar}$ (blue 
          %long-dashed line) and fracdeV (red solid line) as a function of stellar mass.  
          %Running medians are shown as the thicker lines, with the 16 and 84 percentiles (dotted and short-dashed lines). 
          As discussed in the text, $p_\mathrm{bar}$ and fracdeV appear to transition at 
          a similar colour and mass scale: $g-r\approx0.6$ and $M_\ast\approx2\times10^{10}\,M_\odot$.} 
            %0.3 dex bins incremented by 0.05 dex.
 \label{fig5}
\end{figure*}

%Next, we show the distribution of $p_\mathrm{bar}$ and fracdeV as a function of $g-r$ colour and stellar mass in Figure~\ref{fig5}. 
Next, in Figure~\ref{fig5}, we show the median (and $^1/_2$ and 1-$\sigma$ ranges) 
of $p_\mathrm{bar}$ and fracdeV as a function of colour and stellar mass. 
The majority of blue galaxies and low-mass galaxies are disc-dominated, while most 
red and massive galaxies are bulge-dominated.  
The bar probability is also positively correlated with colour and stellar mass, 
consistent with other studies (e.g., Sheth et al.\ 2008; Nair \& Abraham 2010; M11), such that 
redder and more massive galaxies are more likely to have bars. 
Some studies have found a bimodal distribution of bars, such that blue, low-mass, or Sc/Sd-type galaxies also often have bars (Barazza et al.\ 2008; Nair \& Abraham 2010), in contrast to our finding that the median $p_\mathrm{bar}<0.2$ at bluer colours and lower stellar masses.  It is possible that in Galaxy Zoo, a large fraction of weak or short bars are missed %not resolved? 
in these galaxies.

Compared to the correlation with fracdeV, the correlations with $p_\mathrm{bar}$ in Figure~\ref{fig5} are 
not as strong and have more scatter, especially at the red and massive end.  
In other words, red and massive galaxies are more likely to have bars than blue and less-massive 
galaxies, but nonetheless there are many red and massive galaxies 
that lack bars. 
Either these galaxies never formed bars, or perhaps more likely, it is 
possible that they had bars in the past that were weakened (so that they 
are no longer detectable by GZ2) or destroyed; 
%[on issue of bar lifetimes, mention Bournaud \& Combes and Ellison et al.]
some galaxies may even have multiple episodes of bar formation in their lifetime (Bournaud \& Combes 2002).  
% Ellison: bc of bar timescales, (not) related to merging?
% Bournaud F., Combes F., 2002, A\&A, 392, 83
% Bournaud F., Combes F., Semelin B., 2005, MNRAS, 364, L18
% so the bar fraction observed today may be indicative of the duty cycle of multiple periods of bar formation and destruction in different types of galaxies.

%\textbf{mention that the the Pbar-color and fracdeV-color correlations transition at similar 
%colors}, sort of in the ``green valley''.  It also occurs at the transition stellar mass of a 
%few times $10^{10}M_\odot$ (which is like Kauffmann et al. 2004's transition mass). 
It is interesting that the transition from mostly unbarred to mostly barred galaxies and from 
disc-dominated to bulge-dominated galaxies occurs at similar colours and stellar masses. 
The colour transition occurs at extinction-corrected $g-r\approx0.6$, in the 
``green valley" of the colour-magnitude distribution (e.g., Wyder et al.\ 2007), %Loh+ 2010
between the blue and red peaks of the distribution (see Figure~\ref{fig4}). 
The stellar mass transition occurs at $M_\ast\approx2\times10^{10}\,M_\odot$, 
and is similar to the mass scale identified by Kauffmann et al.\ (2003; see also 
Schiminovich et al.\ 2007), above which galaxies have high stellar mass surface 
densities, high concentration indices typical of bulges, old stellar populations, 
and low star formation rates and gas masses.
% Bob thinks this is an important result.

\section{Results: Clustering of Galaxies with Bars and Bulges}\label{sec:clustering}

We now explore the environmental dependence of disc galaxies with bars and bulges 
by measuring marked projected correlation functions (described in Section~\ref{sec:markstats}). 
For the marks, we use 
$p_\mathrm{bar}$ and fracdeV, which indicate the presence or lack of a bar 
and bulge, respectively.  We first present the total environmental dependence 
of these two galaxy properties in Section~\ref{MCFs}, and then we attempt to 
separate their environmental dependences in Section~\ref{disentangleMCFs}.

\subsection{Environmental Correlations of Bar and Bulge Probability}\label{MCFs}

\begin{figure*}
 %\includegraphics[width=\hsize]{wp_Mp_Mr194_PbarfracdeVmarks_newerrors.ps}
 %\caption{Upper panel: projected correlation function (circle points) and weighted projected correlation functions with $p_\mathrm{bar}$ and fracdeV marks (blue squares and red triangles, respectively).  Lower panel: marked projected correlation functions with $p_\mathrm{bar}$ 
 %and fracdeV marks. \textbf{Shall we separate these into two plots (like Fig.~\ref{fig7})?}}
 \includegraphics[width=0.497\hsize]{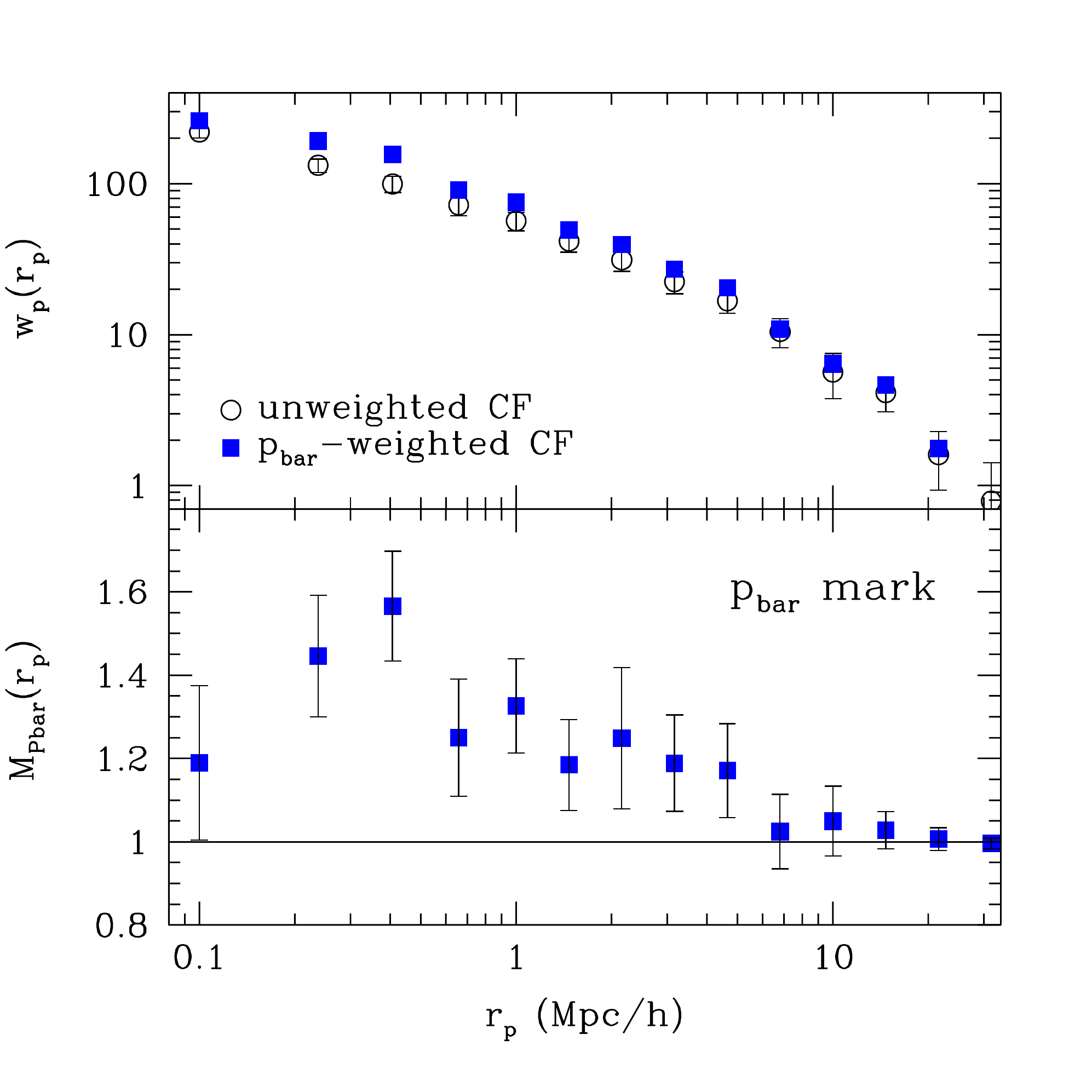} %{wp_Mp_Mr194Galextcorr_Pbarmark_new2_N30err.ps} %{wp_Mp_Mr194_Pbarmark_newerrors.ps} %maybe show wp_Mp_Mr194Galextcorr_Pbarmark_new.ps instead?
 \includegraphics[width=0.497\hsize]{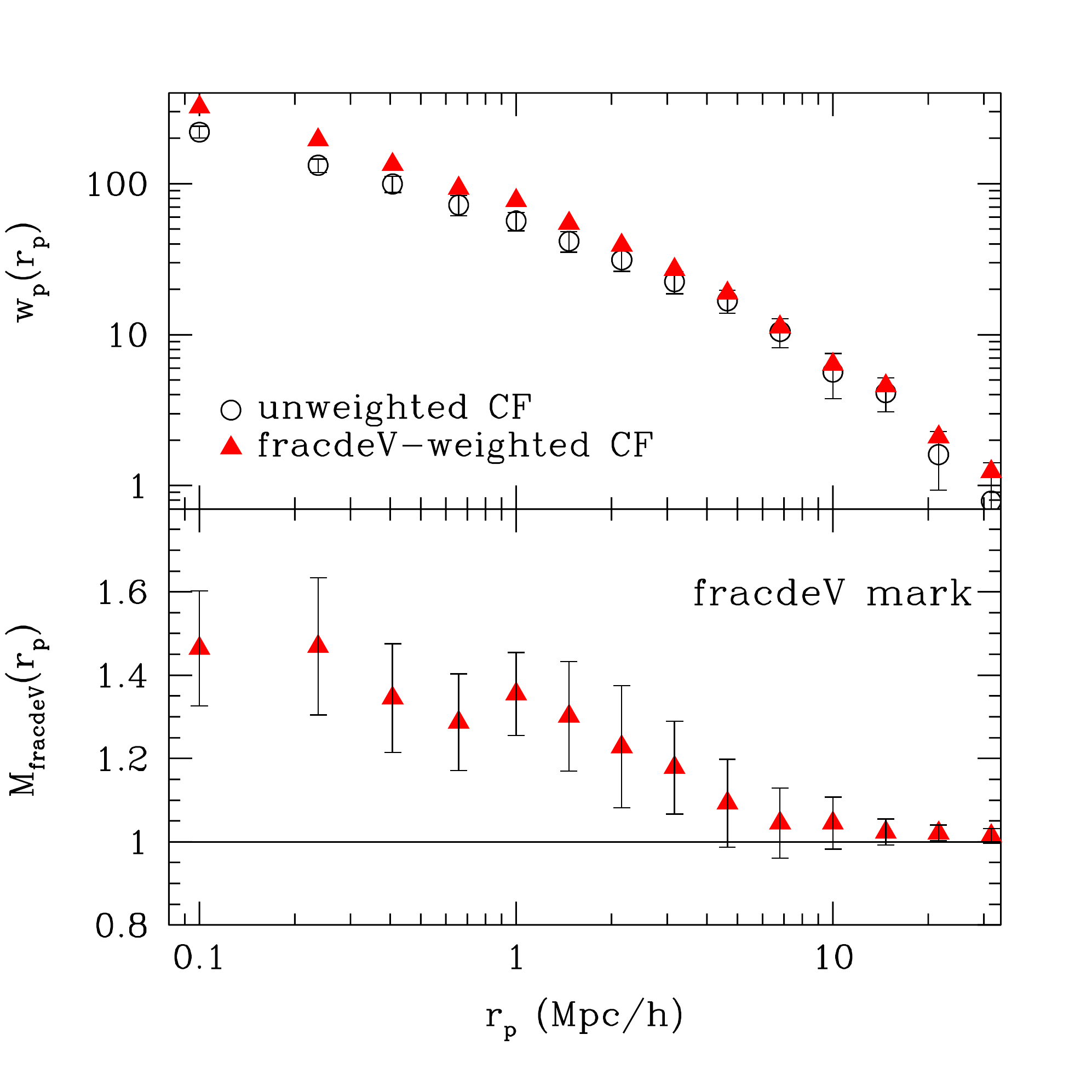} %{wp_Mp_Mr194Galextcorr_fracdeVmark_new2_N30err.ps} %{wp_Mp_Mr194_fracdeVmark_newerrors.ps}
 \caption{Upper panel: projected correlation function $w_p(r_p)$ (circle points) and weighted 
          projected correlation functions $W_p(r_p)$. Lower panel: marked projected correlation 
          functions, using $p_\mathrm{bar}$ mark (left) and fracdeV mark (right). 
          Recall that the marked correlation function is defined as $M(r_p)\equiv[1+W_p(r_p)/r_p]/[1+w_p(r_p)/r_p]$ (Eqn.~\ref{markedwp}).}
 \label{fig6}
\end{figure*}

In the upper panels of Figure~\ref{fig6}, we show the projected clustering 
of the galaxies in our (disc-dominated) catalogue. 
In the lower panels, we show the the marked correlation functions, quantifying the environmental 
correlations of $p_\mathrm{bar}$ and fracdeV across a wide range of scales. 
%For the error analyses, we used $N_\mathrm{rand}=30N_\mathrm{dat}$ and 30 jack-knife samples. 
The errors of the measurements are estimated using jack-knife resampling, and 
are analyzed in more detail in Appendix~\ref{app:JK}. 
As noted in the appendix, a significant fraction of the error estimates is due to a 
single outlying jack-knife subsample.  
%\textbf{[Maybe we should be plotting the (smaller) errors, with this outlier excluded, in the figures?]}
%which is excluded from the calculations of these error bars.

We see a statistically significant environmental correlation in Figure~\ref{fig6} 
for both $p_\mathrm{bar}$ and fracdeV, which means that disc galaxies with bars and those with bulges tend to reside in denser environments on average. 
The environmental correlation is especially strong at small scales, at $r_p\leq2\,\mathrm{Mpc}/h$. 
At these scales, the clustering signal is dominated by the ``one-halo term" (pairs of galaxies within dark matter haloes), while at larger spatial scales the ``two-halo term" (pairs of galaxies in separate haloes) dominates the clustering (e.g., Zehavi et al.\ 2004).  In Section~\ref{sec:clustering}, we will interpret these correlation functions with the halo model of galaxy clustering (see Cooray \& Sheth 2002 for a review). 

%A simple way to quantify the statistical significance of the $M(r_p)$ measurements 
%is to compute the probability that the mark correlation occurs by chance, assuming 
%Gaussian and uncorrelated errors:
%\begin{equation}
% P_M \,=\, \prod_{r_p} \biggl[ 1 - \mathrm{erf}\biggl( \frac{M(r_p)-1}{\delta M(r_p)\,\sqrt{2}} \biggr) \biggr] ,
% \label{probM}
%\end{equation}
%where $\delta M(r_p)$ is the error on $M(r_p)$. 
%For both the $p_\mathrm{bar}$ and fracdeV mark correlations, these probabilities are extremely 
%small, because $M(r_p)$ is significantly larger than unity in most of the $r_p$ bins, yielding 
%better than $10\sigma$ significance for both marks. 
%
It is important to quantify the statistical significance of the $M(r_p)$ measurements, the degree 
to which they are inconsistent with unity.  (A result of unity occurs when the weighted and unweighted correlation functions are the same, i.e., when the weight is not correlated with the environment.)  
Since the errors are correlated, the statistical significance should be quantified using the 
covariance matrices (see Eqn.~\ref{covar} and Appendix~\ref{app:JK}):
\begin{equation}
 \sigma_M^2 \,=\, (\pmb{M}-\mathbf{1})^T \,\mathbf{Covar}^{-1}\, (\pmb{M}-\mathbf{1}) ,
 \label{sigmaM}
\end{equation}
where $\pmb{M}$ is the $p_\mathrm{bar}$ or fracdeV mark correlation function, 
and $\pmb{M}-\mathbf{1}$ is its deviation from unity. 
(This is similar to the way one would compute the $\chi^2$ of a fit, where in this 
case a good ``fit'' with low $\sigma_M^2$ would be a measurement consistent with unity, or no environmental correlation).  The $p_\mathrm{bar}$ and fracdeV mark correlation functions yield 
$\sigma_M^2=39.9$ and $43.4$, respectively, which could be interpreted as $6.3$ and $6.6\sigma$ significance for the marks.  
%$\sigma_M^2=65.4$ and $70.9$, respectively, which could be interpreted as greater than $8\sigma$ significance for both marks.  
%Even if the outlying jack-knife subsample were included in the error analysis, the mark correlations would still have $6\sigma$ significance. %8sigma if excluded
%
%$\sigma_M^2=39.9$ and $43.4$, respectively, which could be interpreted as $>6\sigma$ significance for both marks. 
%When the outlying jack-knife subsample is excluded, as is done for the error bars in Figure~\ref{fig6}, the resulting values for the $p_\mathrm{bar}$ and fracdeV mark correlations are $\sigma_M^2=65.4$ and $70.9$, respectively, which have $>8\sigma$ significance.
% get_inverse_covar.c
%%%%% quote a number for statistical significance %%%%% 
% could simply try this...if indep.: sigma^2 = [sum over rp] (M/dM)**2; if indep., use covariance
% one possibility is to invert the covariance matrix and estimate chi^2 w.r.t. unity (see Sec. 3.3 of Norberg+ 2009; & Eqn. 8 of Zehavi+ 2011). use gsl_linalg_LU_decomp & gsl_linalg_LU_invert (p.127)
Note that this significance estimate is not highly dependent on the binning: narrower 
$r_p$ bins, for example, would yield more measurements with $M(r_p)>1$, but they would have larger and more correlated errors.

The positive correlation at particular spatial separations $r_p$ implies that 
\textit{galaxies with larger values of $p_\mathrm{bar}$ and fracdeV 
tend to be located in denser environments at these scales}. 
This is one of the main results of the paper. 
This can also be seen in the upper panels of the figures, in which 
the weighted correlation functions are larger than the unweighted ones. 
% shall we mention DM haloes yet?

%The $p_\mathrm{bar}$ and fracdeV mark correlation functions appear to be of similar strength. 
%The mark distributions of the two properties are not the same (Fig.~\ref{fig2}), 
%but when we rescale the fracdeV distribution to match that of the $p_\mathrm{bar}$ distribution, 
%the resulting mark correlation is similar to the one in Figure~\ref{fig6}. 

The environmental dependence of bulges is not surprising; 
because of the ``morphology-density relation", in which dominant bulge components 
are associated with earlier-type morphologies, it is expected that bulge-dominated 
galaxies tend to be located in denser environments, in groups and clusters 
(e.g., Postman \& Geller 1984; Bamford et al.\ 2009). 
Similar two-point clustering analysis have also clearly shown that 
bulge-dominated disc galaxies tend to be more strongly clustered than 
disc-dominated ones, on scales of up to a few Mpc (Croft et al.\ 2009; S09). 
Semi-analytic models, using stellar bulge-to-total ratios, %which recently better resolved blah...,
have similarly predicted that bulge-dominated disc galaxies tend to form 
in more massive dark matter haloes %(in the range $10^{12}<M_\mathrm{halo}<10^{14}M_\odot$) 
(e.g., Baugh et al.\ 1996; Benson \& Devereux 2010; De Lucia et al.\ 2011). 
%maybe also mention age-enviro, color-enviro, or SFR-enviro, vs morph-density (Kauffmann+ 2004; Blanton+ 2005; Wolf+ 2007; S09 
Many theorists have argued that bulge formation is linked to minor and major galaxy mergers, and mergers and interactions tend to be more common in denser environments, especially in galaxy groups (e.g., Hopkins et al.\ 2009; Hopkins et al.\ 2010a; Martig et al.\ 2012); %maybe could also cite Narayanan et al. 2010 
there is some observational evidence in favor of the link between bulges and mergers as well (e.g., Ellison et al.\ 2010).

On the other hand, %as opposed to bulge-dominated galaxies, 
one might not expect barred galaxies to be correlated with the environment, 
if bars form entirely by internal secular processes. 
Some recent studies have argued that barred and unbarred galaxies are located in 
similar environments, or have only weak evidence that barred galaxies 
are more strongly clustered at small scales (Marinova et al.\ 2009; Li et al.\ 2009; 
Barazza et al.\ 2009; Mart{\'i}nez \& Muriel 2011; Wilman \& Erwin 2012). %and Lee et al.\ 2011
% Karen's suggestion: ...you might comment that the SAB classification in RC3 includes objects which even in RC3 were thought to possibly not have bars (it's weak bars and/or unsure classifications) so very dangerous to use in the way this paper does.  Also remember we saw many examples of RC3 barred galaxies with low pbar from Galaxy Zoo (and no bar in Nair & Abraham) so I'm starting to think the RC3 bar IDs should be taken very seriously.
%
Nevertheless, our volume-limited GZ2 catalogue is much larger than those of these 
studies (which except for Mart{\'i}nez \& Muriel %and Lee et al.
consist of $\sim1000$ galaxies, or $\approx1/17^\mathrm{th}$ as many as ours), 
and as stated in Section~\ref{sec:markstats}, an advantage of mark clustering 
statistics is that one can analyze the entire catalogue, without splitting it 
and without requiring a classification of ``cluster", ``group", and ``field" environments. 
We are thus able to quantify the correlation between bars and the environment 
with greater statistical significance. 

%
% sparse sampling tests I tried:
%    1. draw N galaxies randomly from the sample
%    2. try narrower redshift range
%    3. try splitting the sample with n_jack
To test the effect of small number statistics, we performed a number of sparse sampling measurements (randomly selecting galaxies, or only selecting galaxies in small subregions or redshift slices of the sample). 
%
%In general, we find that if there are fewer than $\sim2000$ galaxies used for the clustering measurements, the result is only a hint of an environmental dependence of $p_\mathrm{bar}$, which is not statistically significant; the unmarked correlation function is noisy as well, for such a small number of galaxies.
In general, we find that if fewer than $\sim2000$ galaxies were used for our clustering measurements, we would not have a statistically significant detection of the environmental dependence of $p_\mathrm{bar}$, and the unmarked correlation functions would be too noisy. 
% Bob: Shouldn't we add this test - and the consequences - to the conclusions. This allows us to quantitatively say the lack of trends before were likely small sample sizes (as demonstrated). We are missing this in the abstract as well?
%\textbf{[mention sparse sampling tests in conclusions \& abstract?]} 
This may explain why previous studies of smaller galaxy catalogs did not detect a significant bar-environment correlation. 

The $p_\mathrm{bar}$ marked correlation function in Figure~\ref{fig6}, $M_\mathrm{Pbar}(r_p)$, 
increases in strength with decreasing spatial separation. 
Such a trend is expected when the mark is positively correlated with the environment. 
An exception to this correlation occurs at small separations ($r_p\sim100~\mathrm{kpc}/h$), 
where the result is consistent with no correlation at all. 
%mention fiber collisions again? 
Fiber collisions could affect the lack of correlation at these scales, but at most 
$7\%$ of the targeted galaxies lack measured redshifts, and we obtain 
$M_{p_\mathrm{bar}}(100\,\mathrm{kpc}/h)\approx1$ for all jack-knife subsamples, 
so this low mark correlation is likely a real effect. 
The weakening environmental correlation at small separations suggests that whatever 
conditions that make bars more likely at larger scales are removed in close pairs 
of galaxies.  For example, close pairs are more likely to experience mergers, 
and bars may be weakened or destroyed immediately following merger activity, 
although new bars may form later (Romano-D\'{i}az et al.\ 2008). 
Nonetheless, Marinova et al.\ (2009) and Barazza et al.\ (2009) find 
that bar fractions are slightly larger in cluster cores, although this is of weak statistical significance according to the authors. 
%
% see e-mail exchange with Karen, Sara, and Preethi
More recently, Nair \& Ellison (in prep.) find that the bar fraction of disc 
galaxies decreases as pair separation decreases, consistent with our results.
%
%\textbf{discuss this: are the conditions that make bars more likely at larger scales 
%removed in close pairs of galaxies (maybe Marinova; cf., Barazza)?  or perhaps the bar probability is higher among satellites than centrals?} 
% NB: still need to define `central' and `satellite' galaxies

It is interesting that $M_\mathrm{Pbar}(r_p)$ peaks at approximately $400~\mathrm{kpc}/h$ 
(more precisely, the bin's range is $316<r_p<525~\mathrm{kpc}/h$). 
Many of the galaxies contributing to the signal at these scales are likely 
``satellite" galaxies in groups, rather than the central galaxies.  In fact, 
considering that these are disc galaxies and that $p_\mathrm{bar}$ is correlated 
with colour, it is likely that many of these are the same objects as the 
``red spirals" discussed in S09 (most of which have bars, according to Masters et al.\ 2010b; M11), 
%Karen: Comment that most red spirals have bars where you talk about how the barred galaxies could be the same as the red spirals. 
a relatively large fraction of which are satellites ($f_\mathrm{sat}\approx1/3$). 

% maybe just hint at the interpretation of MCFs here, since this will be discussed more in Section~\ref{HODmodeling{.  more speculative comments should be left for the discussion.

We also show (unmarked) clustering of barred versus non-barred galaxies 
($p_\mathrm{bar}>0.2$ and $<0.2$), and bulge-dominated versus disc-dominated ones 
(fracdeV$>0.5$ and $<0.5$), in the upper panels of Figure~\ref{fig7}. 
At large scales ($r_p>2\,\mathrm{Mpc}/h$), their clustering strength is the same. 
At smaller separations, however, barred galaxies tend to be more strongly 
clustered than unbarred ones and bulge-dominated galaxies tend to be more strongly 
clustered than disc-dominated ones.  The scale at which the correlation functions 
diverge corresponds to the scale of the transition from the ``one-halo term"  
(pairs of galaxies within haloes) to the ``two-halo term" (galaxies in 
separate haloes).  These clustering measurements then suggest that barred 
and unbarred galaxies may reside in the same dark matter haloes, but the 
former are more likely to be central galaxies than the latter.  The same 
applies for the presence/absence of bulges in central/satellite galaxies.  
We will return to this issue when we apply halo occupation modeling to the 
measured correlation functions, in Section~\ref{sec:HODmodels}.

\subsection{Disentangling the Environmental Correlations}\label{disentangleMCFs}

As we have shown in previous sections, disc galaxies with large bulges are 
more likely to have a bar (see Fig.~\ref{fig3}).  We have also shown that 
both bulge-dominated discs and discs with bars are more strongly clustered 
than average (Fig.~\ref{fig6}).  We address in this Section the question 
of whether one of these two galaxy properties is more dependent on the 
environment, or whether their environmental correlations are independent.  
That is to say, we will determine whether bulge-dominated galaxies with 
bars are more strongly clustered than bulge-dominated galaxies without bars, and 
whether barred galaxies with bulges are more strongly clustered than barred 
galaxies with no or small bulges. 

In addition, we know that disc galaxies hosting bars tend to be redder and have 
higher stellar masses than those with weak or no bars (Fig.~\ref{fig5}). 
We will later address in Section~\ref{sec:MassColor} whether the environmental 
correlations of galaxy colour or stellar mass (e.g., Skibba \& Sheth 2009; 
Li \& White 2009) can account for the environmental correlation we have observed of bars. 

%%% The following is the old intro to this Section: %%%
%Considering that barred galaxies also often have bulges (see Fig.~\ref{fig3}), 
%it is important to determine whether the environmental correlation of fracdeV 
%is more `fundamental' than that of $p_\mathrm{bar}$ and is causing the latter, 
%or whether these environmental correlations are independent.  
%In Section~\ref{sec:MassColor}, we also address the issue as to whether they are due to 
%an `underlying' correlation with another galaxy property, such as stellar mass. 

\begin{figure*}
 \includegraphics[width=0.497\hsize]{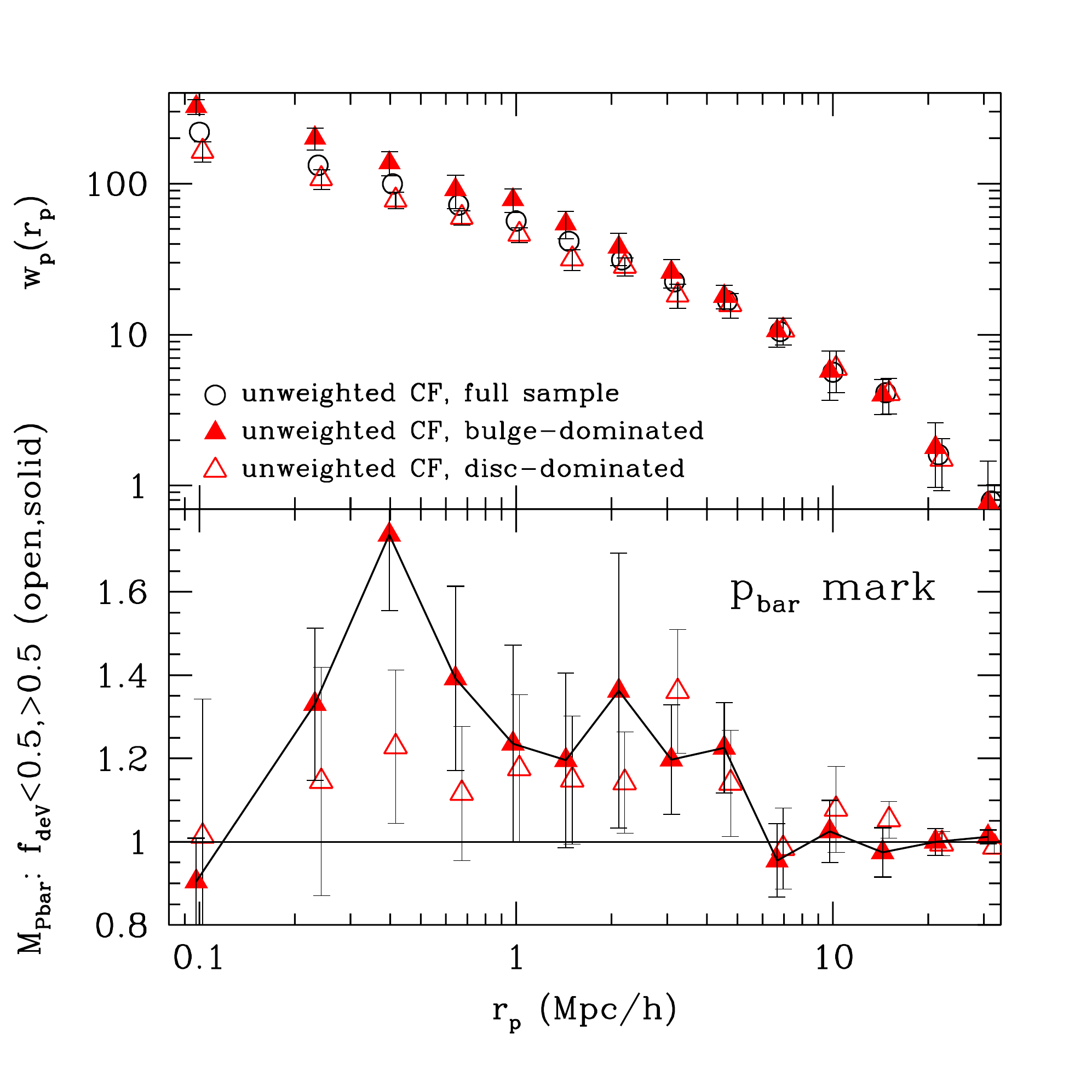} %{wp_Mp_Mr194Galextcorr_Pbarrescaled_fracdeVcuts_new_N30err.ps} %wp_Mp_Mr194_Pbarrescaled_fracdeVcuts_1err.ps %wp_Mp_Mr194_Pbar_fracdeVcuts_errors.ps 
 \includegraphics[width=0.497\hsize]{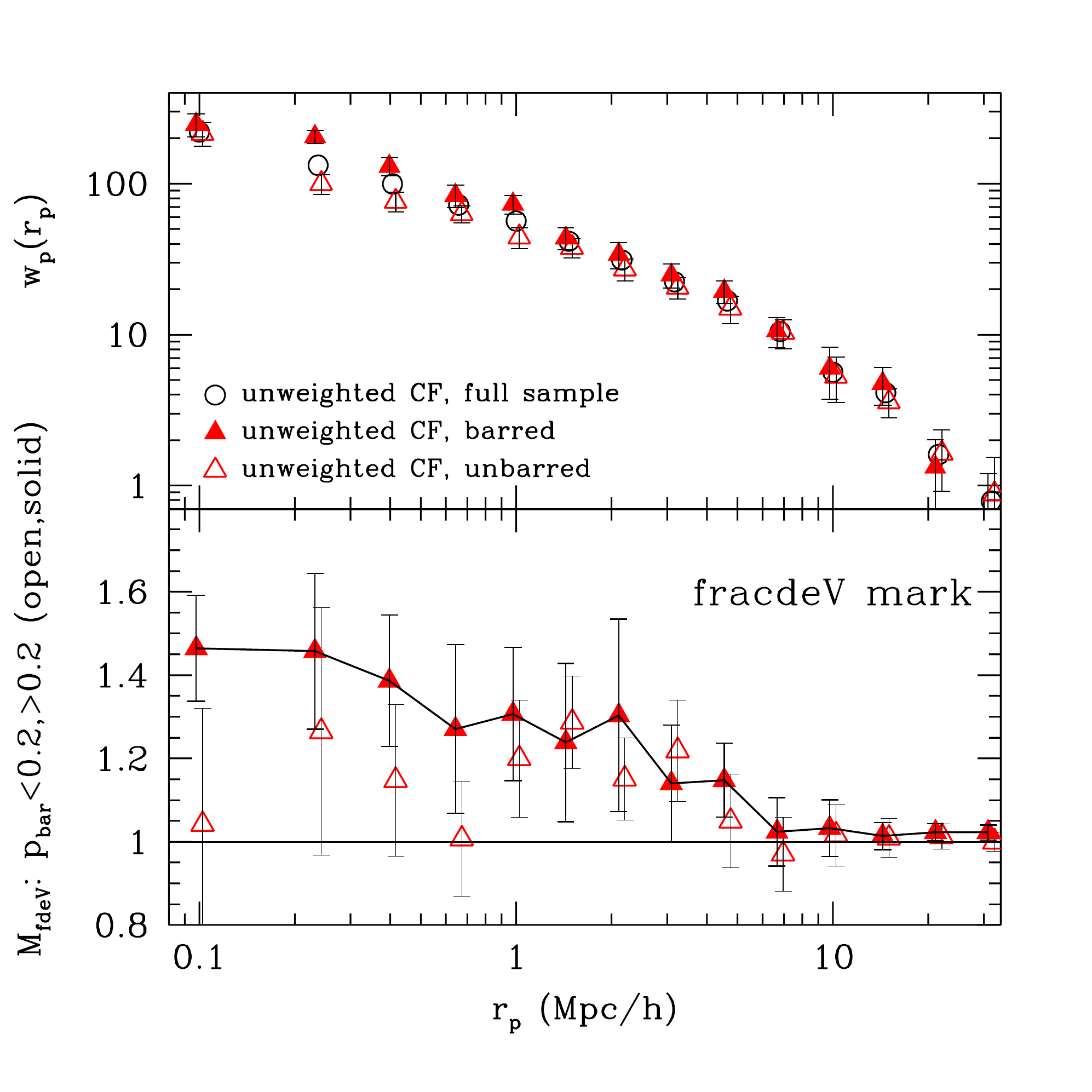} %{wp_Mp_Mr194Galextcorr_fracdeVrescaled_Pbarcuts02_new_N30err.ps} %wp_Mp_Mr194_fracdeVrescaled_Pbarcuts02_1err.ps %wp_Mp_Mr194_fracdeV_Pbarcuts02_new100.ps 
 \caption{Left plot, lower panel: $p_\mathrm{bar}$ mark correlation functions with 
          fracdeV$<0.5$ (disc-dominated; open triangle points) and $>0.5$ (bulge-dominated; 
          solid triangles).  Right plot, lower panel: fracdeV mark correlation functions 
          with $p_\mathrm{bar}<0.2$ (unbarred; open triangles) and $>0.2$ (barred; solid 
          triangles); the split is done at {0.2} in order to have a similar number of 
          galaxies in each subsample.  Mark distributions have been rescaled to match 
          distribution of whole sample (Fig.~\ref{fig2}).  For clarity, the points are 
          slightly offset from each other, %error bars are omitted for the mark correlation 
          %functions of disc-dominated and unbarred galaxies, 
          and the points are connected for 
          bulge-dominated and barred galaxies, to guide the eye.  Upper panels show the 
          (unweighted) correlation functions, for the full sample (same as in Fig.~\ref{fig6}) 
          and the subsamples, indicated by the legends in the figures.}
 \label{fig7}
\end{figure*}

In the lower panels of Figure~\ref{fig7}, we show the $p_\mathrm{bar}$ mark 
correlation functions for bulge-dominated and disc-dominated galaxies (fracdeV$>0.5$ and $<0.5$), 
as well as the fracdeV mark correlation functions for barred and unbarred galaxies 
($p_\mathrm{bar}>0.2$ and $<0.2$).  Using the fracdeV$>0.5$ threshold, $44\%$ of our (disc) galaxy 
catalogue is bulge-dominated, and using $p_\mathrm{bar}>0.2$, $49\%$ of it is barred. 
Following the procedure described in the appendix of S09, the mark correlations 
are shown when the marks are rescaled so that they have the same distribution 
as that of the whole sample (see Fig.~\ref{fig2}).  Such a rescaling is necessary 
in order to compare the mark correlation functions.  (In this case, the 
mark correlation measurements are similar, within $\sim$10\%, when the mark 
distributions are not rescaled.) 
%
% show these plots instead (and I don't think appendix is necessary for this):
% wp_Mp_Mr194_Pbarrescaled_fracdeVcuts_1err.ps & wp_Mp_Mr194_fracdeVrescaled_Pbarcuts02.ps
% rather than wp_Mp_Mr194_Pbar_fracdeVcuts_errors.ps & wp_Mp_Mr194_fracdeV_Pbarcuts02_new100.ps

The $p_\mathrm{bar}$ and fracdeV mark correlation functions are all still above unity, 
but they are statistically significant only for bulge-dominated (fracdeV$>0.5$) and 
barred ($p_\mathrm{bar}>0.2$) galaxies, respectively. 
Using Eqn.~\ref{sigmaM}, these $p_\mathrm{bar}$ and fracdeV mark correlations 
both have a statistical significance of $6\sigma$. %$10\sigma$ w/o outlying JK subsample
%why do these have such high statistical significance?  because e.g. the Mpbar(400kpc) point already has a high significance alone.
% should I quote statistical significance of fdeV<0.5 & pbar>0.2 measurements too?
%
%%%%% Bob thinks the following is a strong statement, and should be in abstract! %%%%%
In other words, bulge-dominated galaxies exhibit a significant bar-environment correlation, 
and barred galaxies exhibit a bulge-environment correlation.  
%(On the other hand, unbulged galaxies lack a strong bar-environment correlation, and unbarred galaxies lack a bulge-environment correlation.)  
Considering that these residual environmental correlations are so significant, it appears 
that the environmental dependencies of barred and bulge-dominated galaxies are somewhat 
independent of each other: the bar-environment correlation is not due to the 
bulge-environment correlation, and vice versa. 
%We conclude from this that \textit{the environmental dependencies of barred and bulged galaxies appear to be mostly independent of each other.} 
(The environmental dependencies of bars and \textit{pseudo}-bulges, %(e.g., Drory \& Fisher 2007), 
however, may be more closely related, as discussed in the introduction.) 
Lastly, we point out that though S0s are to some extent environmentally dependent (Hoyle et al.\ 2011; Wilman \& Erwin 2012), they are not likely to be driving the bar-environment correlation in the left panels of Fig.~\ref{fig7}.  S0s do not have a particularly large bar fraction compared to their spiral counterparts (Laurikainen et al.\ 2009; Buta et al.\ 2010), and their bar fraction does not exhibit a significant environmental dependence (Barway et al.\ 2011; Marinova et al.\ 2012). 
%and in any case, our ``bulge-dominated'' galaxies include S0s and bulge-dominated spirals.

\section{Interpretation of the $p_\mathrm{bar}$-environment correlation}\label{sec:interpretation}

%[let's try to emphasize \textit{physical} interpretations of our clustering results, although some of that could go in the concluding section.]

%\textbf{need better segue.} 
In the previous section, we quantified the environmental dependence of barred galaxies, 
using projected clustering measurements with the largest catalogue of galaxies with bar classifications to date.  
Here we perform tests and analyses of these results, in order to better understand 
the origin of these environmental correlations. 

We also quantified the environmental dependence of galaxy bulges, but as stated in 
Section~\ref{MCFs}, this has been thoroughly studied already and is closely related 
to the morphology-density relation.  Furthermore, the colour and stellar mass 
dependence of the morphology-density relation has been studied elsewhere 
(e.g., Kauffmann et al.\ 2004; Blanton et al.\ 2005; Park et al.\ 2007; van der Wel et al.\ 2010), 
including with Galaxy Zoo data (Bamford et al.\ 2009; Skibba et al.\ 2009), 
so we will not study it further here. 

In Section~\ref{sec:MassColor}, we examine the stellar mass and colour 
dependence of the measured $p_\mathrm{bar}$-environment correlation.  
Then in Section~\ref{sec:mocktest}, we use mock galaxy catalogues to predict the strength of the 
$p_\mathrm{bar}$-environment correlation if it were entirely due to redder galaxies occupying more massive dark matter haloes.  
Lastly, we apply halo occupation models to the clustering of barred and unbarred galaxies in Section~\ref{sec:HODmodels}.

\subsection{Dependence of the environmental correlation on stellar mass and colour}\label{sec:MassColor}

The probability of a galaxy being barred is correlated with its stellar mass 
(see Fig.~\ref{fig5}b; Nair \& Abraham 2010), so it is important to ask whether the environmental 
dependence of barred galaxies measured in Section~\ref{MCFs} is %simply 
due to the environmental dependence of stellar mass. 
Li et al.\ (2009) argue that in their catalogue, at fixed stellar mass, the projected clustering 
of barred and unbarred galaxies is similar. %, which is evidence in favor of this statement. 
With a much larger volume-limited catalogue, we can now 
analyze the stellar mass and colour dependence with greater accuracy. 
%When we split the catalogue by stellar mass and measure the $p_\mathrm{bar}$ mark 
%correlation function in each stellar mass bin (not shown), we do obtain a correlation between 
%$p_\mathrm{bar}$ and the environment, contrary to the Li et al.\ (2009) result; 
%however, it is of weak statistical significance ($2\sigma$ at most). 
%We have also measured mark \textit{cross}-correlation functions, by correlating this 
%catalogue with a larger catalogue (in which we did not apply the inclination cut), 
%but the resulting mark correlations are still of weak significance. 

%and mark (cross?)correlations at fixed stellar mass (and/or color).  
%(the problem with the mark cross-corrs is that we only get a gain of about 1.5x.) 
%the mark correlations are still noisy (noisier than Fig.~\ref{fig7}), 
%and i'm not sure how to improve them. 
%measurements with $M_\ast$ bins would be useful though (even if noisy), to compare to Li et al.\ (2009) in particular. 
%
%\begin{figure}
% \includegraphics[width=\hsize]{wp_Mp_Mr194_Pbar_Mstarcuts_new3.ps}
% \caption{Pbar mark, $9.75\leq M_\ast<10.25$ and $10.25\leq M_\ast<10.75$.}
% \label{fig8}
%\end{figure}
% wp_Mp_Mr194_Pbar_Mstarcuts_new2.ps wp_Mp_Mr194_fracdeV_Mstarcuts_new2.ps
% 9.75<=logM*<10.25 & 10.25<=logM*<10.75.   These plots still aren't that great though.

\subsubsection{Mark shuffling test}\label{sec:shuffle}

Rather than splitting our catalogue by stellar mass and then measuring the $p_\mathrm{bar}$ mark correlation function of the subcatalogues, we can take better advantage of 
the number statistics with a different test. %of the stellar mass dependence. 
%We use the distribution of bar probability as a function of stellar mass 
%($p(p_\mathrm{bar}|M_\ast)$ in Fig.~\ref{fig5}b), 
%and randomly shuffle the marks at a given mass and repeat the clustering measurement.  
Our procedure is as follows. 
We randomly shuffle the $p_\mathrm{bar}$ marks at a given stellar mass, 
and then repeat the clustering measurement. 
The distribution $p(p_\mathrm{bar}|M_\ast)$ does depend on stellar mass, but 
there is substantial scatter, so it is unclear a priori what the resulting 
mark correlation function would look like. 
The result can be directly compared to the original measurement (Fig.~\ref{fig6}a), because 
by merely shuffling the marks we are not changing the overall mark distribution.  
If the resulting mark correlation function is similar to the original (unshuffled) one, 
then this could be interpreted as evidence that the $p_\mathrm{bar}$ mark correlation 
is due to the environmental correlation of stellar mass.  If the resulting mark correlation 
is weak but significant, then the environmental correlation is partly due to stellar mass, and if there is no mark correlation, then it is not due to stellar mass at all.

\begin{figure*}
 \includegraphics[angle=270,width=0.497\hsize]{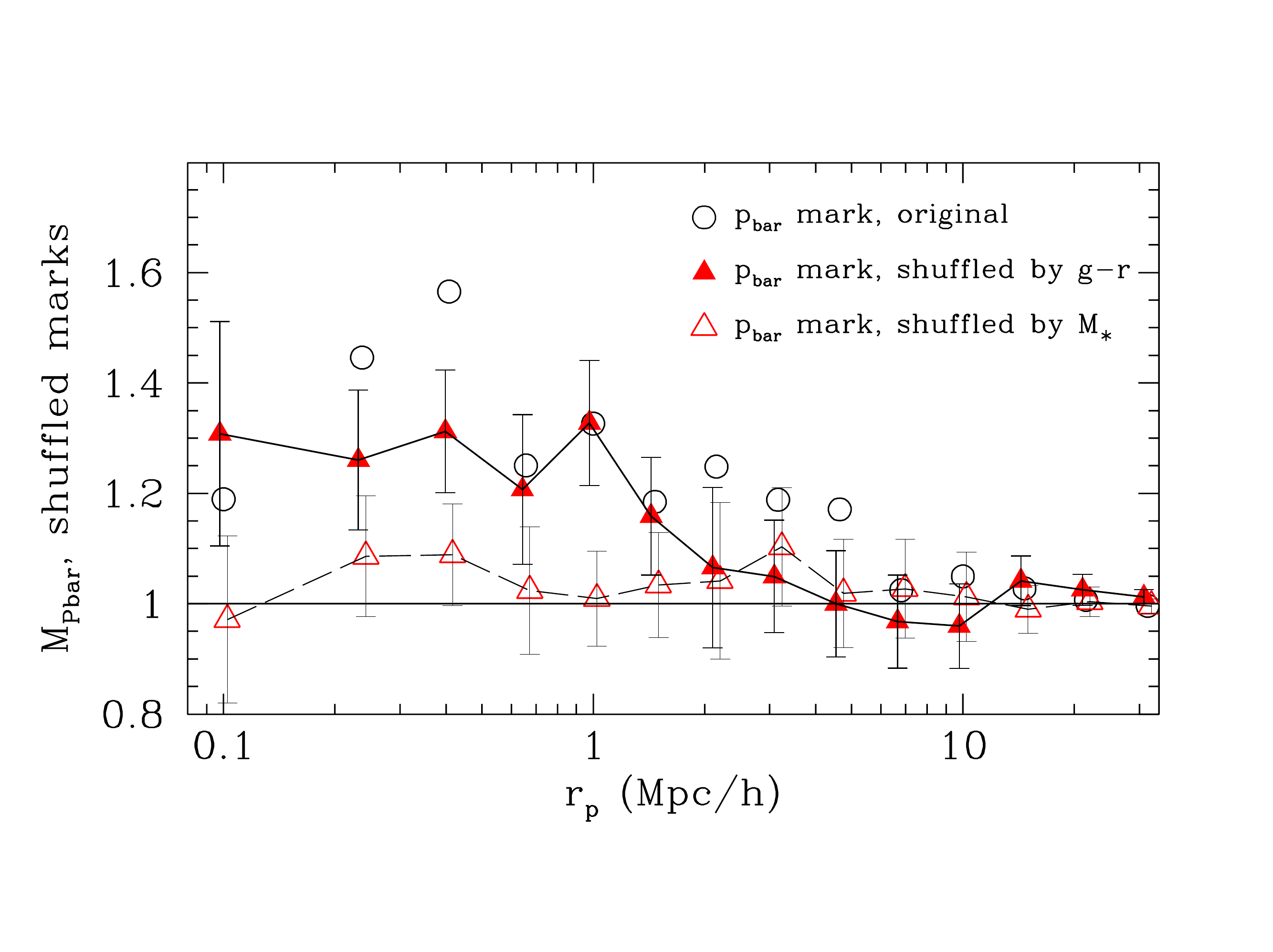} %{Mp_Mr194_Pbar_shuffled2b_new_N30err.ps} %new3d.ps
 \includegraphics[angle=270,width=0.497\hsize]{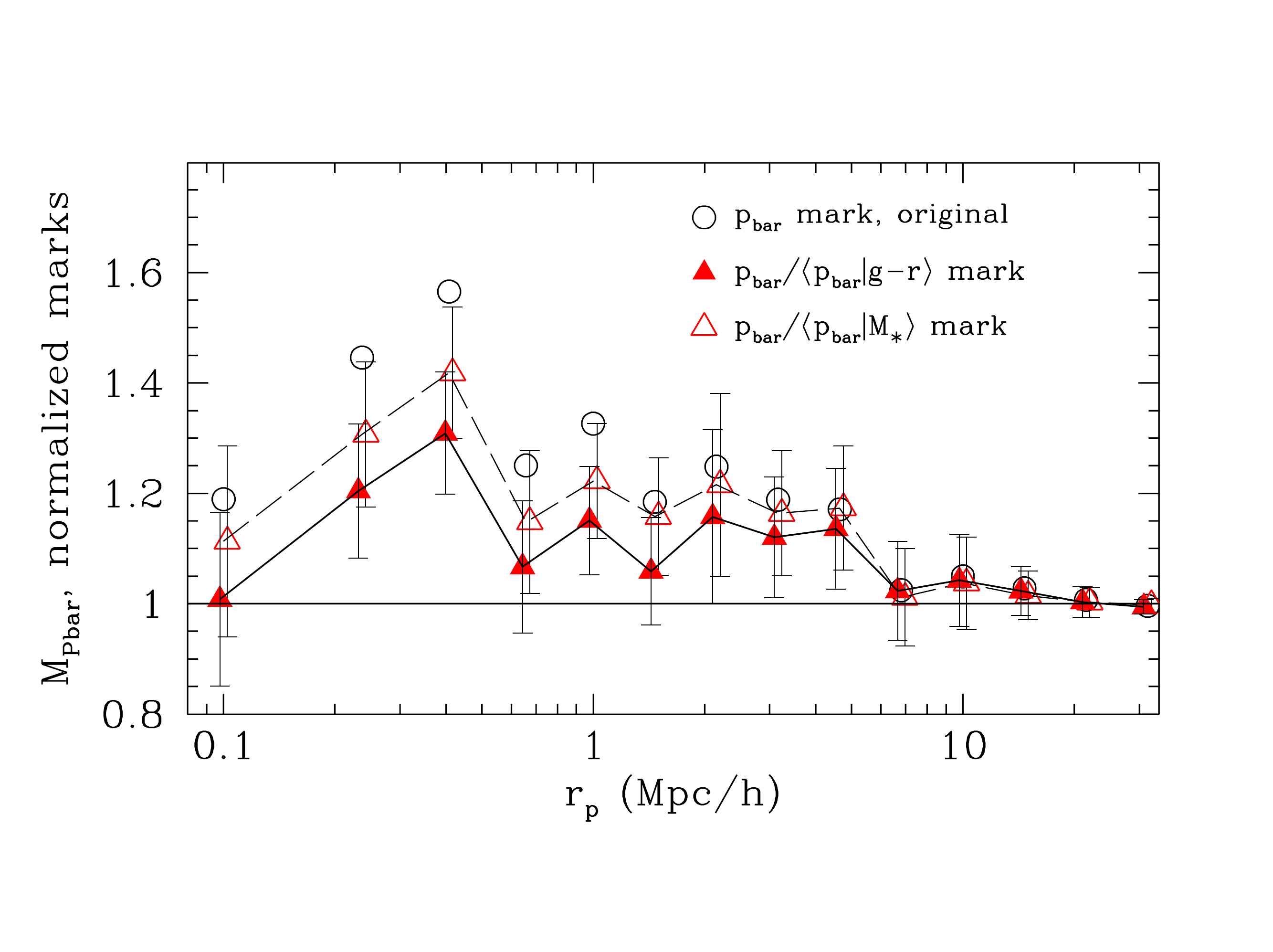} %{Mp_Mr194_Pbar_normalized_new_N30err.ps} %new3d.ps
 %I used dev postlandfile for these plots
 \caption{Left: $p_\mathrm{bar}$ mark correlation function, shuffled as a function of $g-r$ colour (solid triangles) and stellar mass (open triangles).  
          Right: $p_\mathrm{bar}/\langle p_\mathrm{bar}|g-r\rangle$ mark (solid triangles) and $p_\mathrm{bar}/\langle p_\mathrm{bar}|M_\ast\rangle$ mark (open triangles) correlation functions. 
          For comparison, the original $p_\mathrm{bar}$ 
mark correlation function is also shown (open circles, same as Fig.~\ref{fig6}a). 
          Ten bins were used, most with $\sim$1000-1400 galaxies/bin; using fewer or more bins yields similar results.  The unweighted correlation function (upper panel in previous two figures) is omitted, because the full sample is used for all three mark correlation measurements.}
 \label{fig9}
\end{figure*}

We performed this test using ten bins of stellar mass (of width 0.15~dex), and the result 
yields no significant environmental correlation, as shown in Figure~\ref{fig9}a (open triangles in the lower panel). 
Shuffling by stellar mass appears to nearly completely wash out the correlation, 
as $M_{p_\mathrm{bar}}$ is consistent with unity in every $r_p$ bin. 
%This is surprising, because from the Li et al.\ (2009) result, one might expect 
%that the clustering dependence of $p_\mathrm{bar}$ is mostly due to that of stellar mass. 
%On the contrary, we conclude that the environmental dependence of $p_\mathrm{bar}$ is \textit{not} 
%due to that of stellar mass.  
%This result is independent of the width of the stellar mass bins. 
%\textbf{Downplay this, because $M_\ast$ errors $\sim$bin width.}
Note, however, that the stellar mass uncertainties are comparable or larger than the bin widths, 
which could artificially weaken the correlation.  
%(We will perform another test below, which does not involve binning.)

%need to change height.  or edit device postencap

%Karen: We should comment on the two open circles in bottom panel of Fig 8 which are above the (g-r) colour shuffled correlation. This indicates that at these scales (but only these scales) the bar-environment correlation cannot simply be explained by the colour-environment and colour-bar correlations.
%
We have also performed the same test by shuffling the $p_\mathrm{bar}$ marks as a 
function of %extinction-corrected 
$g-r$ colour (see the distribution in Fig.~\ref{fig5}a), 
and in this case, the $p_\mathrm{bar}$ mark correlation (solid triangles in 
Fig.~\ref{fig9}a) is nearly as strong as the original mark correlation measurement. 
%with an overall $6\sigma$ significance. 
This suggests that the environmental dependence of %extinction-corrected 
colour partially explains that of $p_\mathrm{bar}$. 
The colour-shuffled correlation function does not reproduce either the upturn at 
$400~\mathrm{kpc}/h$ or the downturn at $100~\mathrm{kpc}/h$; however, we attribute 
this to the shuffling process. 

By taking the ratio of the marked correlation functions in Figure~\ref{fig9}a,  
we can make an approximate estimate of the fraction of the environmental 
dependence of $p_\mathrm{bar}$ that is accounted for by colour and stellar mass.  
In particular, we use all of the jack-knife subsamples (not just the measurements 
in the figure) to estimate this as robustly as possible, and we use the mean and 
variance of the ratio $(M'-1)/(M_{P\mathrm{bar}}-1)$, where $M'$ is either the 
colour-shuffled or mass-shuffled mark correlation function.  We use the 
measurements over the range $0.1\le r_p\le2.2~\mathrm{Mpc}/h$, which encompasses 
the environmental correlations within dark matter haloes.  We find that colour 
accounts for $60\pm10~\%$ of the environmental dependence of $p_\mathrm{bar}$, 
while stellar mass accounts for only $15\pm2~\%$. 
%By taking the ratio of the marked correlation functions in Figure~\ref{fig9}a, we 
%estimate that colour accounts for $60\pm30~\%$ of the environmental 
%dependence of $p_\mathrm{bar}$, while stellar mass accounts for only $10\pm10~\%$.  
\textit{This suggests that the environmental dependence of colour explains the majority, but not all, of the environmental dependence of bars.}  
Our results are not consistent with Lee et al.\ (2012a), who claim that the 
environmental dependence of bars disappears at fixed colour or central velocity dispersion, 
and the disagreement may be due to their use of different bar classifications and environment measures; in addition, lenticular galaxies are excluded from their sample, but not from ours.

\subsubsection{Normalized mark test}\label{sec:norm}

%%%%%  NORMALIZED MARK TEST %%%%%
%[\textbf{normalized mark test.}  this purpose of this test is to remove the environmental dependence of stellar mass or colour (see e.g. Cooper et al.\ 2010), %3rd para,Sec.3 
%and consequently determine the residual environmental dependence of $p_\mathrm{bar}$.] 
It is possible that the contribution from stellar mass is larger than estimated above, 
because the masses have larger uncertainties than the colours. 
To address this, we perform another test of the stellar mass and colour contribution 
to the bar-environment correlation, which does not involve binning these parameters. 
The purpose of this test is to remove the environmental dependence of stellar mass or 
colour (see e.g. Cooper et al.\ 2010), and consequently assess the strength of the 
residual environmental dependence of $p_\mathrm{bar}$. 

%%%%%% see GZ2barenviro_v2_newemails.txt %%%%%%

Our procedure is as follows. 
For every galaxy, the $p_\mathrm{bar}$ mark is normalized by the mean 
$p_\mathrm{bar}$ of galaxies with that stellar mass (i.e., we use 
$p_\mathrm{bar}/\langle p_\mathrm{bar}|M_\ast\rangle$ as the mark). 
Note that the mean is slightly larger than the median, which is plotted 
in Figure~\ref{fig5}, and is similarly a smooth function of stellar mass, 
so this normalization is not sensitive to the mass uncertainties. 
Then the mark distribution is rescaled so that it matches the overall 
$p_\mathrm{bar}$ distribution (as was done in Sec.~\ref{disentangleMCFs}), 
because consistent mark distributions are required in order to 
compare mark correlations.  Now the new mark correlation function is 
measured, and can be compared to the original one.  With this test, if 
the mark correlation function were close to unity, it would mean that 
$M_\ast$ accounts for most of the environmental correlation.  The same 
test is also done to assess the contribution of the $g-r$ 
colour-environment correlation, with the analogous 
$p_\mathrm{bar}/\langle p_\mathrm{bar}|g-r\rangle$ mark.

The result is shown in Figure~\ref{fig9}b. 
The colour-normalized mark correlation function is closer to unity, 
and therefore accounts for more of the bar-environment correlation. 
As in the previous section, we can estimate the relative contribution of 
colour and mass to this correlation, now using the ratio 
$(M_{P\mathrm{bar}}-M')/(M_{P\mathrm{bar}}-1)$, where $M'$ is the colour- 
or mass-normalized mark correlation function. This yields an estimate of 
$60\pm5~\%$ of the bar-environment correlation accounted for by colour, 
consistent with the mark shuffling test in Section~\ref{sec:shuffle}. 
Stellar mass now accounts for $25\pm10~\%$, a larger contribution than 
estimated above, but still less significant than colour. 
%and taking the ratio of the mark correlations yields an estimate of 
%$60\pm30~\%$ of the bar-environment correlation accounted for by colour, 
%consistent with the mark shuffling test in Section~\ref{sec:shuffle}. 
%Stellar mass now accounts for $25\pm15~\%$, a larger contribution than estimated above, but still less significant than colour. 
%
%\textbf{Address Li et al.\ clustering result.} 
%This is a surprising result, because from the Li et al.\ (2009) clustering analysis, 
%one might expect that the clustering dependence of $p_\mathrm{bar}$ is 
%\textit{mostly} due to that of stellar mass. On the contrary, we conclude that 
We conclude that the environmental dependence of $p_\mathrm{bar}$ is \textit{not} primarily due to that of stellar mass.  

Perhaps more than stellar mass, the colour is a better tracer of 
star formation (and dust content) in disc galaxies, %(e.g., Masters et al.\ 2010a,b), 
which in turn is expected to be related to the likelihood of the galaxies 
having a bar (e.g., Scannapieco et al.\ 2010; Masters et al.\ 2010b). 
Redder disc galaxies with older stellar populations and in more massive haloes 
are more likely to have formed a stable bar; however, mergers/interactions can 
disrupt a bar, which could explain why the original $p_\mathrm{bar}$ mark 
correlation function (circle points in Fig.~\ref{fig9}), unlike the 
colour-shuffled one, turns toward unity at small separations. 
%
%\textbf{how is color important here but stellar mass isn't?}  is extinction-corrected color a better tracer of dust (and gas?) content than SFR?  or is there a simpler explanation? 
%\textbf{or does anyone have a simpler explanation?}

%[shall we perform the same test for the fracdeV mark?  I'm not sure it's necessary.  if we do do it, maybe it should go in an appendix, rather than here.]

\subsection{Colour dependence in mock galaxy catalogues}\label{sec:mocktest}

To add to the interpretation of the colour dependence of the $p_\mathrm{bar}$-environment 
correlation in the previous section, we analyze the clustering of galaxies in a mock 
galaxy catalogue, in which we add bar likelihoods with a prescription based on galaxy colour. 

We use the mock catalogue of Muldrew et al.\ (2012), which was constructed by 
populating dark matter haloes of the Millennium Simulation (Springel et al.\ 2005) using 
the halo occupation model of Skibba \& Sheth (2009).  The catalogue reproduces the 
observed luminosity function, colour-magnitude distribution, and the luminosity 
and colour dependence of galaxy clustering in the SDSS (Skibba et al.\ 2006; Skibba \& Sheth 2009). 
Central galaxies in haloes are distinguished from satellite galaxies, which are 
distributed around them and are assumed to follow a Navarro, Frenk \& White (1996) 
profile with the mass-concentration relation from Macci\`{o}, Dutton \& van den Bosch (2008). 

%describe the three step process by which we created this mock.  
%the mock approximately reproduces $p(M_r)$, $p(g-r|M_r)$ (Fig.~\ref{fig4}), $p(p_\mathrm{bar})$ (Fig.~\ref{fig2}), and $p(p_\mathrm{bar}|g-r)$ (Fig.~\ref{fig5})...
For the purposes of this work, which is focused on disc galaxies, we construct a 
sub-catalogue from this mock, by selecting galaxies from the colour-magnitude distribution.  
In particular, we first select galaxies with $M_r-5\mathrm{log}(h)\leq -19.4$.  We require that the luminosity 
function is consistent with the data.  (Since the catalogue was constrained with 
absolute magnitudes $K$-corrected to $z=0.1$, we use the $^{0.1}M_r$ luminosity function.)  
We randomly select galaxies (independently of halo mass or central/satellite status) 
in absolute magnitude bins until the consistent luminosity function is obtained.  
Secondly, we similarly require that the colour-magnitude distribution, $p(g-r|M_r)$, 
is consistent with the data, using bins of 0.25 mag (see e.g. Skibba \& Sheth 2009). 
%although a double-Gaussian fit was not applied here.  
The selection of disc galaxies in the GZ catalogue means that 
the red sequence is under-represented (M11).  
Finally, we use $p(p_\mathrm{bar})$ (Fig.~\ref{fig2}), and $p(p_\mathrm{bar}|g-r)$ 
(Fig.~\ref{fig5}) distributions to generate ``$p_\mathrm{bar}$" for the mock galaxies.  
That is, we assume that the environmental dependence of $p_\mathrm{bar}$ is due 
to that of $g-r$ colour, which in turn is due to more massive haloes in dense environments.

\begin{figure}
 \includegraphics[width=\hsize]{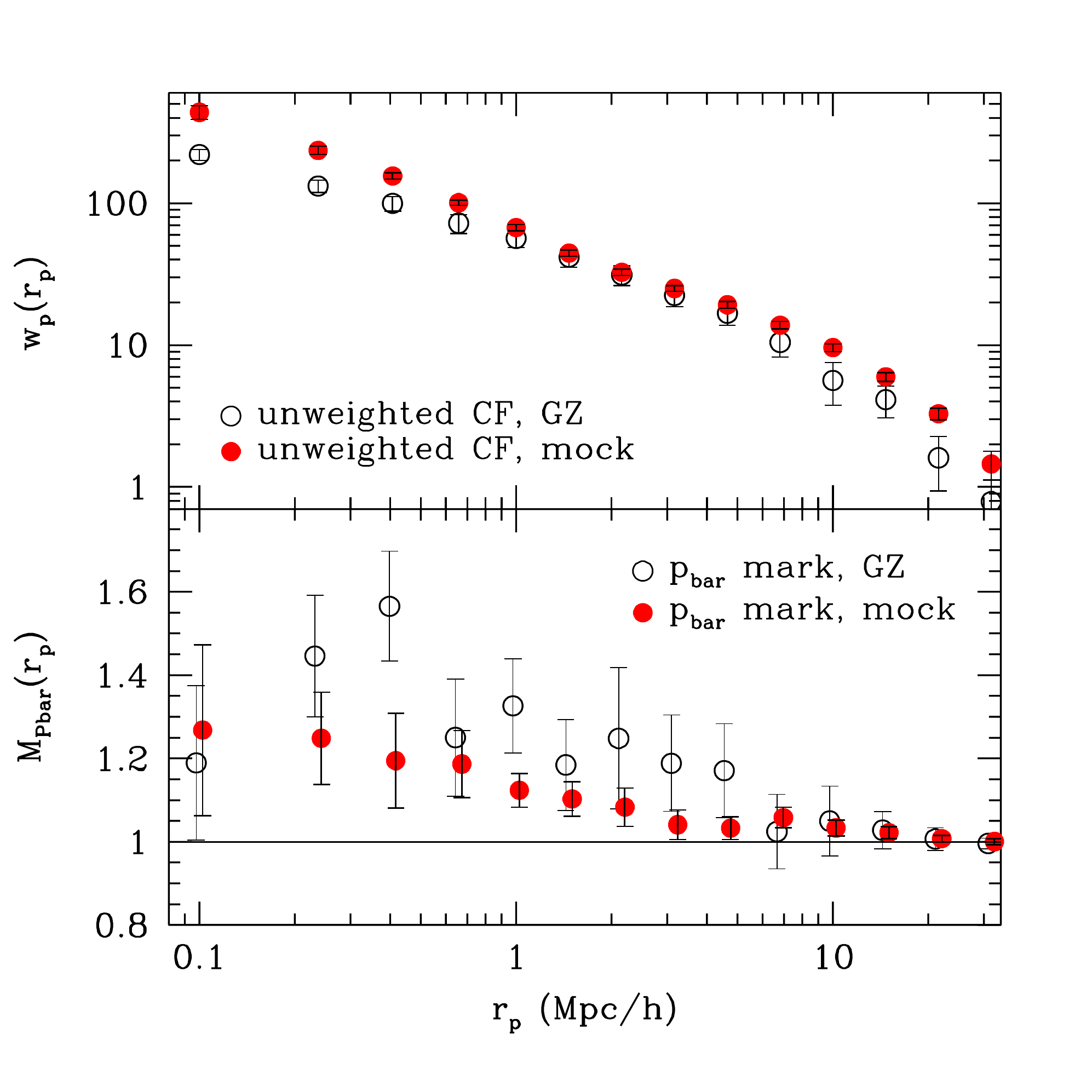} %{wpMp_mock_mean_JKerrors_labels2b.ps} %{wpMp_mock_mean.ps} %wpMp_mock_test121.ps
 \caption{Clustering comparison between the Galaxy Zoo and mock catalogues.  
          Solid red circles indicate the projected correlation function and mark correlation 
          function of the mock catalogue, using $p_\mathrm{bar}(g-r)$ 
          as the mark (see distribution in Fig.~\ref{fig5}a, and text for details). 
          For comparison, the original GZ measurements are also shown here (open circles, same as Fig.~\ref{fig6}a); the points are slightly offset in the lower panel, for clarity. 
          %\textbf{[I still need to add JK error bars; the variance among the eight mocks is too small for an error estimate.]} 
         }
 \label{fig10}
\end{figure}

We can now measure the projected correlation function and $p_\mathrm{bar}$ marked 
correlation function of the mock catalogue, in order to compare to the GZ measurements in Figure~\ref{fig6}.  
The result (averaged over eight realizations) is shown in Figure~\ref{fig10}. 
As with the observational measurements, the errors are estimated using jack-knife resampling; the variance of the eight mocks is much smaller. 
If we were to apply the observed errors instead (and account for the different size of the 
GZ and mock catalogues), we obtain similar error bars at large scales but smaller ones at small scales ($r_p<\mathrm{few}\,\mathrm{Mpc}/h$).

In the upper panel, the discrepancy between these projected correlation functions at large scales 
has been previously observed and is not statistically significant (see Zehavi et al.\ 2005; Skibba et al.\ 2006); it is likely due to cosmic variance. 
The discrepancy at small scales, however, is significant.  
The fact that the correlation functions are consistent at scales of $r_p\geq1\,\mathrm{Mpc}/h$, 
%but not at smaller separations might mean that the satellite fraction of the mock 
%catalogue ($f_\mathrm{sat}\approx23\%$) is inconsistent with the data.  
%The suppressed small-scale clustering of the GZ (disc galaxy) catalogue is likely 
%due to a slightly different than expected satellite distribution as a function of halo mass, 
%which the colour-magnitude selection procedure does not reproduce. 
but the small-scale clustering of the GZ catalogue is suppressed, could mean that 
the satellite distribution as a function of halo mass is slightly different in the 
real universe, and is not reproduced with the colour-magnitude selection procedure. 
%
% I could also measure the clustering of "barred" and "unbarred" galaxies in the mock,  like in the upper panel of Fig.6b, but this won't be useful since the total wp(rp) doesn't match the data.  let's stick to HOD modeling those wp(rp) in the next section.

The $p_\mathrm{bar}(g-r)$ mark correlation function of the mock is weaker than 
the GZ measurement, but similar to the ($g-r$)-shuffled mark measurement in Figure~\ref{fig9}a. 
This suggests that part, but not all, of the environmental dependence of 
$p_\mathrm{bar}$ is due to more massive haloes hosting redder galaxies, which 
are more likely than average to be barred. 
By taking the ratio of the marked correlation functions, 
$(M_\mathrm{mock}-1)/(M_\mathrm{GZ}-1)$, we estimate that the 
colour-halo mass correlation accounts for $50\pm10~\%$ of the environmental dependence 
of $p_\mathrm{bar}$, which is slightly lower than, but consistent with the estimate in 
Section~\ref{sec:MassColor}; conversely, the rest (also $50\pm10~\%$) 
%By taking the ratio of the marked correlation functions, we estimate that the 
%colour-halo mass correlation accounts for $50\pm20~\%$ of the environmental dependence 
%of $p_\mathrm{bar}$, consistent with the estimate in Section~\ref{sec:MassColor}; 
%conversely, the rest (also $50\pm20~\%$) 
is due to other %secular? 
processes unrelated to colour or stellar mass, perhaps involving the gas 
content (see Masters et al.\ 2012) or angular momentum distribution.

Also note that, as in Figure~\ref{fig9}, the $p_\mathrm{bar}(g-r)$ mark correlation 
function in Figure~\ref{fig10} lacks a drop in strength at $r_p\sim100\,\mathrm{kpc}/h$, 
which we see in the original clustering measurement. %(Fig.~\ref{fig6}a). 
This implies that, in the real universe, although galaxies at small separations (usually center-satellite galaxy 
pairs) tend to be redder in more massive haloes, this does not entail a higher bar 
fraction; the lack of a $p_\mathrm{bar}$-environment correlation at small separations 
in Figure~\ref{fig6}a is not related to galaxy colour. 

%Bob: would prefer HOD instead i.e., avg. number of galaxies per Mhalo.
%%%%% I SHOULD CONSIDER PLOTTING THIS LIKE AN HOD PLOT. i.e., <Ncen|M> & <Nsat|M> & <Nbar|M>?
%
% measure the $M_\mathrm{vir}$ distribution of the mock (which is hopefully $\sim$similar 
% to the HOD result), and $f_\mathrm{sat}$ (which is probably too low here).
% mass distributions here mock_Mr194_halomassdist_v111new.ps (& v222) or  ...v111_Pbar2.ps
% here are the raw numbers & fractions: mock_Mr194_halomassdist_v111new.dat (& v222)
\begin{figure}
 \includegraphics[width=\hsize]{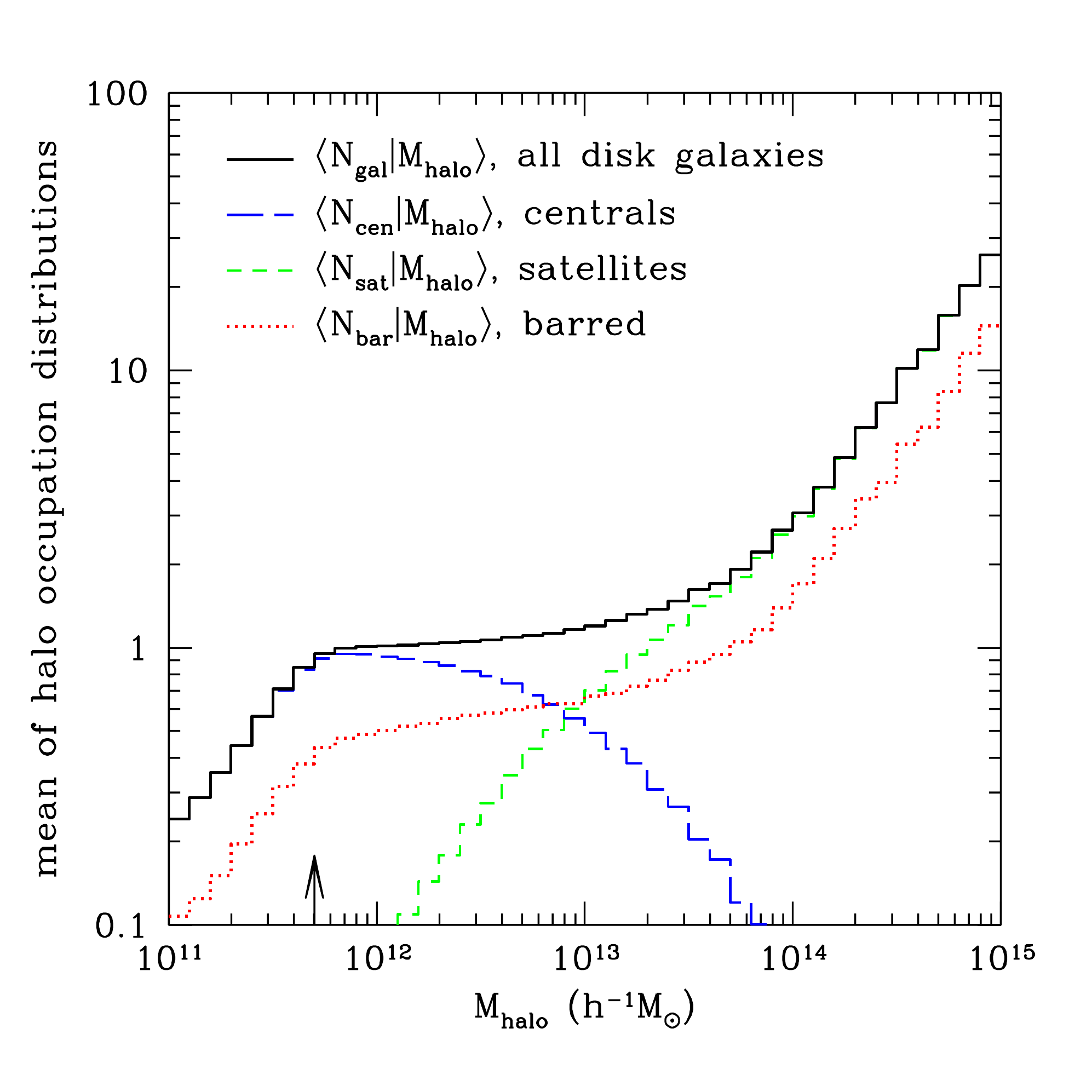} %{mock_Mr194_HOD_v111_Pbar2_labels_v2.ps} %{mock_Mr194_halomassdist_v111_Pbar2_labels.ps}
 \caption{Mean of halo occupation distributions in the mock catalogue. 
          Black solid histogram: mean occupation number of all (colour-magnitude selected) 
          disc galaxies in the mock; blue dashed histogram: mean occupation number 
          of central galaxies; green short-dashed histogram: mean occupation 
          number of satellite galaxies. 
          The peak of the halo mass distribution is indicated by the arrow. 
          The red dotted histogram shows the mean occupation number for galaxies 
          with $p_\mathrm{bar}>0.2$ (where the bar likelihood is computed from the 
          $p(p_\mathrm{bar}|g-r)$ distribution), which are a combination of 
          central galaxies in low-mass haloes and satellites in massive haloes.  
          A larger $p_\mathrm{bar}$ threshold yields a lower 
          $\langle N_\mathrm{bar}|M_\mathrm{halo}\rangle$, but of similar shape.
          %Halo mass distributions of galaxies in the mock. 
          %Black solid histogram: total distribution; blue dashed histogram: distribution for central galaxies; 
          %green short-dashed histogram: distribution for satellite galaxies.
          %The red dotted histogram shows the distribution for galaxies with $p_\mathrm{bar}>0.2$, 
          %which are a combination of central galaxies in low-mass haloes and satellites 
          %in massive haloes.  A larger $p_\mathrm{bar}$ threshold yields a lower 
          %distribution, but of similar shape.
         }
 \label{fig10b}
\end{figure}

Finally, we have computed the halo mass distribution and halo occupation 
distribution of galaxies in the mock catalogue.  The halo occupation 
distribution (HOD) is the number distribution of galaxies occupying haloes of a 
given mass, and of particular importance for galaxy clustering is the mean 
occupation function, $\langle N|M\rangle$ (which is described further in 
Section~\ref{sec:HODmodels}).  The mean occupation functions of galaxies in 
the mock are shown in Figure~\ref{fig10b}.  The mock galaxies are mostly hosted 
by haloes with masses $M_\mathrm{vir}\geq4\times10^{11}\,h^{-1}\,M_\odot$; 
there are fewer haloes less massive than this, due to the luminosity threshold 
($M_r=-19.4$).  The central galaxy HOD drops off at high masses because 
the central galaxies of these haloes rarely meet the CMD 
selection criteria of our GZ catalogue; to wit, many centrals in massive haloes 
are elliptical, not disc, galaxies (Skibba et al.\ 2009; Guo et al.\ 2009; De Lucia et al.\ 2011). %maybe also cite Villalobos et al. 2012 
Satellite galaxies dominate in number at 
masses of $M_\mathrm{vir}\geq10^{13}\,h^{-1}\,M_\odot$.  In the mock, the 
``barred" galaxies (determined from the $p(p_\mathrm{bar}|g-r)$ distribution), 
indicated by the dotted histogram, consist of a combination of central 
galaxies in low-mass haloes and satellites in massive haloes.  The fraction of 
barred galaxies in the mock is not strongly halo mass dependent, but it is 
highest between $10^{12.5}<M_\mathrm{vir}<10^{14.3}\,h^{-1}\,M_\odot$, in the 
haloes that typically host galaxy groups. 
The HOD statistics %halo mass distribution and satellite fraction 
of the mock catalogue will be compared to the results of halo occupation 
modeling, in the following section. 

%Finally, we have also computed the halo mass distribution of galaxies in the mock, shown in Figure~\ref{fig10b}. 
%(Note that in these distributions each \textit{galaxy} has one count, so haloes 
%hosting multiple galaxies are counted multiple times.) 
%The mock galaxies are mostly hosted by haloes with masses 
%$M_\mathrm{vir}\geq4\times10^{11}\,h^{-1}\,M_\odot$. 
%The central galaxy halo distribution drops off rapidly at high masses 
%because there are fewer massive haloes and their central galaxies rarely 
%meet the CMD selection criteria of our GZ catalogue; to wit, 
%many centrals in massive haloes are elliptical galaxies. 
%In the mock, the ``barred'' galaxies (determined from the $p(p_\mathrm{bar}|g-r)$ 
%distribution) consist of a combination of central galaxies in low-mass haloes and 
%satellite galaxies in massive haloes. 
%The fraction of barred galaxies in the mock is not strongly halo mass dependent, 
%but it is highest between $10^{12.5}<M_\mathrm{vir}<10^{14.3}\,h^{-1}\,M_\odot$. 
%The halo mass distribution and satellite fraction of the mock catalogue 
%can be compared to the results of halo occupation modeling, 
%in the following section. 

\subsection{Halo occupation modeling of the clustering measurements}\label{sec:HODmodels}

%\textbf{Put in context better.} %make clear that we're using real (not mock) data. 
In this section, complementary to the mock catalogue analysis of the previous section, 
we apply dark matter halo models to the measured projected 
correlation functions, $w_p(r_p)$, of the whole volume-limited sample of (disc) galaxies, 
and of the subsamples of barred and unbarred galaxies, plotted in the upper panels of 
Figure~\ref{fig6} and Figure~\ref{fig7}b.  
Since there are only small differences between these measurements for barred and unbarred galaxies, 
%and the error bars are not very small, 
one can expect small differences between the well-fitting models. 
%\textbf{[Is it worthwhile to apply HOD models to the (un)bulged subsamples too?  I think that's probably not necessary for the paper.]} 
% should we mention a possible dependence of the HOD fitting on the clustering errors, or even correlated errors (see Zehavi et al. 2005).
The purpose of the halo model analysis is to constrain the types of haloes 
that host barred and unbarred disc galaxies.

% this was from Skibba et al. (2009), from the red spirals section
We use a halo occupation model of galaxy clustering, (e.g., %Zehavi et al. 2005; 
Zheng et al.\ 2007; Zehavi et al.\ 2011), in which the halo occupation distribution, 
$P(N|M)$, of central and satellite galaxies depends on halo mass, $M$, 
and the luminosity threshold, $L_\mathrm{min}$.
In this case the luminosity threshold is $M_r\leq-19.4$, corresponding to 
an approximate halo mass threshold of $M_\mathrm{min}\approx4-5\times10^{11}\,h^{-1}\,M_\odot$ 
(which is consistent with the mock catalogues in Section~\ref{sec:mocktest}). 

Haloes of mass $M$ are occupied by $N_\mathrm{gal}$ galaxies, consisting of 
a single central galaxy and $N_\mathrm{sat}$ satellite galaxies, 
such that the mean occupation function is described as the following: 
%In halo occupation models, for galaxies more luminous than some threshold
%($M_r<-19.4$ in this case), corresponding to an approximate halo mass threshold ($M_\mathrm{min}\approx5\times10^{11}\,h^{-1}\,M_\odot$),
%haloes are occupied by a single central galaxy and $N_\mathrm{sat}$ satellite
%galaxies, where $\langle N_\mathrm{sat}|M,L_\mathrm{min}\rangle \approx (M/M_1)^\alpha$,
%$M_1\propto M_\mathrm{min}(L_\mathrm{min})$ and $\alpha$ increases with $L_\mathrm{min}$.
%In practice, we account for the fact that there is significant scatter
%in the relation between central galaxy luminosity and halo mass, and 
%that the satellite halo occupation function drops off more rapidly
%than a power-law at low masses just above $M_\mathrm{min}$...
%
% this is from SS09/Muldrew+
%...By construction, the number of galaxies consists of the 
%number of central galaxies plus the number of satellites, such that 
\begin{equation}
 \langle N_\mathrm{gal}|M,L_\mathrm{min}\rangle \,=\,
    \langle N_{\mathrm {cen}}|M,L_\mathrm{min}\rangle\,
    \Bigl[1 + \langle N_{\mathrm {sat}}|M,L_\mathrm{min}\rangle\Bigr]
 \label{censat}
\end{equation}
\noindent where,
\begin{equation}
  \langle N_\mathrm{cen}|M\rangle \,=\, \frac{1}{2}\Biggl[1\,+\,\mathrm{erf}\Biggl(\frac{\mathrm{log}(M/M_\mathrm{min})}{\sigma_{\mathrm{log}M}}\Biggr)\Biggr]
 \label{NcenM}
\end{equation}
\noindent and
\begin{equation}
  \langle N_\mathrm{sat}|M\rangle = 
    \Biggl(\frac{M-M_0}{M_1^{ ' }}\Biggr)^\alpha .
 \label{NsatM}
\end{equation}
%\noindent (See Appendix A2 of SS09 for details).  All of the free parameters depend on luminosity. 
In practice, we account for the fact that there is significant scatter
in the relation between central galaxy luminosity and halo mass, and 
that the satellite halo occupation function drops off more rapidly
than a power-law at low masses just above $M_\mathrm{min}$. 
See Appendix A2 of Skibba \& Sheth (2009) for details. %and Zheng+

% from S09: 
%The satellite galaxy fraction is then
%\begin{equation}
% f_\mathrm{sat} \,=\, \frac{
%   \int_{M_\mathrm{min}} dM\,(dn/dM)\,\langle N_\mathrm{sat}|M,L_\mathrm{min} \rangle}{
%   \int_{M_\mathrm{min}} dM\,(dn/dM)\,(\langle N_\mathrm{cen}|M\rangle +
%     \langle N_\mathrm{sat}|M,L_\mathrm{min} \rangle)},
%\end{equation}
%where $dn/dM$ is the halo mass function.
%The range of models that approximately fit the measurement (with $\chi_\mathrm{d.o.f.}^2<1$) 
%have a satellite fraction of $f_\mathrm{sat}\approx32\%$.
%This is fifty per cent larger than the satellite fraction of all galaxies with $M_r<-19.5$,
%$f_\mathrm{sat}\approx24\%$.
We will also use the halo occupation models to compare the fraction of 
satellite galaxies of barred and of unbarred galaxies. 
The satellite fraction is given by 
\begin{equation}
 f_\mathrm{sat} \,=\, \frac{
   \int_{M_\mathrm{min}} dM\,(dn/dM)\,\langle N_\mathrm{sat}|M\rangle}{
   \int_{M_\mathrm{min}} dM\,(dn/dM)\,(\langle N_\mathrm{cen}|M\rangle +
     \langle N_\mathrm{sat}|M\rangle)},
     %or just say \langle N_\mathrm{gal}|M\rangle, rather than Ncen+Nsat
 \label{fsatM}
\end{equation}
where $dn/dM$ is the halo mass function (Sheth \& Tormen 1999; Tinker et al.\ 2008b). 
Note that we will not attempt to account for the fact that central galaxies in 
massive haloes will often not meet the selection criteria for disc galaxies, 
because these galaxies will be dominated in number by satellites (see Fig.~\ref{fig10b}), 
whose abundance we can constrain.

\begin{figure}
 \includegraphics[width=\hsize]{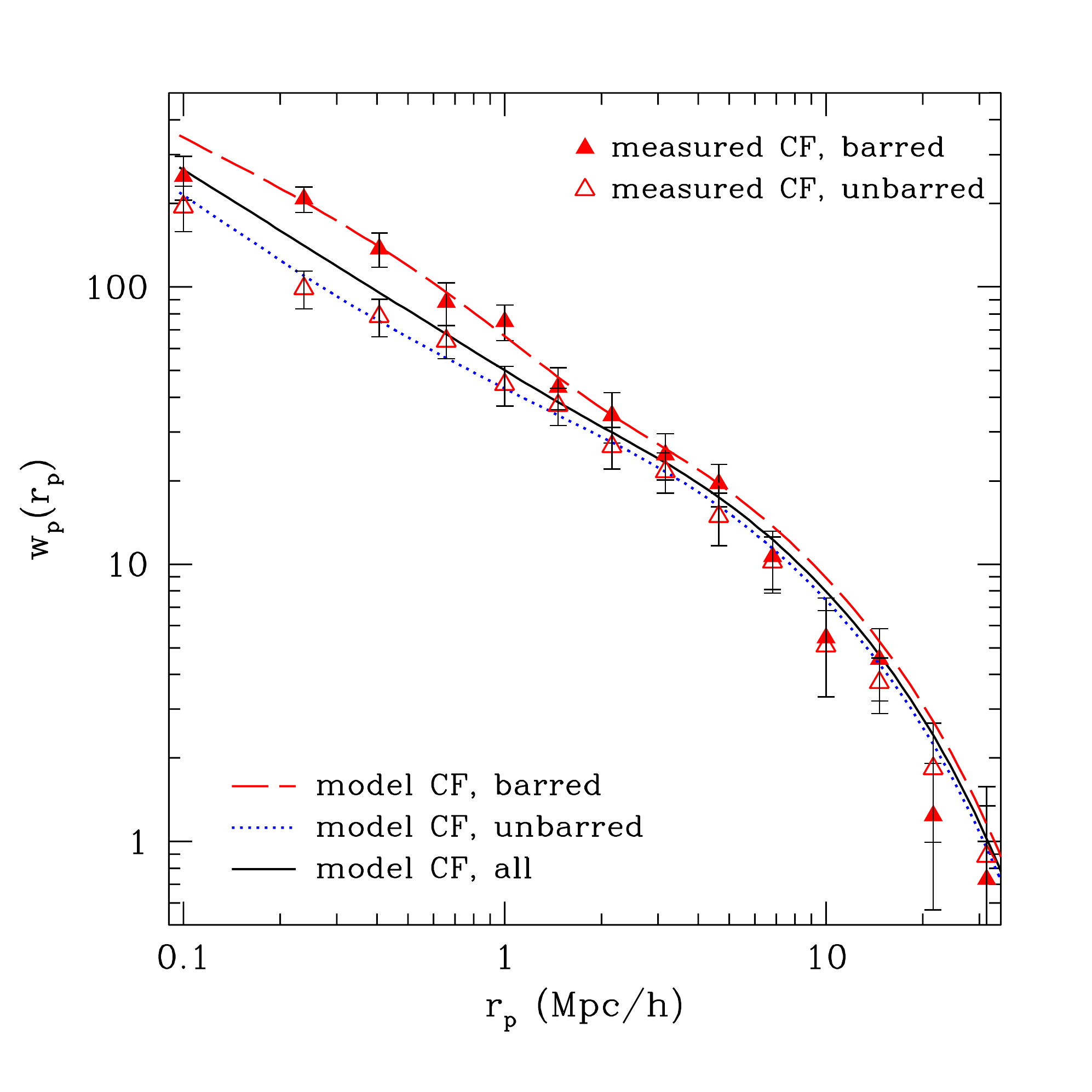} %{wprpbar_HODcomparison_labels.ps} %wprpbar_HODcomparison_test2.ps
 \caption{Halo occupation models of clustering of all galaxies in the sample 
          (solid black curve), the barred subsample of galaxies (red dashed curve), 
          and the unbarred subsample (blue dotted curve).  
          The corresponding measured correlation functions are also shown, for barred 
          and unbarred galaxies (solid and open triangle, respectively).}
 \label{fig11}
\end{figure}

% Idit: shall we show a table with the best-fit HOD parameters?
In Figure~\ref{fig11}, we show the results of the halo occupation modeling, 
applied to the whole catalogue and to the subsamples of barred and unbarred galaxies. 
The parameters of these HOD models are listed in Table~\ref{table1}. 
We fixed the parameters $\sigma_{\mathrm{log}M}=0.26$ and $M_0=10^{11.60}M_\odot/h$, because they are not constrained well by HOD models of clustering (see Zheng et al.\ 2007; Zehavi et al.\ 2011);  $\sigma_{\mathrm{log}M}$, which quantifies the scatter between central galaxy luminosity and halo mass, can be constrained by satellite kinematics and the conditional luminosity function (More et al.\ 2009; Cacciato et al.\ 2009). 
\begin{table}
  \begin{center}
    \begin{tabular}[h!]{ l | c c c }
      sample & log~$M_\mathrm{min}/h^{-1}M_\odot$ & $M_1^{'}/M_\mathrm{min}$ & $\alpha$ \\
      \hline
      all      & $11.65\pm0.05$ & $16.0\pm1.0$ & $0.90\pm0.04$ \\
      barred   & $11.65\pm0.05$ & $16.0\pm1.0$ & $0.95\pm0.06$ \\
      unbarred & $11.65\pm0.05$ & $19.0\pm1.0$ & $0.80\pm0.06$ \\
      \hline
    \end{tabular}
  \end{center}
  \caption{Halo occupation distribution parameters for the model (\ref{censat}-\ref{NsatM}) fitted to the clustering measurements of the whole sample and the barred and unbarred subsamples, shown in Fig.~\ref{fig11}.}
  \label{table1}
\end{table}

As stated above, the halo mass threshold of the three measurements is %approximately 
the same.  Nonetheless, because of differences in the small-scale clustering, 
there are differences in the satellite HOD, $\langle N_\mathrm{sat}|M\rangle$ (Eqn.~\ref{NsatM}). 
In particular, firstly, the fraction of satellite galaxies varies.  
The whole catalogue has $f_\mathrm{sat}\approx23\%$, %which is also 
consistent with the mock catalogue analysis in Section~\ref{sec:mocktest},
which yielded a similar fraction (also $\approx23\%$). 
For comparison, the barred and unbarred subsamples have $f_\mathrm{sat}\approx25\%$ and $20\%$, 
respectively.

Secondly, the key difference between the well-fitting models for the barred subsample 
is that they have a steeper slope $\alpha$ (compared to the slope for the full sample 
and for unbarred galaxies), which means that the larger satellite fraction of barred 
galaxies is due to more satellites \textit{in more massive haloes}. 
In contrast, the well-fitting models for the unbarred subsample have a shallower slope, 
%($\alpha\approx0.8$; the value of $M_1^{'}/M_\mathrm{min}$ is slightly larger as well); 
so not only is the unbarred sample dominated by central galaxies in lower mass 
haloes, but the small fraction of unbarred satellites is not in the most massive haloes either.

\section{Conclusions and Discussion}\label{sec:discuss}

%\textbf{[add more weight to obs results.]} e.g. shuffling, colour/mass transition 

%Idit:  \S7  Needs a first summary parag ("We looked at...  We did this and that").
%Right now it's a mix of bulleted and paragraphs all discussing similar points. I actually prefer not having bullets (but rather a short parag on each), for a consistent look.

We selected a volume-limited catalogue of 15810 nearly face-on disc galaxies in the 
SDSS, which have visual morphology classifications from Galaxy Zoo~2. We analyzed the 
properties of galaxies with bars and bulges, characterizing bar and bulge likelihood 
with the $p_\mathrm{bar}$ and fracdeV parameters.  Using ``marked" two-point 
correlation functions, we quantified the environmental dependence of bar and bulge 
likelihood as a function of the projected separation between galaxies.

To conclude, the following are the main results of our paper:

% Bob: the text is too vague. For example, in the
%first bullet you say there is a strong correlation but do not give the sense of the
%correlation e.g. are there more bars with greater stellar mass? It would be good to
%give the reader that information - including any quantitative evidence if available.
%
%Bob: I'd also ensure you add any quantitative evidence you have e.g. quote the percentage of how much we think the color correlations are responsible for the bar-correlations. I think one of the cool conclusions is that ~50\% of the bar mark correlation is probably due to another parameter, which could be "internal" e.g. gas content. This is very trendy right now, as Karen will tell you, so it's really interesting that not all of this is due to "known" environmentally-dependent parameters like color, mass, etc. Also, we see a significant drop at very small scales which we can report as likely real and probably due to interactions killing discs.

\begin{itemize}
\item \textbf{Correlations of bars and bulges with colour and stellar mass}: 
We find a strong correlation 
between the bar likelihood ($p_\mathrm{bar}$) and optical colour and stellar mass, 
such that redder and more massive disc galaxies are up to twice as likely to have bars 
than their bluer low-mass counterparts, although there is considerable scatter in the 
correlation, especially at the red (high-mass) end.  We find similar correlations with 
bulge strength (fracdeV), but with less scatter.  The quantities 
$p_\mathrm{bar}$ and fracdeV appear to have a transition at the same stellar mass and 
colour ($M_\ast\approx2\times10^{10}\,M_\odot$, $g-r\approx0.6$). 

\item \textbf{Environmental dependence of bars and bulges}: We clearly detect and 
quantify the environmental dependence of barred galaxies and of bulge-dominated 
galaxies, such that barred and bulge-dominated disc galaxies tend to be found in denser 
environments than their unbarred and disc-dominated counterparts.  In particular, by 
analyzing $p_\mathrm{bar}$ and fracdeV marked correlation functions, we obtained 
\textit{environmental correlations that are statistically significant (at a level of 
${>6}\sigma$) on scales of 150 kpc to a few Mpc}.  %In addition, the $p_\mathrm{bar}$-environment correlation peaks at $r_p\sim400~\mathrm{kpc}/h$. 
From sparse sampling tests with our catalogue, we argue that the small number statistics of previous studies inhibited their detection of a bar-environment correlation.

\item \textbf{Contribution from colour and stellar mass to bar-environment 
correlation}: By accounting for the environmental dependence of colour and stellar 
mass, we argue that they contribute approximately $60\pm5~\%$ and $25\pm10~\%$, 
of the $p_\mathrm{bar}$-environment correlation, respectively.  
From a similar analysis of a mock galaxy catalogue, we argue that  
%We observed that 
the environmental dependence of $p_\mathrm{bar}$ appears to be partially ($50\pm10~\%$) 
due to the fact that redder galaxies, 
which are often barred, tend to be hosted by more massive haloes.  
Conversely, up to half of the bar-environment correlation is \textit{not} due to colour 
or stellar mass, and must be due to environmental influences or to another independent parameter (possibly gas content, or angular momentum distribution).
%By shuffling $p_\mathrm{bar}$ at a given optical colour or stellar mass and then measuring the clustering dependence on $p_\mathrm{bar}$, we observed that the $p_\mathrm{bar}$-environment correlation appears to be partially due to that of colour, but not to stellar mass, contrary to other studies. 

%\item \textbf{Mock catalogue analysis}: By analyzing a mock galaxy catalogue with similar selection criteria as the data, we similarly argue that much of the bar-environment correlation can be explained by that of colour.  Redder galaxies, which are often barred, tend to be hosted by more massive haloes, which in turn tend to reside in overdense environments. 

\item \textbf{Halo model analysis of clustering of barred galaxies}: Our analyses with a mock galaxy catalogue and 
halo occupation models suggest that barred galaxies are often either central galaxies 
in low-mass dark matter haloes ($M\sim10^{12}\,M_\odot$) or satellite galaxies in 
more massive haloes ($M\sim10^{13-14}\,M_\odot$, hosting galaxy groups). 
%Nonetheless, although barred galaxies appear to be common in groups and clusters, relatively close pairs of galaxies ($r_p<100\,\mathrm{kpc}/h$) are not more likely than average to be barred.
\end{itemize}

%\textbf{[here we can discuss our results and their implications for galaxy evolution.  for example, how is it that the formation of a bar can depend on the galaxy's environment?  what does it mean that so many satellites in groups \& clusters form bars (or that their bars aren't destroyed)?  and why is the environmental dependence of bars more due to colour than stellar mass?]}
%
We argue that the environmental dependence of galaxy colours can account for approximately a half 
($50\pm20~\%$) of the environmental correlation of $p_\mathrm{bar}$. The optical 
colours are correlated with star formation rate and age, as well as gas and dust content, 
all of which may be related to the presence of disc instabilities such as bars.  
We find that a galaxy's stellar mass and bulge component, on the other hand, do not 
appear to be strongly related to its likelihood of having a bar. 
This suggests that it is primarily older disc galaxies with lower star formation rates 
(which often reside in denser environments) 
that are able to form and maintain a stellar bar (see Masters et al.\ 2010b).  
%provided that they are not ``harassed" by neighbouring galaxies. 
%
%Nonetheless, although the group environment may be conducive to bar formation, the effects 
%of the environment should not be overemphasized, since bar formation is expected to be 
%closely related to dynamical instabilities and the angular momentum distribution within 
%the disc, which may be only indirectly affected by the galaxy's environment. 

%Conversely, we also argue that $\approx30-70~\%$ of the environmental correlation 
Conversely, this means that \textit{the remaining half of 
the environmental correlation of $p_\mathrm{bar}$ is not explained by 
the environmental dependence of colour or stellar mass}, suggesting that bar 
formation (or the lack of bar destruction) is likely influenced by the galaxy's 
environment, in addition to the effects described above. 
Bulge formation, which is to some extent independently correlated with 
the environment (see Section~\ref{disentangleMCFs}), is also expected to be 
affected by interactions and merger activity (Hopkins et al.\ 2009; Kannan et al., in prep.). 
%Toomre 1977; and Scannapieco \& Tissera 2003
% for effect on disk, see Kannan, Maccio', et al., 2012, ApJ, 746. 10

%\textbf{[move to introduction?]} 
%Our detection of a statistically significant correlation between bar likelihood and 
%the environment appears to be in conflict with some recent studies that have found 
%weak or no evidence of an environmental dependence of barred galaxies (Marinova et 
%al.\ 2009; Aguerri et al.\ 2009; Barazza et al.\ 2009; and another clustering study, 
%Li et al.\ 2009).  Nevertheless, as stated in Section~\ref{sec:clustering}, our 
%volume-limited catalogue (consisting of 15810 galaxies) is much larger than those of 
%these studies, and an advantage of our mark clustering statistics 
%An advantage of our clustering statistics is that they are 
%sensitive to environmental correlations %include a reference? 
%and one can analyze the entire catalogue at once, without attempting the difficult 
%classification of different environments (e.g., ``cluster", ``group", and ``field" 
%environments) with environmental measures that often have considerable scatter 
%between them (see Muldrew et al.\ 2011).  
%
%\textbf{[moved from intro.]} 
During the final stages of this work, Mart{\'i}nez \& Muriel (2011) in a related study 
found that the bar fraction does not significantly depend on group mass or 
luminosity, or on the distance to the nearest neighbour.  Their sample is 
smaller than ours, however, and is apparent magnitude-limited rather than volume-limited. 
In addition, they use bar classifications from Nair \& Abraham (2010), which 
include somewhat weaker bars than Galaxy Zoo~2 (see M11), which are bars that tend 
to be found in bluer galaxies (and hence in less dense environments). 
% Karen: It's not true I think that NA10 have weaker bars than those found using pbar>0.2 (as in the Appendix A of the gas paper).
In another recent paper, Lee et al.\ (2012a) also analyze the environmental dependence of 
bars, using bar classifications consistent with Nair \& Abraham (2010), and 
claim that the bar fraction does not depend on the environment at fixed colour or 
central velocity dispersion, contrary to our results.  
However, a crucial difference between these two analyses and ours is that they use 
environment measures that mix environments at different scales, while we analyze 
the environmental correlations as a function of galaxy separation.
%\textbf{[mention sparse sampling tests.]} 
% also mention Giordano et al. (2011) here or in intro?

%\textbf{[dependence on environment at different scales (mention Blanton \& Berlind; Muldrew et al.; Wilman et al.).]}
%
A particularly interesting result of this paper is the \textit{scale dependence 
of the environmental correlations} of bar and bulge likelihood (see Fig.~\ref{fig6}, 
Sec.~\ref{sec:clustering}). 
Environmental correlations should be interpreted differently at different scales 
(Blanton \& Berlind 2007; Wilman et al.\ 2010; Muldrew et al.\ 2012), 
as %such that 
galaxies at small separations ($r_p<2~\mathrm{Mpc}/h$) are often hosted by the same 
dark matter halo, while galaxies at larger separations are often hosted by separate 
haloes. We see that more massive haloes, which tend to reside in relatively dense 
environments, tend to host more disc galaxies with bars and bulges.

Moreover, the $p_\mathrm{bar}$-environment correlation peaks at 
$r_p\sim400~\mathrm{kpc}/h$, which suggests that many barred galaxies are central 
or satellite galaxies in groups and clusters.  That is, \textit{some aspect of group 
environment triggers the formation of bars}, 
%is conducive to the formation of (stable) bars, 
in spite of the fact that bars are often thought to form by 
internally driven secular processes (e.g., Kormendy \& Kennicutt 2004). 
Secular processes may sometimes be externally driven. 
For example, cosmological simulations predict that tidal interactions with dark matter substructures, which are common in such environments, could induce bar formation and growth 
(Romano-D\'{i}az et al.\ 2008; Kazantzidis et al.\ 2008). %cf., Weinberg \& Katz 2007?
On the other hand, the $p_\mathrm{bar}$-environment correlation is not significant for 
closer pairs, suggesting that the enhanced likelihood of galaxies being barred is erased 
if the galaxies are merging with each other; however, this measurement at 
small separations ($r_p<100~\mathrm{kpc}/h$) has large uncertainty and may be 
affected by fiber collisions (see Sec.~\ref{sec:cat}), so it should be 
viewed with caution.  Analyses of bars in close pairs of galaxies 
(e.g., Nair \& Ellison, in prep.) could shed more light on this issue.  

%\textbf{[effect of \textit{minor} mergers \& interactions (Noguchi 1996; Keel et al.\ 1996).]}  %recall Berkeley conversations
In general, we can at least conclude that group environments increase the likelihood 
of bar formation in disc galaxies.  
Minor mergers and interactions are relatively common in galaxy groups (Hopkins et al.\ 2010b), 
and tidal interactions with neighboring galaxies can trigger disc instabilities and subsequent 
bar formation (Noguchi 1996; Berentzen et al.\ 2004);  %see also Combes 2004
there is observational evidence for this as well (Elmegreen et al.\ 1990; Keel et al.\ 1996). %and Nair et al. in prep.
Tidal interactions can also affect the bar's pattern speed and other properties (Miwa \& Noguchi 1998). 
%as can interactions between the bar and dark matter halo (Miwa \& Noguchi 1998; Rautiainen et al.\ 2008). 
%Dimitri: So around 400 kpc the environment seems to be dense enough to trigger
%the disc instability and I think you could stress that this is
%important to understand how bars form. Perhaps you can come with a way
%to quantify the perturbation in the potential at this scale and this
%can be very useful for the theory of bar formation.
The evolution of barred galaxies in group environments and in minor mergers/interactions is clearly in need of further study.

%Karen: I very much like the discussion/conclusions, but I wonder if we want
%to end with a more general paragraph on the wider implications. For
%example are we suggesting here that secular and external processes are
%quite strongly linked, so that the latter can trigger the former, and
%even if more galaxy evolution than had previously been thought happens
%through secular evolution, this is often triggered by environment... ?
%We don't have to add this, but I just sometimes like that more general
%round up at the end.
%\textbf{[general comments about wider implications.  perhaps secular and external processes are linked, and some supposedly secular processes can actually be triggered by a galaxy's environment!]}
%We can also conclude that, 
Considering that bar formation does appear to depend 
on the host galaxy's environment, and that bars form by secular evolution, 
our results suggest that the dichotomy between internal secular processes and external 
environmental processes is not as strict as previously thought.  It is possible that some 
structural changes in galaxy discs may be triggered or influenced by the galaxy's environment. 
%\textbf{[clarify discussion of secular evol.]} %see GZ2barenviro_v4_secularevolution.txt
% Karen: Harassment sometimes described as secular evolution. Not sure if correct term. Ask Bill?
For example, Kormendy \& Bender (2012) recently argued that, ``harassment", the cumulative effect of encounters with satellite galaxies, may influence secular evolution. 
%cite Moore et al. 1996 for harassment?
Furthermore, ``strangulation", in which the hot diffuse gas around newly accreted satellites is stripped, removes the fuel for future star formation (Larson et al.\ 1980), and could contribute to more stable or growing bars (Berentzen et al.\ 2007; Masters et al.\ 2012).

%\textbf{relation between enviro dep of bars \& nuclear activity (McKernan, Hopkins, Quataert).}
In addition, our results could also indicate a link between bars and active galactic nuclei (AGN) activity.  We have shown that barred galaxies tend to reside in dense group environments, while galaxies hosting AGN also tend to be found in such environments (e.g., Mandelbaum et al.\ 2009; Pasquali et al.\ 2009). %also Li et al. 2006; Hopkins et al. 2007; Hickox et al. 2009
In some models, AGN are assumed to be fueled by recent mergers; however, some have also argued that bars and disc instabilities may be an internal mechanism through which low angular momentum gas is driven towards the nucleus (Bower et al.\ 2006; Hopkins \& Quataert 2010; McKernan et al.\ 2010).  
Nonetheless, a correlation between barred galaxies and AGN activity has not been detected observationally (Lee et al.\ 2012b; Cardamone et al., in prep.). 
%maybe mention Lee et al. (2012)
%also maybe De Lucia et al. 2011, maybe Schawinski+, and Hopkins et al. 2010, MNRAS, 401, 1131

%\textbf{[we also should mention constraints on galaxy formation models, such as Benson \& Devereux and De Lucia et al.]}
%
Lastly, we note that our results can be used to constrain galaxy formation models, 
such as the semi-analytic models of Benson \& Devereux (2010) and De Lucia et al.\ 
(2011), and the hydrodynamic simulations of Heller et al.\ (2007), Croft et al.\ (2009), Sales et al.\ (2012). 
%and the simulations of Debattista et al.\ (2006) and Heller et al.\ (2007). 
Marked correlation functions, and marked clustering statistics in general, are 
sensitive to environmental correlations at different scales, such that small changes 
in model parameters could yield environmental dependencies of galaxy bars and bulges 
that can be compared to %are consistent with or ruled out by our 
measurements with Galaxy Zoo (see Figures~\ref{fig6}-\ref{fig9}). 
In addition, a result from our halo model analysis is that barred galaxies tend to be central 
galaxies in lower mass haloes ($M_\mathrm{halo}\sim10^{12}~M_\odot$) and satellite galaxies 
in more massive haloes ($M_\mathrm{halo}\sim10^{13}~M_\odot$), which can also be compared to 
other models.

\section*{Acknowledgments}

We thank Sara Ellison, Preethi Nair, and Dimitri Gadotti for valuable discussions about our results. 
KLM acknowledges funding from The Leverhulme Trust as a 2010 Early Career Fellow.  
RCN and EME acknowledge STFC rolling grant ST/I001204/1 ``Survey Cosmology and Astrophysics" for support. 
IZ acknowledges support by NSF grant AST-0907947. 
BH acknowledges grant number FP7-PEOPLE- 2007- 4-3-IRG n 20218. 
KS acknowledges support provided by NASA through Einstein Postdoctoral Fellowship grant number PF9-00069 issued by the Chandra X-ray Observatory Center, which is operated by the Smithsonian Astrophysical Observatory for and on behalf of NASA under contract NAS8-03060.

We thank Jeffrey Gardner, Andrew Connolly, and Cameron McBride
for assistance with their $N$tropy code, which was used 
to measure all of the correlation functions presented here.
$N$tropy was funded by the NASA Advanced Information Systems 
Research Program grant NNG05GA60G.

%Galaxy Zoo acknowledgement.
This publication has been made possible by the participation of more than 200000 volunteers in the Galaxy Zoo project. Their contributions are individually acknowledged at 
\texttt{http://www.galaxyzoo.org/volunteers}. %http://www.galaxyzoo.org/Volunteers.aspx 
Galaxy Zoo is supported by The Leverhulme Trust.

%\textbf{Also include SDSS acknowledgement?}
Funding for the SDSS and SDSS-II has been provided by the 
Alfred P. Sloan Foundation, the Participating Institutions, 
the National Science Foundation, the U.S. Department of Energy, 
the National Aeronautics and Space Administration, 
the Japanese Monbukagakusho, the Max Planck Society, 
and the Higher Education Funding Council for England. 
The SDSS was managed by the Astrophysical Research     
Consortium for the Participating Institutions. 
%The SDSS Web Site is http://www.sdss.org/.

The SDSS is managed by the Astrophysical Research Consortium for the 
Participating Institutions. The Participating Institutions are the 
American Museum of Natural History, Astrophysical Institute Potsdam, 
University of Basel, University of Cambridge, Case Western Reserve 
University, University of Chicago, Drexel University, Fermilab, the 
Institute for Advanced Study, the Japan Participation Group, Johns 
Hopkins University, the Joint Institute for Nuclear Astrophysics, the 
Kavli Institute for Particle Astrophysics and Cosmology, the Korean 
Scientist Group, the Chinese Academy of Sciences (LAMOST), Los Alamos 
National Laboratory, the Max-Planck-Institute for Astronomy (MPIA), 
the Max-Planck-Institute for Astrophysics (MPA), New Mexico State 
University, Ohio State University, University of Pittsburgh, 
University of Portsmouth, Princeton University, the United States 
Naval Observatory, and the University of Washington.

\appendix

\renewcommand{\thefigure}{\Alph{appfig}\arabic{figure}}
\setcounter{appfig}{1}

\section{Errors of the Clustering Measurements}\label{app:JK}

As discussed in Section~\ref{sec:markstats}, we use jack-knife resampling to estimate 
the statistical errors of our clustering measurements $w_p(r_p)$ and $M_p(r_p)$ 
(Figures~\ref{fig6}-\ref{fig9}). 
The jack-knife errors of our marked projected correlation functions tend to be larger 
than the Poisson errors (not shown). 
We note that Norberg et al. (2009) has shown that the jack-knife method does not 
recover the scale dependence of errors of the (unmarked) correlation function, often 
overestimating the errors at small scales, and the results are sensitive to the 
number of sub-catalogues into which the data is split.
Although our uncertainty estimates are important, our primary results are not 
particularly sensitive to the precise value of the errors.  
We have performed additional error analyses with twice as many 
jack-knife sub-catalogues, and obtained similar errors (within 10\%) at all scales 
for both the correlation functions and marked correlation functions.

\begin{figure}
 %\centering
 \includegraphics[width=\hsize]{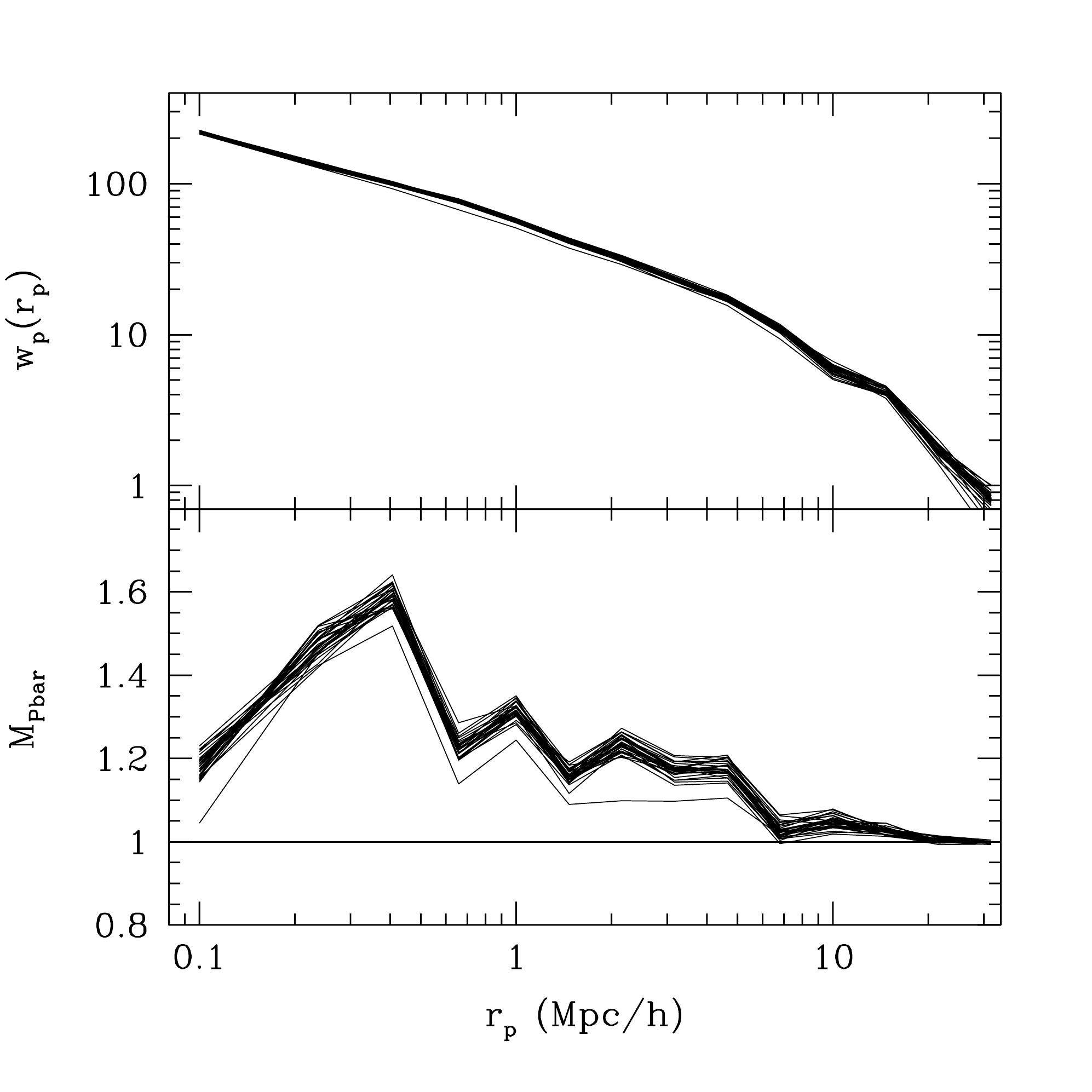} %{newsample_Mr194_PbarMark_JKsubsamples.ps}
 \caption{Jack-knife clustering measurements for Fig.~\ref{fig6}a.}
 \label{figA1}
\end{figure}

We estimate the error in each $r_p$ (projected galaxy separation) bin by computing 
the variance of the measurements of the jack-knife sub-catalogues. 
For example, clustering measurements of the thirty jack-knife sub-catalogues 
used to estimate the errors of the measurements in Figure~\ref{fig6}a 
are shown in Figure~\ref{figA1}. 
%
% newsample_Mr194_(Pbar/fracdeV)Mark_JKsubsamples.ps
Most of the jack-knife sub-catalogues yield similar measurements, although a single outlying 
measurement is responsible for a significant fraction of the error. 
Outliers may be due to anomolously large structures or voids, or in the case 
of mark clustering, to noise in the marks. 
Nonetheless, the outlying measurement is not an extreme outlier, and the 
Sloan Great Wall, an extremely massive superstructure that can influence clustering measurements 
(Zehavi et al.\ 2011; Norberg et al.\ 2011) is beyond our upper redshift limit ($z<0.060$). 
The jack-knife measurements for the lower panel of Figure~\ref{fig6}b (with the fracdeV mark) are 
similar, also with a single outlier. 
%For the error bars shown in Figures~\ref{fig6}-\ref{fig9}, this outlier was excluded, in order to avoid overestimating the errors. 
If the outlier were excluded, the resulting jack-knife error estimates 
would be lower by $21\%$ on average for $w_p$, by $18\%$ for $M_\mathrm{Pbar}$, and by 
$25\%$ for $M_\mathrm{fracdeV}$.  %In particular, with the outlier, the errors 
%increase substantially on scales of $400\,\mathrm{kpc}/h\leq r_p\leq3\,\mathrm{Mpc}/h$, 
%such that the environmental correlations in Figures~\ref{fig6} are of high 
%statistical significance especially at $180\,\mathrm{kpc}/h\leq r_p\leq1.8\,\mathrm{Mpc}/h$

\begin{figure}
 %\centering
 \includegraphics[width=\hsize]{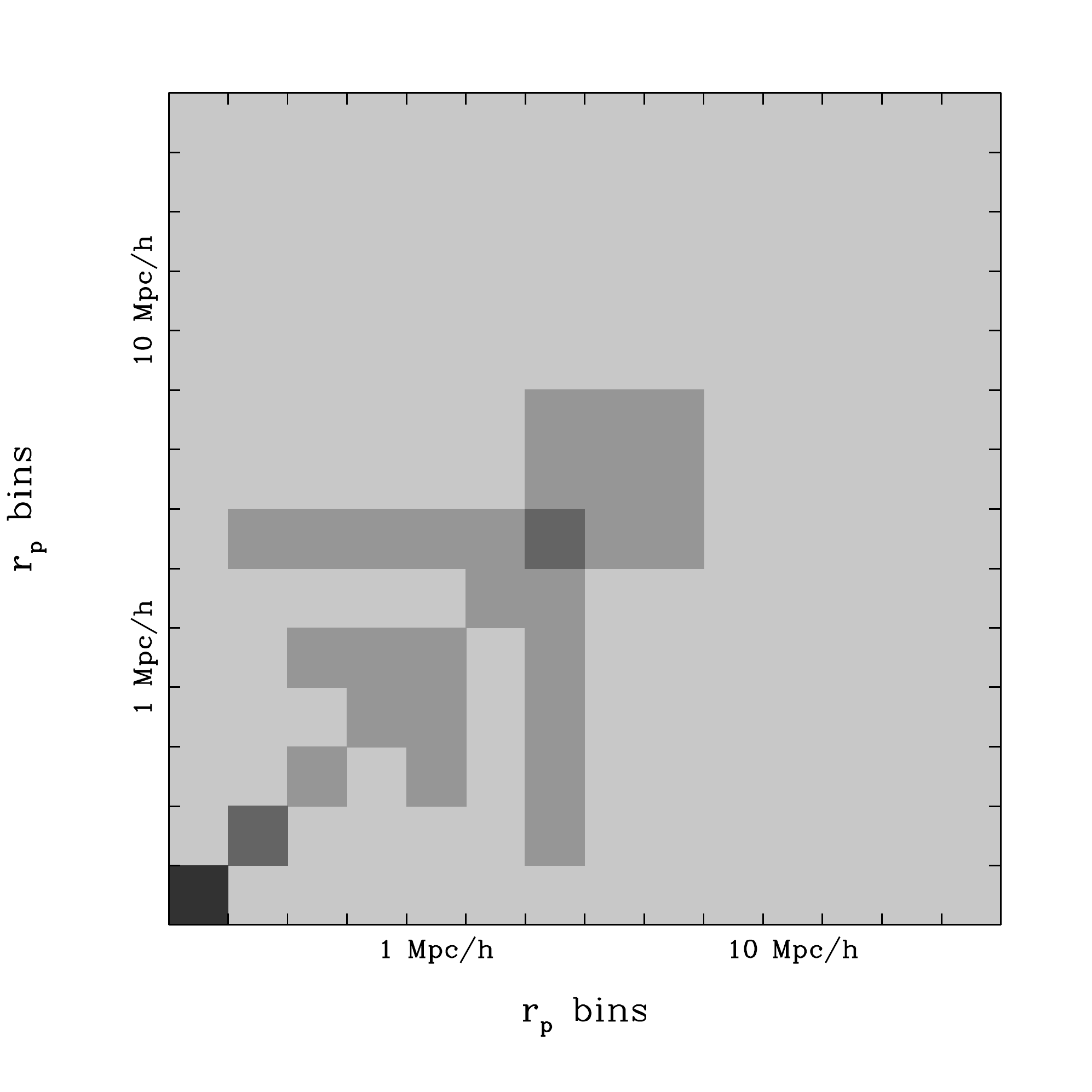} %{Mp_Pbarmark_Mr194_Covar_labels.ps} %Mp_Pbarmark_Mr194_Covar.ps
 \caption{Jack-knife covariance matrix %$\mathrm{Covar}(r_{p,i},r_{p,j})$ 
   for the $M_\mathrm{Pbar}(r_p)$ measurement of Fig.~\ref{fig6}a.  
   The grayscale as a function of galaxy separation indicates regions at which the measurement errors are correlated, such that darker regions have more correlated measurements. }
 \label{figA2}
\end{figure}

We also compute jack-knife covariance matrices (Eqn.~\ref{covar}) for each of the 
clustering measurements.  We show the covariance matrix of the measurement in the 
lower panel of Figure~\ref{fig6}a ($p_\mathrm{bar}$ mark) in Figure~\ref{figA2}. 
%
% Mp_(Pbar/fracdeV)mark_Mr194_Covar.ps
Most of the errors are not strongly correlated, although the $r_p$ bins 
centered at 1 and $2\,\mathrm{Mpc}/h$ are weakly correlated with smaller-scale bins 
(most of which have stronger mark correlations), 
which may explain why the mark correlation measurements of these bins 
are slightly larger than in the neighboring $r_p$ bins.

In any case, the range of jack-knife clustering measurements in Figure~\ref{figA1} 
is not extremely large, and the correlations between the $r_p$ bins in Figure~\ref{figA2} 
are not extremely strong, so we conclude that the measured mark correlation functions 
are robust, as are the inferred environmental correlations.

\label{lastpage}

\end{document}